\documentclass[twocolumn,epjc3]{svjour3}  

\journalname{Eur. Phys. J. C}
\usepackage[utf8]{inputenc} 
\usepackage[T1]{fontenc}    
\usepackage{cite}
\usepackage{amsmath}
\usepackage{enumitem}
\usepackage{float}
\usepackage[T1]{fontenc}
\usepackage{color}
\usepackage{graphicx}
\usepackage[utf8]{inputenc}
\usepackage{xcolor}
\usepackage[colorlinks = true,
            linkcolor = blue,
            urlcolor  = blue,
            citecolor = blue,
            anchorcolor = blue]{hyperref}

\newcommand{\herwig}{\textsc{Herwig}}
\newcommand{\pythia}{\textsc{Pythia}}

\newcommand{\ptcutwb}{$p_T^{\mathrm{cut}}$}
\newcommand{\ptcut}[1]{$p_T^{\mathrm{cut}} = #1 $~GeV}
\newcommand{\instruction}[1]{}

\newcommand{\qgsp}{$\Delta_{[q,g]}$}
\graphicspath{ {./figures/} }

\def\makeheadbox{
}
\date{}
\begin{document}
\title{Novel approach to measure quark/gluon jets at the LHC}
\author{Petr Baro\v{n}\thanksref{e1,addr1}
        \and
        Michael H. Seymour\thanksref{e2,addr2}
        \and
        Andrzej Si\'{o}dmok\thanksref{e3,addr3}
}
\institute{Institute of Nuclear Physics, Polish Academy of Sciences,
        ul. Radzikowskiego 152, 31-342 Krak\'{o}w, Poland 
        \label{addr1}
        \and
          Lancaster-Manchester-Sheffield Consortium for Fundamental Physics, 
        Department of Physics and Astronomy, \\University of Manchester, M13 9PL, U.K.
        \label{addr2}
        \and
        Jagiellonian University, 
        ul. prof. Stanislawa \L{}ojasiewicza 11, 30-348 Krak\'{o}w, Poland
        \label{addr3}
}
\thankstext{e1}{e-mail:  Petr.Baron@ifj.edu.pl}
\thankstext{e2}{e-mail:  Michael.Seymour@manchester.ac.uk}
\thankstext{e3}{e-mail:  Andrzej.Siodmok@uj.edu.pl}

\maketitle
\begin{abstract}
{%
In this paper, we present a new proposal on how to measure quark/gluon 
jet properties at the LHC. The measurement strategy takes advantage of the 
fact that the LHC has collected data at different energies.
Measurements at two or more energies can be combined to yield distributions of any jet property separated into quark and gluon jet samples on a statistical basis, without the need for an independent event-by-event tag.
We illustrate our method with a variety of different angularity observables, and discuss how to narrow down the search for the most useful observables.
}
\end{abstract}
\section{Introduction}
\label{Intro}
Experimentally, we can study partons (quarks and gluons) by analyzing jets 
(narrow, energetic sprays of particles) whose kinematic characteristics mirror those of an 
initiating parton that cannot be directly measured.
By employing an appropriate jet definition, it becomes possible to establish a link between 
jet measurements obtained from clusters of hadrons and calculations performed on clusters of partons.
In a more ambitious approach, it is conceivable to attempt jet tagging with a well-defined flavour label, 
thereby increasing the proportion of, for instance, gluon-tagged jets compared to quark-tagged jets.
The capacity to differentiate quark jets from gluon jets on an event-by-event basis has the potential to considerably increase
the scope and sensitivity of numerous new-physics studies at the Large Hadron Collider (LHC)
~\cite{Gallicchio:2011xq,FerreiradeLima:2016gcz,Whitmore:2019cri,Bhattacherjee:2016bpy,Li:2023tcr,Wong:2023vpx}.
This is because Beyond the Standard Model signals are often dominated by quarks 
while the corresponding Standard Model backgrounds are dominated by gluons~\cite{Gallicchio:2012ez,Sakaki:2018opq}. 

As well as proposing an observable that can distinguish quark jets and gluon jets~\cite{Fodor:1989ir,Pumplin:1991kc,Stewart:2022ari,Metodiev:2018ftz,Frye:2017yrw,Davighi:2017hok,Goncalves:2015prv,Dreyer:2021hhr,Bright-Thonney:2018mxq,Bhattacherjee:2015psa,Komiske:2018vkc,Komiske:2022vxg}, any quantitative analysis must also propose how to calibrate that observable by independently tagging quark and gluon jet samples.
In some studies, this has been done by calibrating against Monte Carlo samples in which the ``truth'' flavour of the jet is known.
However, one might worry about whether event generators make sufficiently reliable predictions of these flavour-dependent properties~\cite{Gras:2017jty,Reichelt:2017hts,Siodmok:2017oyv} and, indeed, this is something one would like to test against the data.
In other studies, another method is used to tag the jet flavour, for example the hard process dependence~\cite{ATLAS:2014vax,Aggleton:2021fcw}, and used to calibrate the measurement of the proposed observable.
Here, one would worry that the two tagging methods are correlated, yielding a biased measurement of the jet property.

In this paper, we study a variety of angularity observables as measures of quark/gluon jet differences.
The main new ingredient we propose is the calibration of those differences using the dependence on the \emph{centre-of mass energy of the Large Hadron Collider}.
The idea is that the properties of jets of a given flavour
and transverse momentum, if suitably defined, are almost entirely independent of
the jet's production mechanism, i.e.\ its rapidity, the energy of the collision,
the colliding beam types, parton distributions, etc\cite{Seymour:1994by}, but
the fraction of jets of a given flavour at fixed transverse momentum does depend
on all those factors, in particular the collision energy. 
However, those jet fractions can be reliably predicted.
Thus, the energy-dependence can be used to extract the flavour-dependent properties on a statistical basis.

This paper is organised as follows: in Section~\ref{sec:strategy}, we present 
the measurement strategy; then, in Section~\ref{sec:syseff}, we discuss important 
systematic effects; in Section~\ref{sec:measures}, we present measures that determine 
the quality of observables; in Section~\ref{sec:results} we provide 
the main results, and finally in Section~\ref{sec:summary} we summarise the study. 
\section{Measurement strategy}
\label{sec:strategy}
The LHC has collected data at many different energies: $900$~GeV, $2.36$~TeV, $5.02$~TeV, $7$~TeV, $8$~TeV and 
$13$~TeV and will also take data at $14$~TeV. There is great potential to significantly 
increase the research potential of LHC by constructing new experimental strategies 
that exploit this unique situation. A measurement strategy based on the flexibility 
of the LHC to run at variable beam energies has already been successfully studied in 
the case of measuring the mass of the $W$ boson~\cite{Krasny:2007cy,Krasny:2010vd} 
and the difference in the mass of the $W^+$ and $W^-$ bosons \cite{Fayette:2009yvt}. 
In these publications, this flexibility was shown to be helpful in defining 
observables that are insensitive to ambiguities in the modelling, as well as in 
minimising the impact of systematic errors in the $W$-boson mass measurement.
This was achieved through the construction of observables that included the ratio of 
physical quantities measured at different energies. 
A second example of this type of measurement that we are aware of is~\cite{Mangano:2012mh}, in which the authors used both the ratios of cross sections, and the ratios of cross-sectional ratios between different centre-of-mass energies at the LHC to study the possibilities for precise measurements and BSM sensitivity.
Sadly, despite many advantages, the idea of using LHC data collected at different
energies to construct new robust observables was not exploited almost at all at LHC. 
The aim of this study is to change the situation and use this unique opportunity to 
construct new observables that are sensitive to the differences between quark and 
gluon jets.

The Tevatron collider also ran at different energies: $630$ GeV and $1.8$ TeV.
The Tevatron experiments, CDF and D0, exploited this to some extent for parton
distribution function measurements, but to our knowledge, only one analysis of
final-state jet properties that combined measurements at the two energies was
published\cite{D0:2001ab,D0:2001nam}. This followed essentially the same method
we will apply to LHC events below, to extract the distribution of subjets within
quark and gluon jets.

In leading-order QCD, the fraction of final-state jets that are of gluon origin increases
with decreasing
\begin{equation}
\label{eq:x}
 x\sim~p_T/\sqrt{s},   
\end{equation}
where $p_T$ is the transverse 
momentum of a jet, $\sqrt{s}$ is the proton-proton collision energy, and $x$ is the momentum
fraction of the initial-state partons within the proton. This is mainly due to the
$x$ dependence of the parton distribution function (PDF). For fixed $p_T$, 
the gluon jet fraction therefore increases when $\sqrt{s}$ is increased.
This suggests an 
experimentally accessible way to define jet samples with different mixtures of quarks 
and gluons by varying $\sqrt{s}$. The main advantage of this approach is that the 
construction of an observable at different energies allows a single set of experimental 
cuts to be used to select jets, keeping all detector parameters unchanged, and, in this 
way, reducing many systematic errors. Let us provide one example of how the measurement 
can be biased when we use different selection criteria for quark and gluon samples%
\cite{OPAL:1991ssr,ATLAS:2014vax,CMS:2021iwu}.
Based on the colour factor we could naively expect the ``quark'' jets to be 
much narrower than the ``gluon'' jets. However, the Monte Carlo simulations 
show that a high $p_T$ quark jet is narrower than a low $p_T$ jet, biasing the entire 
measurement if we define quark- and gluon-enriched samples using
different $p_T$ ranges.
At a more subtle level, even for jets at a given value of $p_T$ and rapidity, it
might be thought that a cut on the rapidity of the recoiling jet in the dijet pair
could be used to vary the quark to gluon mix (the so-called ``same-side opposite-side''
method). However, as shown in \cite{Forshaw:1999iv}, colour-coherence effects in the
hard process mean that the properties of a quark or gluon jet of fixed kinematics
(specifically, the amount of soft radiation into it) respond to the rapidity of
the other jet and only in an inclusive sample are the jet properties of a given flavour
independent of the collider energy, type, etc.
The strategy we present here, by construction, will be almost free from such bias.

There are many ways to define quark and gluon-jet discrimination observables~~\cite{Fodor:1989ir,Pumplin:1991kc,Stewart:2022ari,Metodiev:2018ftz,Frye:2017yrw,Davighi:2017hok,Goncalves:2015prv,Dreyer:2021hhr,Bright-Thonney:2018mxq,Bhattacherjee:2015psa};
however, we follow~\cite{Gras:2017jty} and use five generalised angularities 
$\lambda^{\kappa}_{\beta}$~\cite{Larkoski:2014pca}:
\begin{equation*}
\arraycolsep=5pt
\begin{array}{cccccc}
\label{eq:ang}
(\kappa,\beta)&(0,0) & (2,0) & (1,0.5) & (1,1) & (1,2) \\
\lambda^{\kappa}_{\beta}: 
               & \text{multiplicity} &  p_T^D &  \text{LHA} & \text{width} & \text{mass}
\end{array}
\end{equation*}
Here, multiplicity is the hadron multiplicity within the
jet, $p_T^D$ was defined in
\cite{Pandolfi:1480598,Chatrchyan:2012sn}\footnote{
In \cite{Pandolfi:1480598,Chatrchyan:2012sn},  $p_T^D
\equiv\sqrt{{\frac{\sum_{j \in \text{jet}}p^2_{Tj}}{(\sum_{j \in \text{jet}} p_{Tj})^2}}}=\sqrt{\sum_{i \in \text{jet}} z_i^2}$. 
Therefore, what we call $p_T^D$ in our work is actually its square (in the same way that
what we call mass is actually proportional to the mass-squared of the jet).}, LHA refers to the
``Les Houches Angularity'' (named after the workshop venue where this study was initiated \cite{Badger:2016bpw}),
width is closely related to jet broadening
\cite{Catani:1992jc,Rakow:1981qn,Ellis:1986ig}, and mass is closely
related to jet thrust \cite{Farhi:1977sg}.
In general an angularity is defined as $\lambda^{\kappa}_{\beta} = \sum_{i \in \text{jet}} z_i^\kappa \theta_i^\beta,$
where $i$ runs over the constituents of the jet particles, $z_i \equiv \frac{p_{Ti}}{\sum_{j \in \text{jet}} p_{Tj}}  
\in [0,1]$ is a transverse momentum fraction, $\ \theta_i \equiv \frac{R_{i \hat{n}}}{R} \in [0,1]$ here $R_{i \hat{n}}$
is the rapidity-azimuth distance to the jet axis and $R$ is the jet-radius
parameter.

Let $\lambda$ denote  an angularity in a jet from a mixed sample of quark and gluon jets
and $\lambda_i$ denote the value of one bin of a normalised histogram of $\lambda$.
i.e.\ $\lambda_i$ is the value of $\frac1n\,\frac{\mathrm{d}n}{\mathrm{d}\lambda}$
averaged over the $i$th bin, where $n$ is the number of jets.
We can express 
$\lambda_i$ as a linear combination of the angularity distribution in a gluon $\lambda_{gi}$ and quark 
$\lambda_{qi}$ jet:
\begin{equation}
 \lambda_i = f\lambda_{gi}+(1-f)\lambda_{qi}
\end{equation}
The coefficients are the fractions of gluon $f$ and quark ($1-f$) jets in the mixed 
sample. Let us consider that we measure just two similar samples of jets at two 
different energies $s_1$ and $s_2$ and we assume that $\lambda_{gi}$ and $\lambda_{qi}$
are independent of $\sqrt{s}$ (we will return to this assumption later). We then obtain:
\begin{equation}
 \lambda_{qi}=\frac{f^{s_1}\lambda^{s_2}_i-f^{s_2}\lambda^{s_1}_i}{f^{s_1}-f^{s_2}}
 \label{eq:quarkang}
\end{equation}
and 
\begin{equation}
 \lambda_{gi}=\frac{(1-f^{s_2})\lambda^{s_1}_i-(1-f^{s_1})\lambda^{s_2}_i}{f^{s_1}-f^{s_2}}\,,
 \label{eq:gluonang}
\end{equation}
where $\lambda^{s_1}_i$ and $\lambda^{s_2}_i$ are experimental measurements in mixed jet 
samples at $\sqrt{s_1}$ and $\sqrt{s_2}$, and $f^{s_1}$ and $f^{s_2}$ are fractions 
of gluon jets in the two samples. The jet fraction is provided by 
Monte Carlo simulation. 
\subsection{Event Selection}
\label{sec:cuts}
To prepare the most efficient measurement strategy, we study the production of dijets 
at the LHC at different energies: $900$ GeV, $2.36$ TeV, $7$ TeV and 
$13$ TeV.
The samples were generated using two different Monte Carlo generators: 
\herwig{ 7.2.2}~\cite{Bellm:2015jjp, Bellm:2019zci} with its default settings with 
PDF set MMHT2014lo68cl~\cite{Harland-Lang:2014zoa},
\pythia{ 8.240}~\cite{Sjostrand:2014zea,Bierlich:2022pfr} using its default settings with PDF set NNPDF2.3 QCD+QED LO~\cite{Ball:2013hta} and 
 the jets were reconstructed using the Anti-$k_{T}$ algorithm~\cite{Cacciari:2008gp} implemented 
in the FastJet package~\cite{Cacciari:2005hq,Cacciari:2011ma}. We require exactly two jets that satisfy the following criteria: 
\begin{equation}
    p_{T~\mathrm{sublead}} / p_{T~\mathrm{lead}} > 0.8
\end{equation}
and
\begin{equation}
    (p_{T~\mathrm{lead}}+p_{T~\mathrm{sublead}})/2> p_T^{\mathrm{cut}},
\end{equation}
 where $p_{T~\mathrm{lead}}$ is the transverse momentum of the leading jet
 and $p_{T~\mathrm{sublead}}$ is the transverse momentum of the subleading jet.
We investigated five different jet radii $R = 0.2$, $0.4$, $0.6$, $0.8$, $1.0$, four 
different transverse momentum cuts \ptcutwb{} $=50$, $100$, $200$ and $400$ GeV.
In addition to directly measuring the
angularities, we also want to test the impact of jet grooming (see
e.g.~\cite{Butterworth:2008iy,Ellis:2009su,Ellis:2009me,Krohn:2009th}).
As one grooming example, we use the modified mass drop tagger (MMDT)
with $\mu = 1$ \cite{Butterworth:2008iy,Dasgupta:2013ihk}
(equivalently, soft drop declustering with $\beta = 0$
\cite{Larkoski:2014wba}) and $z_{\rm cut} = 0.1$.

\subsection{Example of deriving q/g multiplicity $\lambda_{0}^{0}$ ($R = 0.4$, \ptcut{100}) 
 using $\sqrt{s}=$ 900--13000~GeV}
\label{sec:example}
To determine the multiplicity of charged particles in gluon and quark jets we perform 
the following steps:
\begin{figure*}[ht!]
\begin{minipage}{\textwidth}
    \centering
    \includegraphics[width=0.495\textwidth]{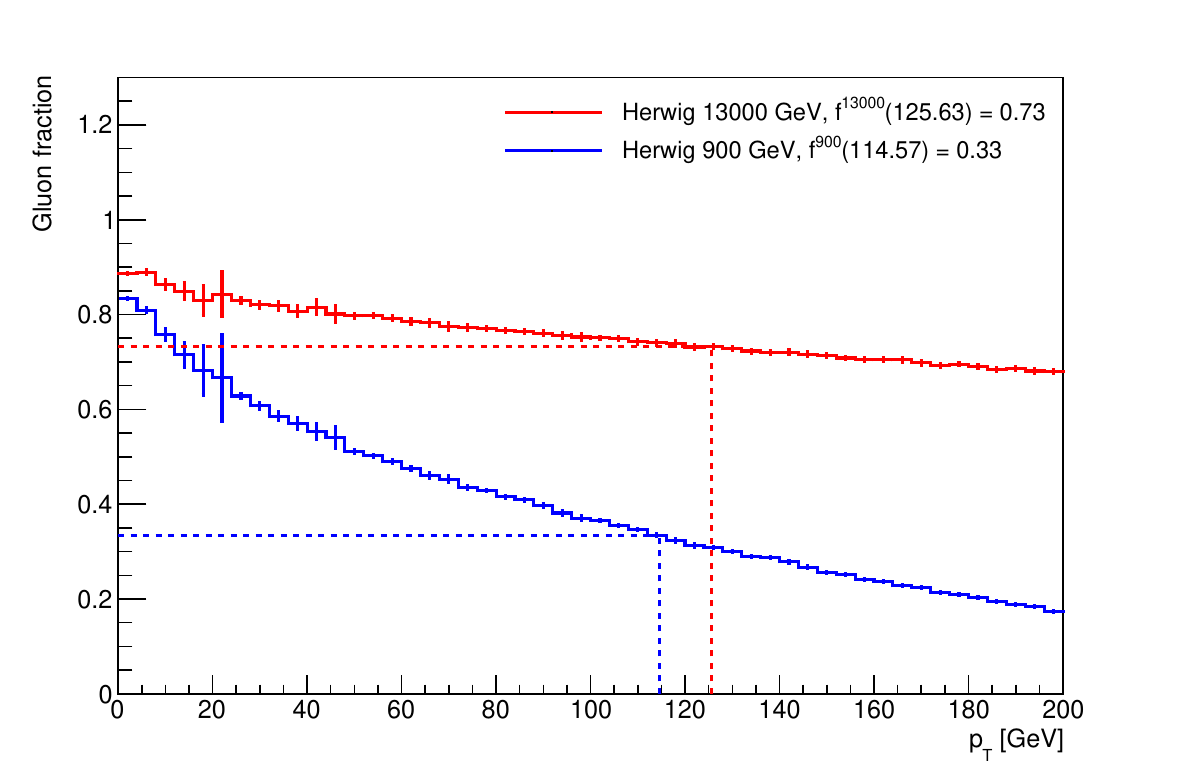}
    \includegraphics[width=0.495\textwidth]{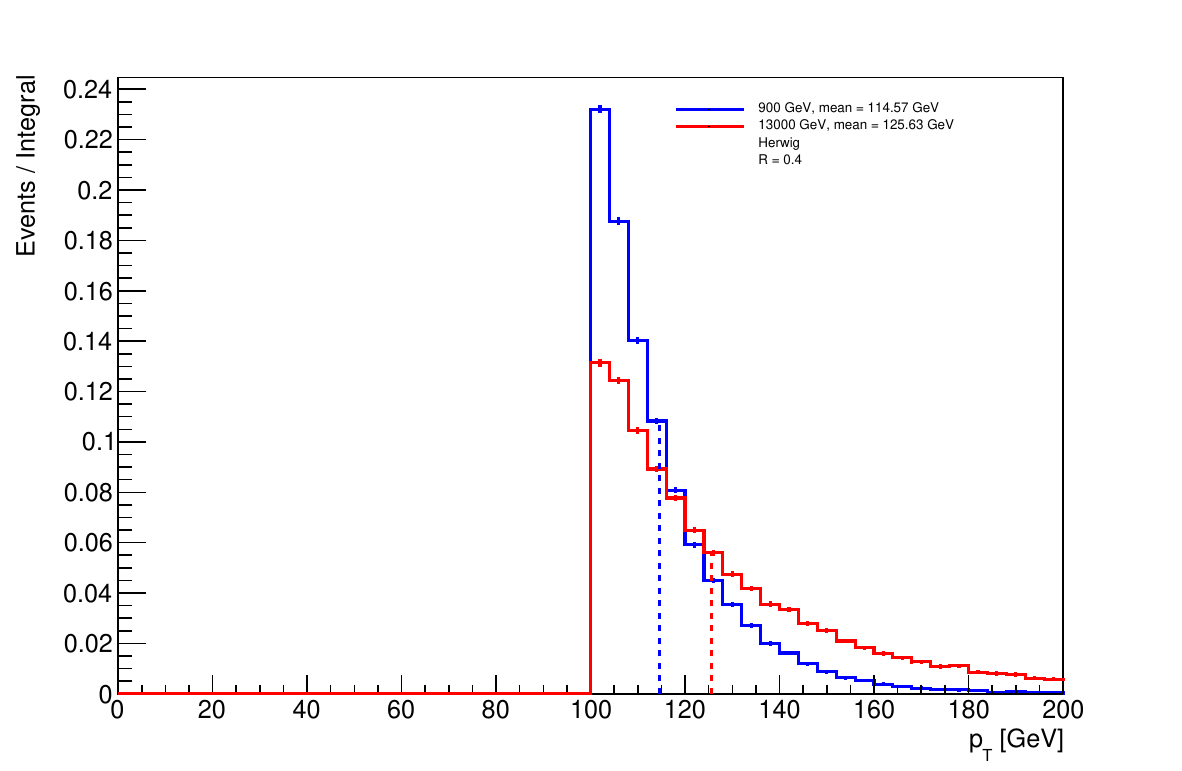}
    \end{minipage}
    \caption{Left panel: gluon fractions obtain from \herwig{'s} simulation of 
    proton-proton dijet process without hadronization and parton showering at 
    $\sqrt{s}= 900$~GeV $f^{900}$ (blue solid line) and  $\sqrt{s}= 13000$~GeV $f^{13000}$ 
    (red solid line).
    Dashed lines show the chosen values $f^{900}$ 
    and $f^{13000}$ for the point at the mean of the jet $p_{T}$ distributions.  
    Right panel: Normalised transverse momentum of the leading and subleading jets 
    at energy 900 and 13000~GeV. Dashed lines represent the mean of the distributions 
    used to evaluate the coefficients of the gluon fraction.}
    \label{fig:pdf}
\end{figure*}
\subsection*{Step 1: Derive gluon fraction -- parton level simulation 
(i.e. without hadronization and parton shower)}
\label{sec:step1}
By disabling hadronization and parton showering in the Monte Carlo generators 
\herwig{ 7} and 
\pythia{ 8}  
the gluon fraction was defined as a function of $p_{T}$
\begin{equation}
f(p_{T}) = \frac{N_{\mathrm{gluons}}(p_{T})}{N_{\mathrm{gluons}}(p_{T}) 
            + N_{\mathrm{quarks}}(p_{T})}
\end{equation}
where $N$ represents the number of partons (quarks or gluons). In the left panel of 
Figure~\ref{fig:pdf} we show examples of gluon fractions as a function of transverse 
momentum $f(p_{T})$ at $\sqrt{s}=$ $900$~GeV and $13000$~GeV of \herwig{} (solid lines). 
We have also performed a more sophisticated approach with the distribution of gluon 
fractions as a 2D map in $p_{T}$ and pseudorapidity $\eta$ of a jet $f(p_{T}, \eta)$. 
However, no significant differences were observed in the resulting quark and gluon 
angularities. Therefore, for simplicity, the strategy of taking the mean of the jet 
$p_{T}$ distributions is used to obtain numerical results in the following sections.
\subsection*{Step 2: Evaluate the scaling coefficients $f^{900}$ and $f^{13000}$}
In the right panel of Figure \ref{fig:pdf} we show the $p_{T}$ distributions of 
jets ($R = 0.4$) that passed the event selection cuts defined in Section~\ref{sec:cuts}
obtained by running a complete Monte Carlo simulation (including hadronization and 
parton shower) at two different collision energies $900$ and $13000$~GeV. 
The transverse momentum mean $\langle p_{T} \rangle$ of the jet distribution for 
the two energies is as follows: 
\begin{itemize}
    \item jet $p_{T}$ ($\sqrt{s}=900$~GeV) $\rightarrow$ $\langle p_{T} 
          \rangle = 114.57$~GeV,
    \item jet $p_{T}$ ($\sqrt{s}=13$~TeV) $\rightarrow 
    \langle p_{T} \rangle = 125.63$~GeV.
\end{itemize}
The scaling coefficients $f^{900}$ and $f^{13000}$, as illustrated by the dashed 
lines in Fig.~\ref{fig:pdf}, are obtained using the gluon fractions of the left 
panel at the $\langle p_{T} \rangle$ derived from the right panel of 
Figure~\ref{fig:pdf}:
\begin{eqnarray}
    f^{900} &=& f^{900}(\langle p_{T} \rangle ) = f^{900}(114.57~\mathrm{GeV})\!=\!0.33,\\
  \!f^{13000} &=& f^{13000}(\langle p_{T} \rangle ) = f^{13000}(125.63~\mathrm{GeV})\!=\!0.73.\phantom{mm}
\end{eqnarray}
\subsection*{Step 3: Derive q/g angularities}
\begin{figure*}[ht!]
\centering
\includegraphics[width=0.8\textwidth]{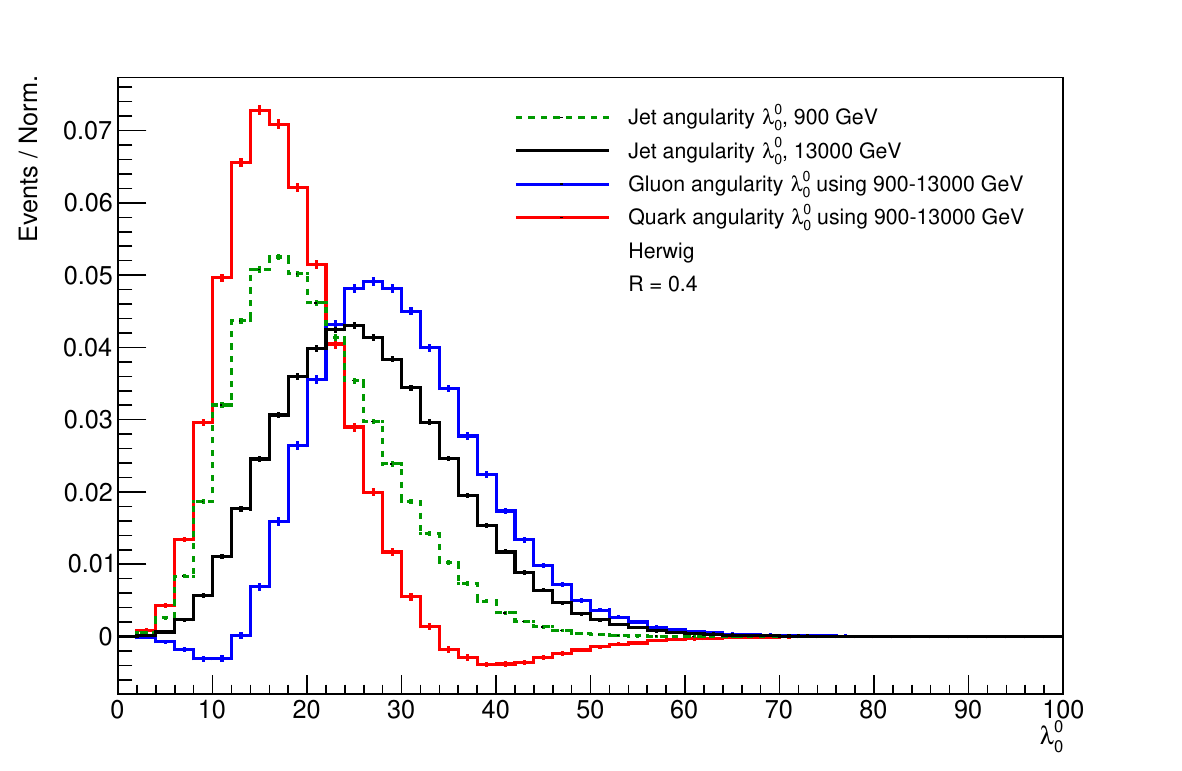}
\caption{Derived distributions of quark and gluon angularity (multiplicity) $\lambda_q$ (red line) 
 and $\lambda_g$ (blue line) as linear combinations of those measured 
 at different energies (green and black lines).}
\label{fig:result}
\end{figure*}

Since jet angularities require the simulation of Monte Carlo events at the hadron level, 
we store them while generating the events needed to obtain the mean transverse momentum 
in Step 2. Then the jet angularities are normalised to the number of jets 
(entries of the distribution). An example of jet angularity $\lambda_{0}^{0}$ 
(multiplicity) at $\sqrt{s}=$~900 (green dashed line) and 13000~GeV (black solid line) 
is shown in Figure~\ref{fig:result}.
Having all ingredients $f^{900}$, $f^{13000}$, $\lambda_{0i}^{0}(900~\mathrm{GeV})$, 
and $\lambda_{0i}^{0}(13000~\mathrm{GeV})$ we are able, using Eqs.~\ref{eq:quarkang} 
and~\ref{eq:gluonang}, to derive quark and gluon multiplicities $\lambda_{0}^{0}$:
\begin{align}
 \lambda_{qi}&=\frac{f^{s_1}\lambda_i^{s_2}-f^{s_2}\lambda_i^{s_1}}{f^{s_1}-f^{s_2}}
           =\frac{f^{13000}\lambda_i^{900}-f^{900}\lambda_i^{13000}}{f^{13000}-f^{900}}  
           \nonumber \\ 
          &=\frac{0.73\lambda_i^{900}-0.33\lambda_i^{13000}}{0.73-0.33} 
           =1.83\lambda_i^{900}-0.83\lambda_i^{13000},
\end{align}
 \begin{align}
 \lambda_{gi}&=\frac{(1-f^{s_2})\lambda_i^{s_1}-(1-f^{s_1})\lambda_i^{s_2}}{f^{s_1}-f^{s_2}} 
            \nonumber \\ 
         &=\frac{(1-f^{900})\lambda_i^{13000}-(1-f^{13000})\lambda_i^{900}}{f^{13000}-f^{900}} 
           \nonumber \\ 
         &=\frac{(1-0.33)\lambda_i^{13000}-(1-0.73)\lambda_i^{900}}{0.73-0.33} 
           \nonumber \\ 
         &=\frac{(0.67)\lambda_i^{13000}-(0.27)\lambda_i^{900}}{0.73-0.33}
           \nonumber \\ 
         &=1.68\lambda_i^{13000}-0.68\lambda_i^{900}.
\end{align}
Figure~\ref{fig:result} illustrates that the result $\lambda_{qi}$ (red line) and 
$\lambda_{gi}$ (blue line) are linear combinations of jet angularities measured at 
different energies.
\section{Robustness of observables to systematic effects}
\label{sec:syseff}
An important question to consider is whether the q/g angularities obtained in our study are robust to the impact of multiparton interactions (MPI) and initial state radiation (ISR). To address this, we performed supplementary Monte Carlo simulations for each observable examined, where we excluded the effects of MPI and ISR. This enabled us to evaluate the robustness of the q/g angularities obtained. 
Another crucial aspect to consider is the extent to which the q/g angularities remain independent of the energy.
To test the energy independence, we use six possible energy combinations ($900-2360$, $900-7000$, $900-13000$, $2360-7000$, $2360-13000$, $7000-13000$ GeV) to derive 
angularities in the same way as described in the example~\ref{sec:example}. Therefore, there are in total 24 distributions (12 for quark and 12 for gluon jet angularities) that we need to analyse per plot.
\begin{figure*}
    \centering
    \includegraphics[width=\textwidth]{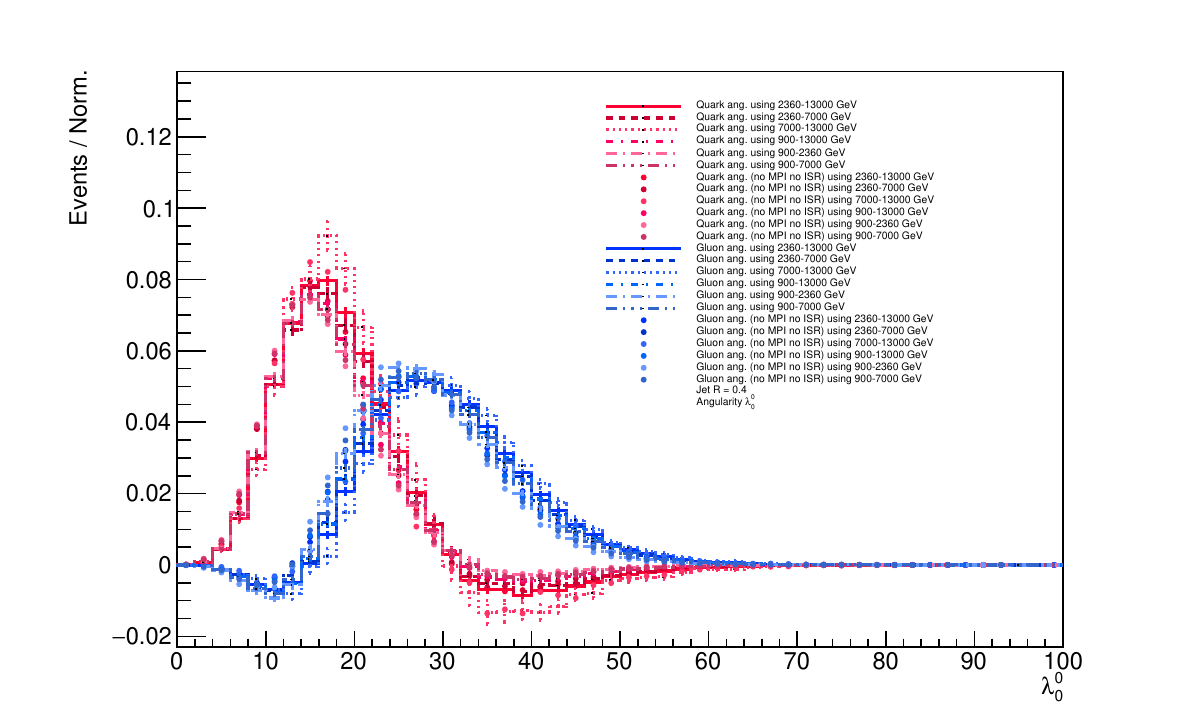} 
    \\
    \includegraphics[width=\textwidth]{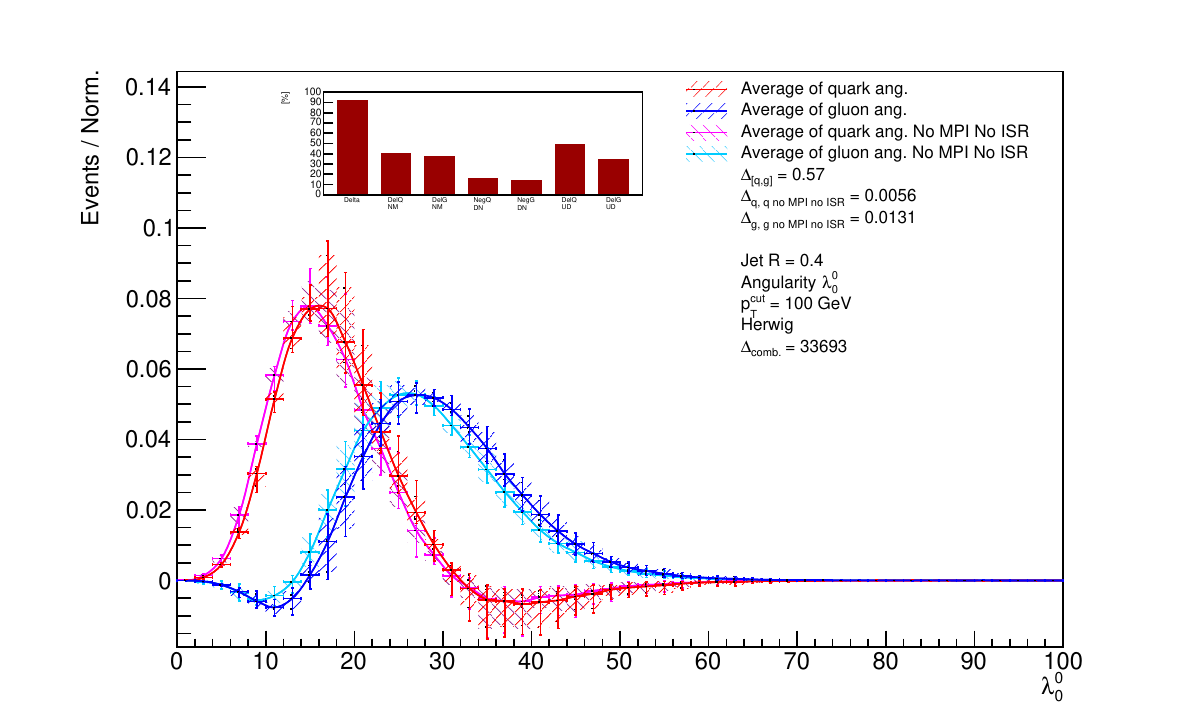}  
    \caption{Quark and gluon multiplicities $\lambda_{0}^{0}$ ($R = 0.4$, \ptcut{100}) 
    for all six energy combinations (above) and averaged plot showing the envelopes of
    the different energy combinations as filled areas and their statistical uncertainties as ticks (below).}
    \label{fig:busy}
\end{figure*}
In the top panel of Figure~\ref{fig:busy} we present an example of this plot showing, similar to Figure~\ref{fig:result}, the multiplicity $\lambda_{0}^{0}$ ($R = 0.4$, \ptcut{100}) of the quark jets (red lines) and the gluon jets (blue lines). This time, the display includes angularities obtained from all energy combinations using full simulation, which are denoted by various types of lines. Additionally, the dots represent the angularities derived with MPI and ISR turned off. In order to simplify this plot, the bottom panel shows the same observable but this time solid lines represent the averaged q/g angularities across different energy combinations, while the filled area depicts the envelope of the different energy combinations, and the ticks represent the envelope of the statistical uncertainties of the angularities. 
By comparing this graph with Figure~\ref{fig:result}, we can gain additional insight into how the observables are robust to these important systematic effects.
The averaging not only simplifies the detailed plots but also allows us to define measures (shown in the subhistogram with 7 bins), which help to sort angularities based on their performance. These measures are discussed in the next section. 
\begin{figure*}[ht!]
\centering
    \includegraphics[width=0.49\textwidth]{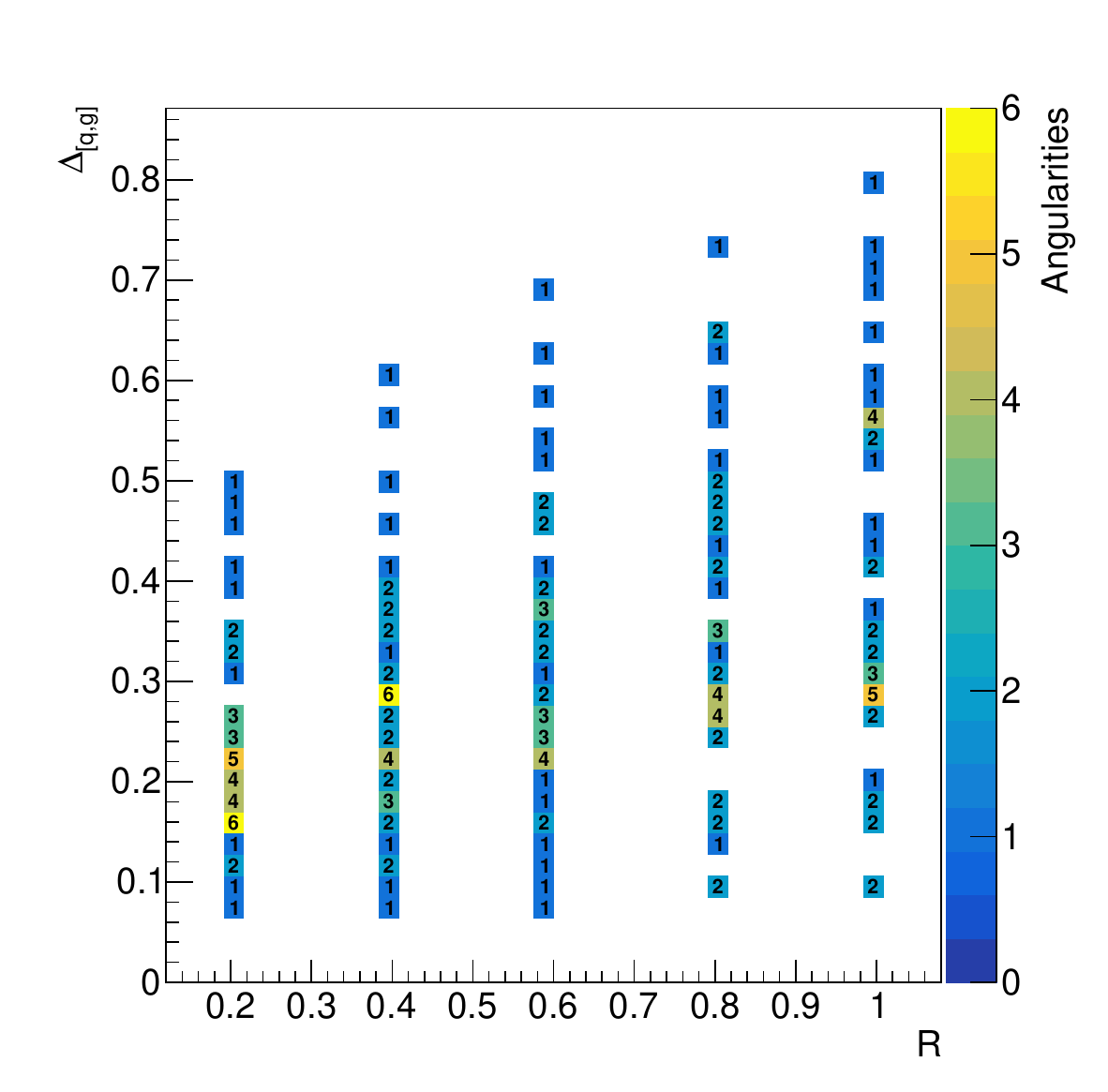} 
    \includegraphics[width=0.49\textwidth]
    {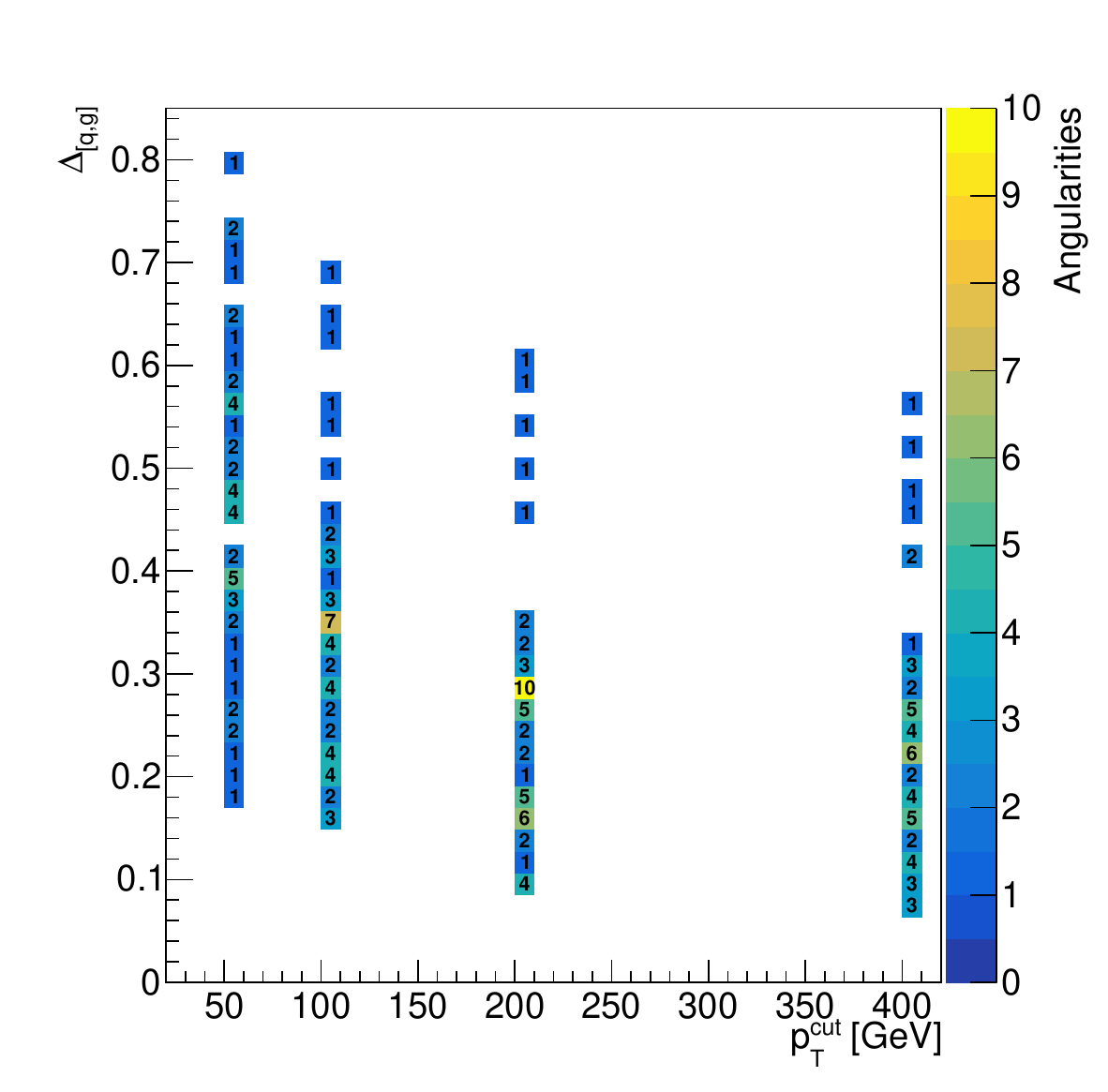} 
    \caption{Classifier separation $\Delta_{[q,g]}$ as a function of the jet radius (left panel) and as a function of $p_{T}^{\mathrm{cut}}$ (right).}
    \label{fig:separation}
\end{figure*}
\section{Quantifying the quark/gluon separation power}
\label{sec:measures}
\begin{figure}[ht!]
    \includegraphics[width=8cm]{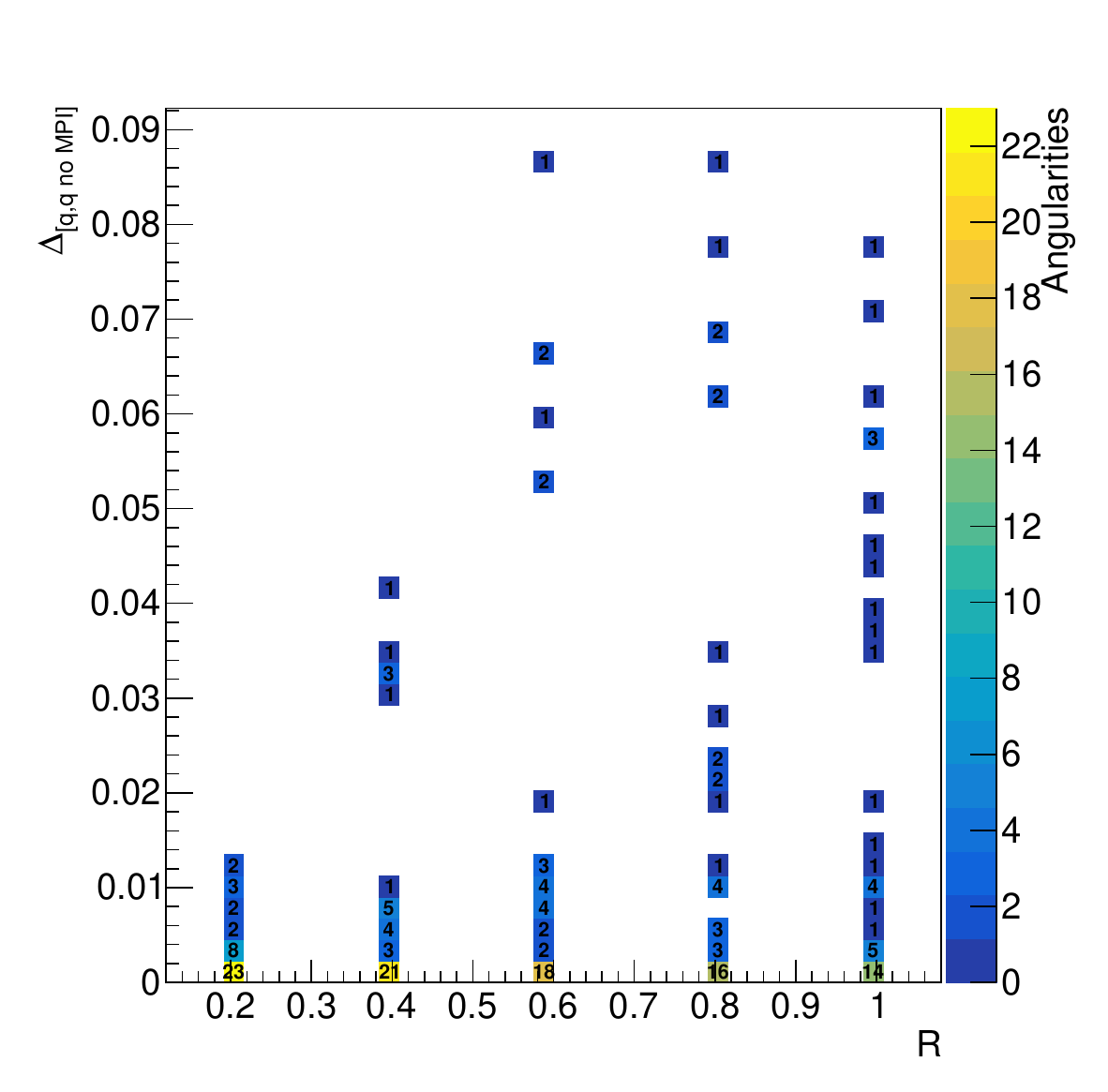}
    \includegraphics[width=8cm]{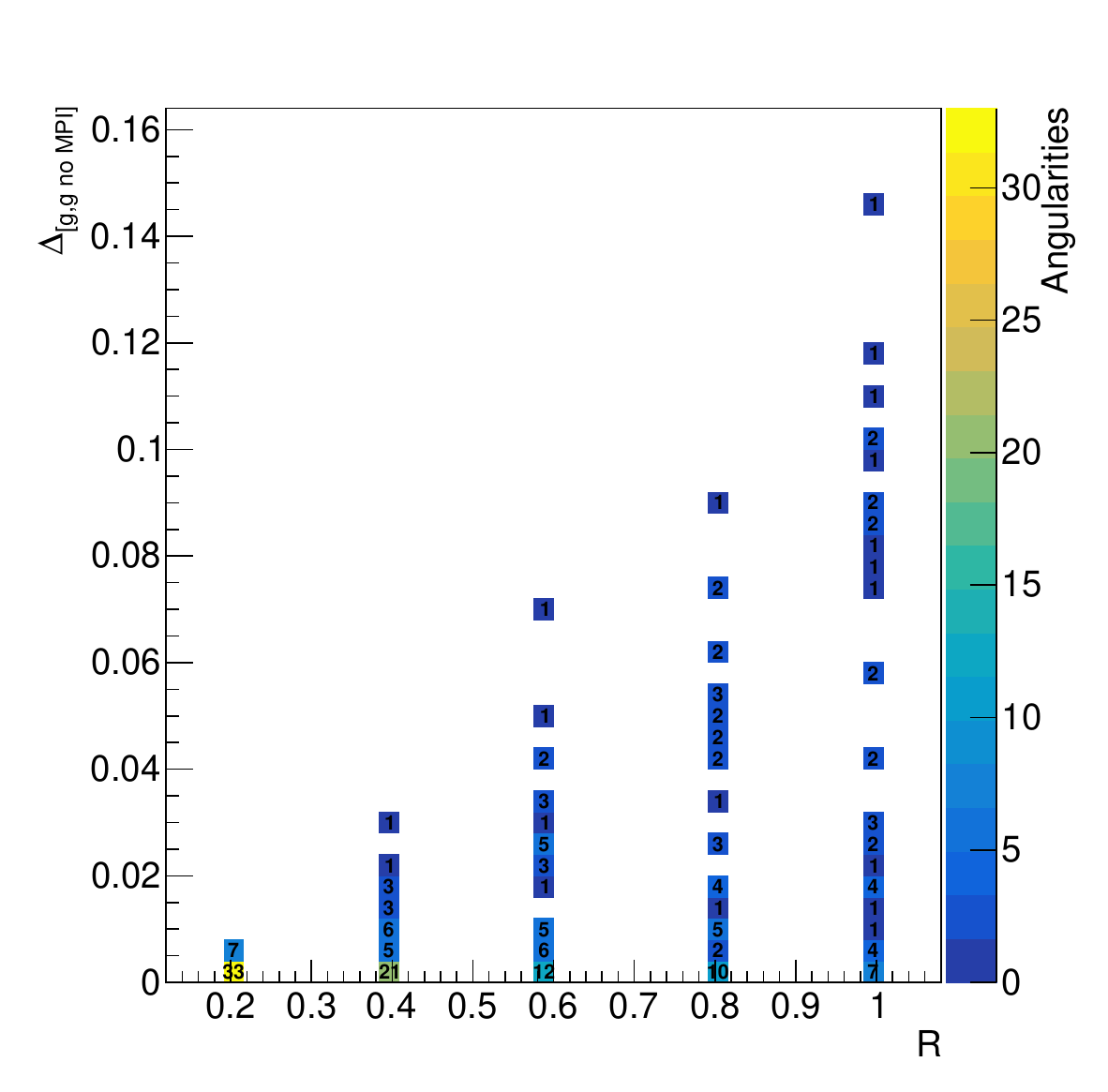}
    \caption{Second/third column quark $\Delta_{[q,q~\mathrm{noMPI}]}$ (top) and gluon $\Delta_{[g,g~\mathrm{noMPI}]}$ (bottom) as a function of jet radius.}
   \label{fig:negativity_R_MPI}
\end{figure}

\begin{figure}[ht!]
    \includegraphics[width=8cm]{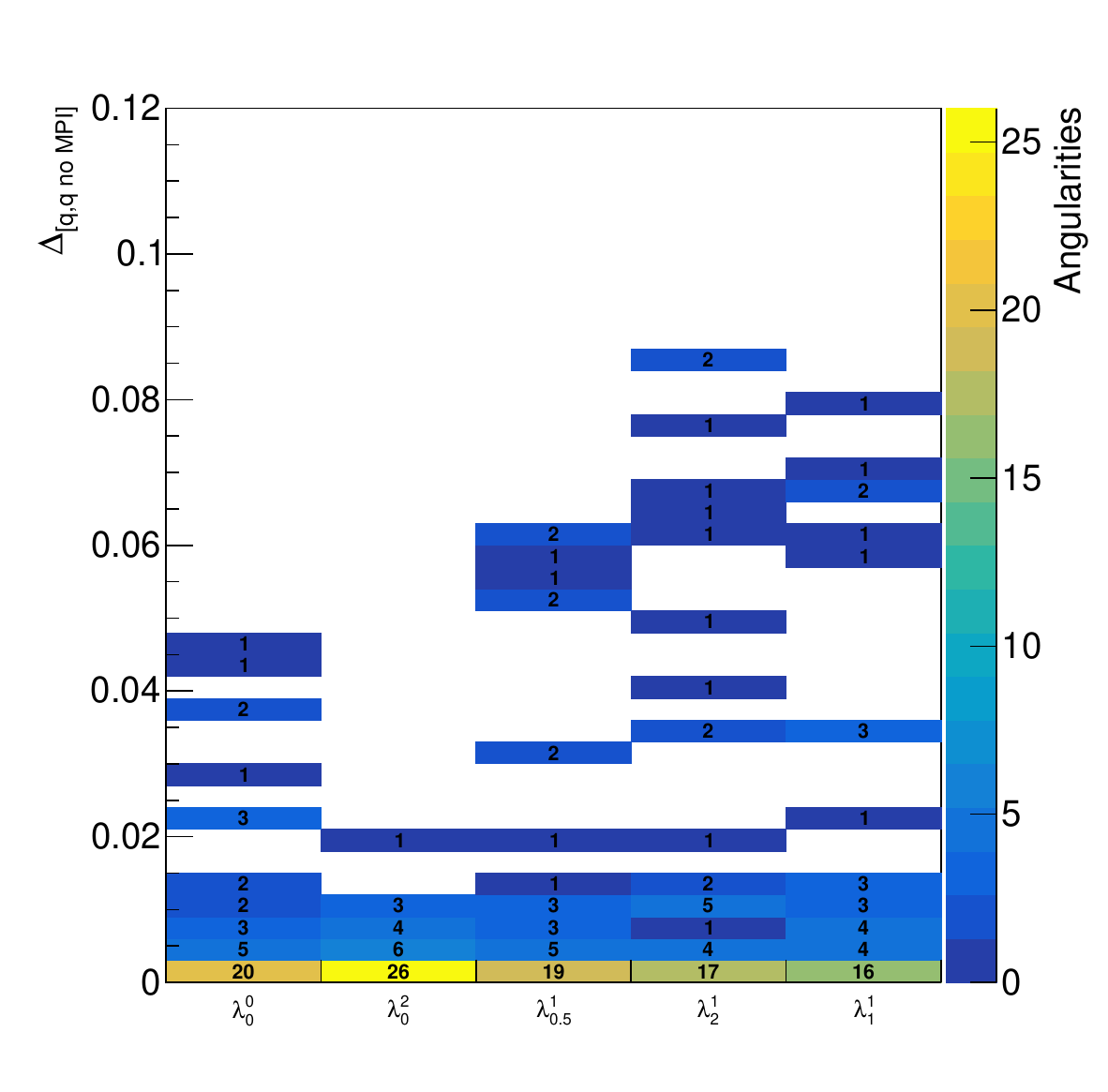}
    \includegraphics[width=8cm]{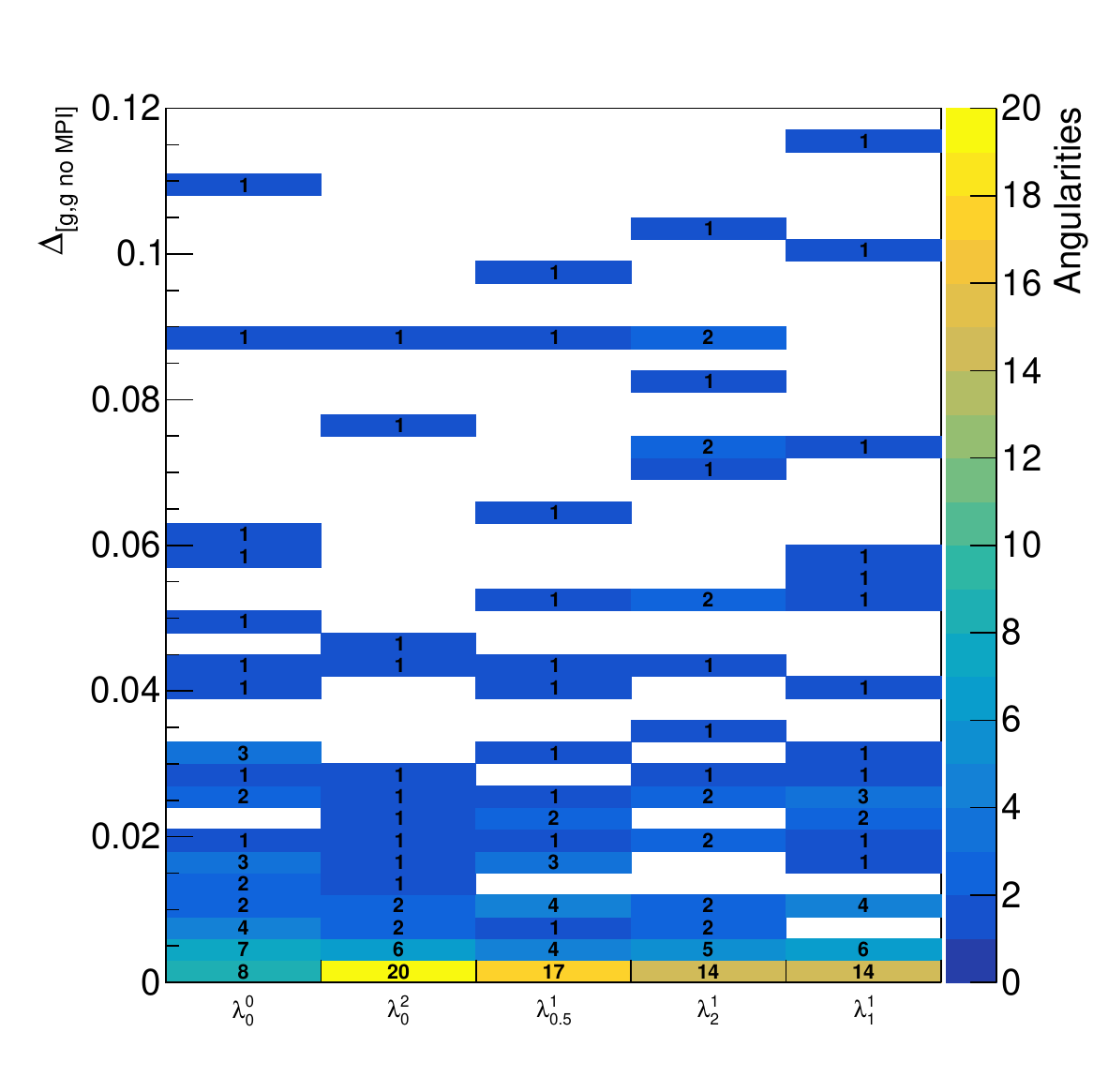}
    \caption{Second/third column quark $\Delta_{[q,q~\mathrm{noMPI}]}$ (top) and gluon $\Delta_{[g,g~\mathrm{noMPI}]}$ (bottom) as a function of angularities.}
    \label{fig:negativity_ang_MPI}
\end{figure}

\begin{figure}[ht!]
    \includegraphics[width=8cm]{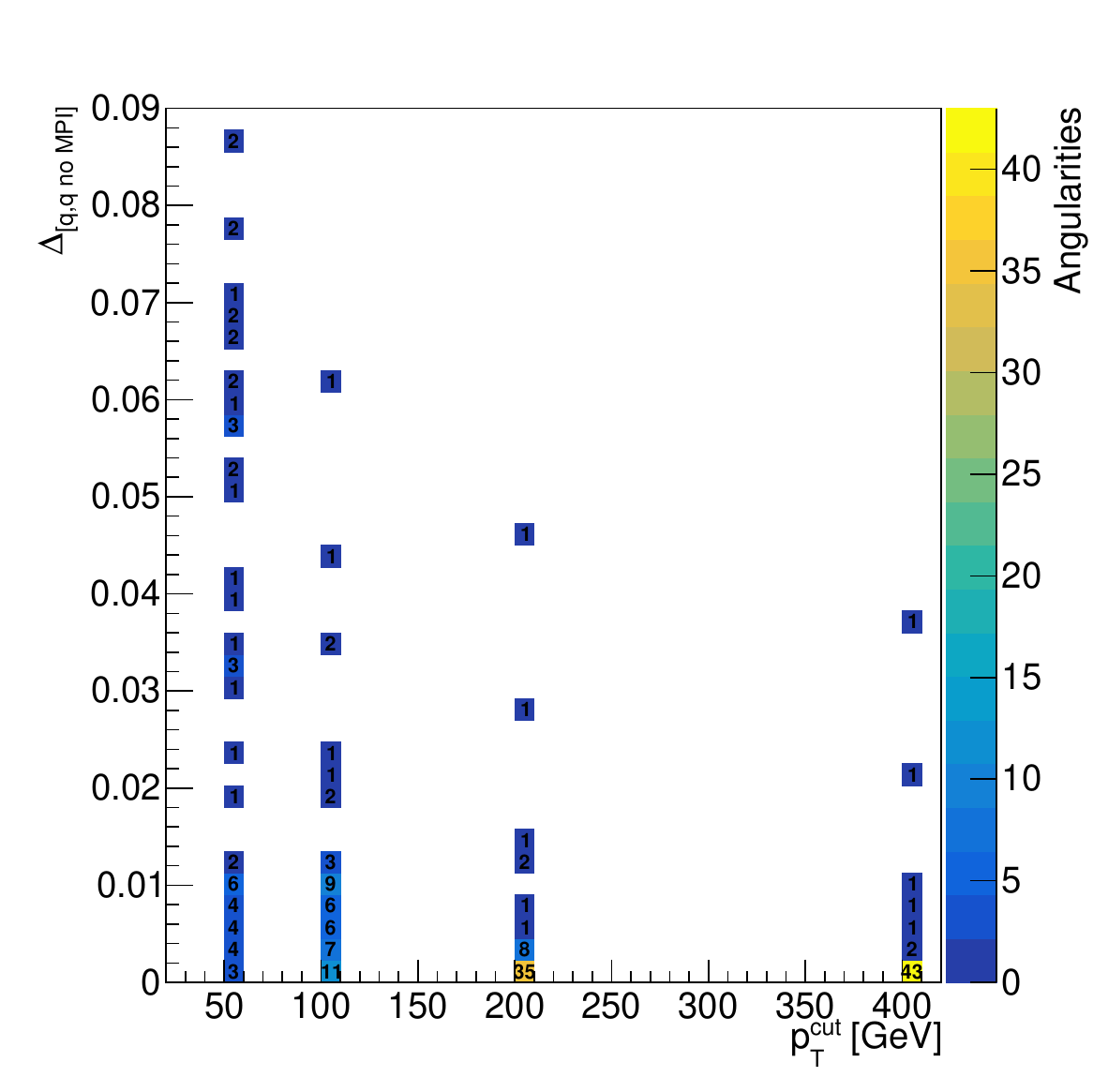}
    \includegraphics[width=8cm]{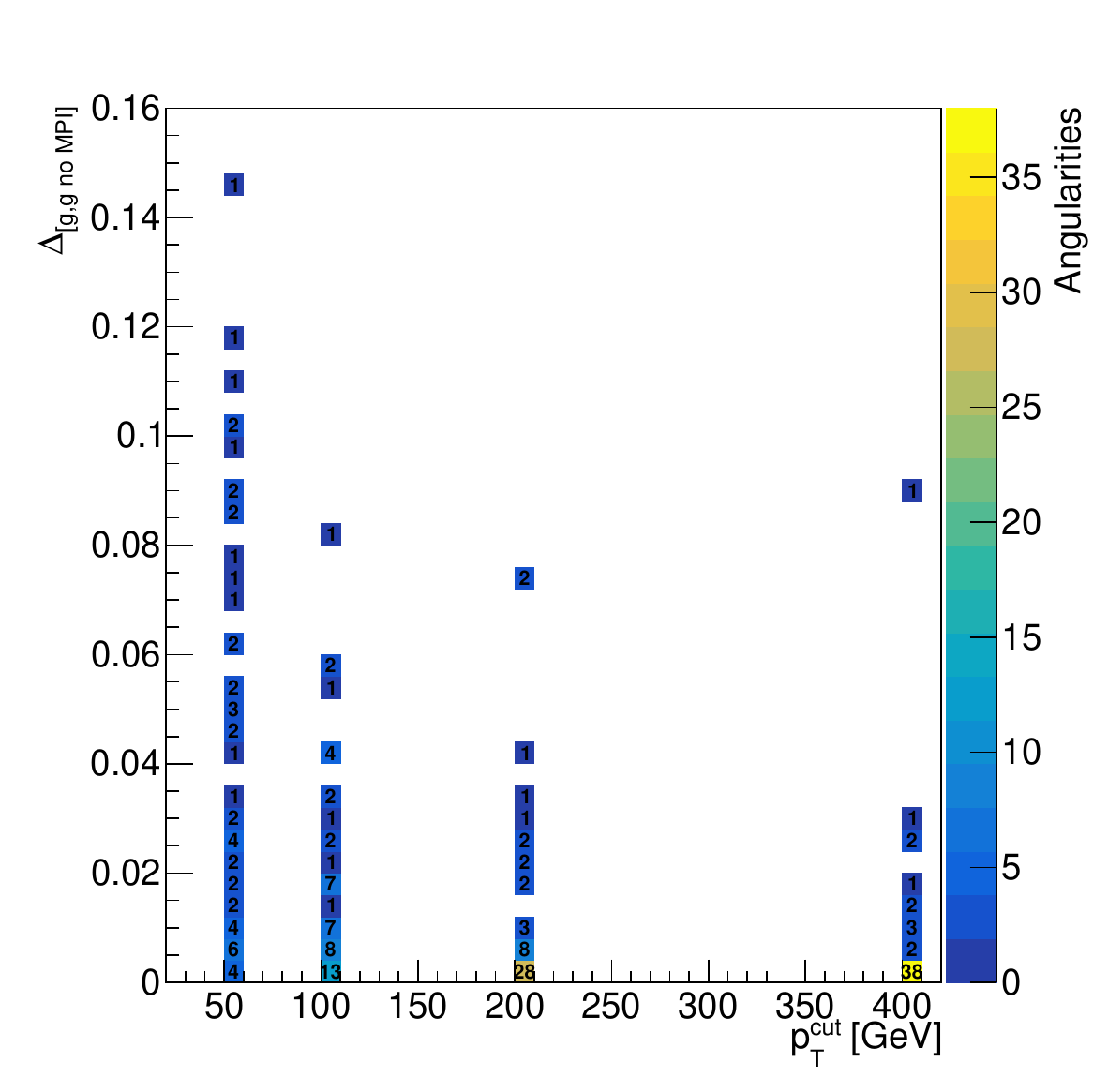}
    \caption{Second/third column quark $\Delta_{[q,q~\mathrm{noMPI}]}$ (top) and gluon $\Delta_{[g,g~\mathrm{noMPI}]}$ (bottom) as a function of $p_{T}^{\mathrm{cut}}$.}
    \label{fig:negativity_Q_MPI}
\end{figure}

\begin{figure}[ht!]
    \includegraphics[width=8cm]{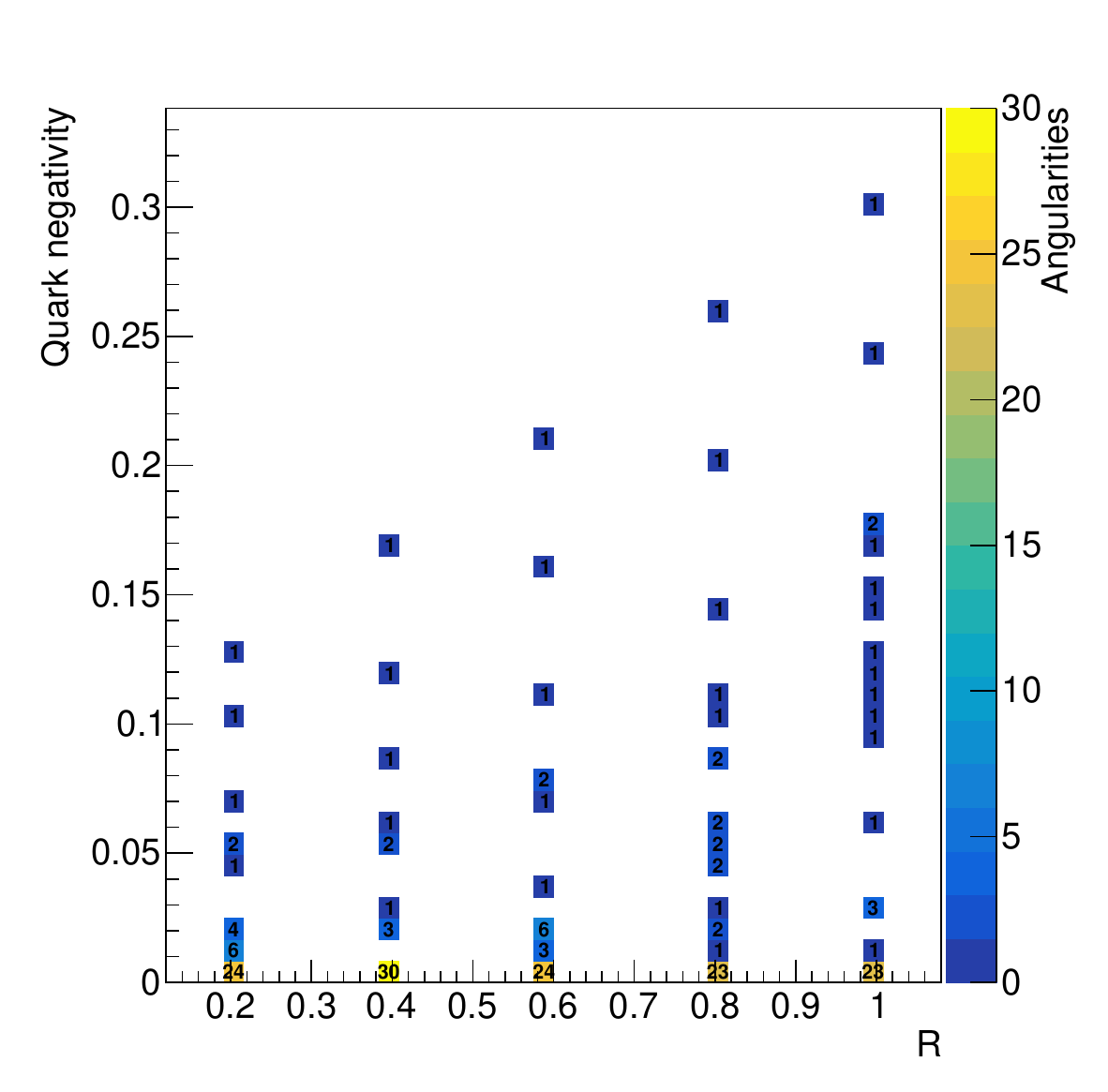}
    \includegraphics[width=8cm]{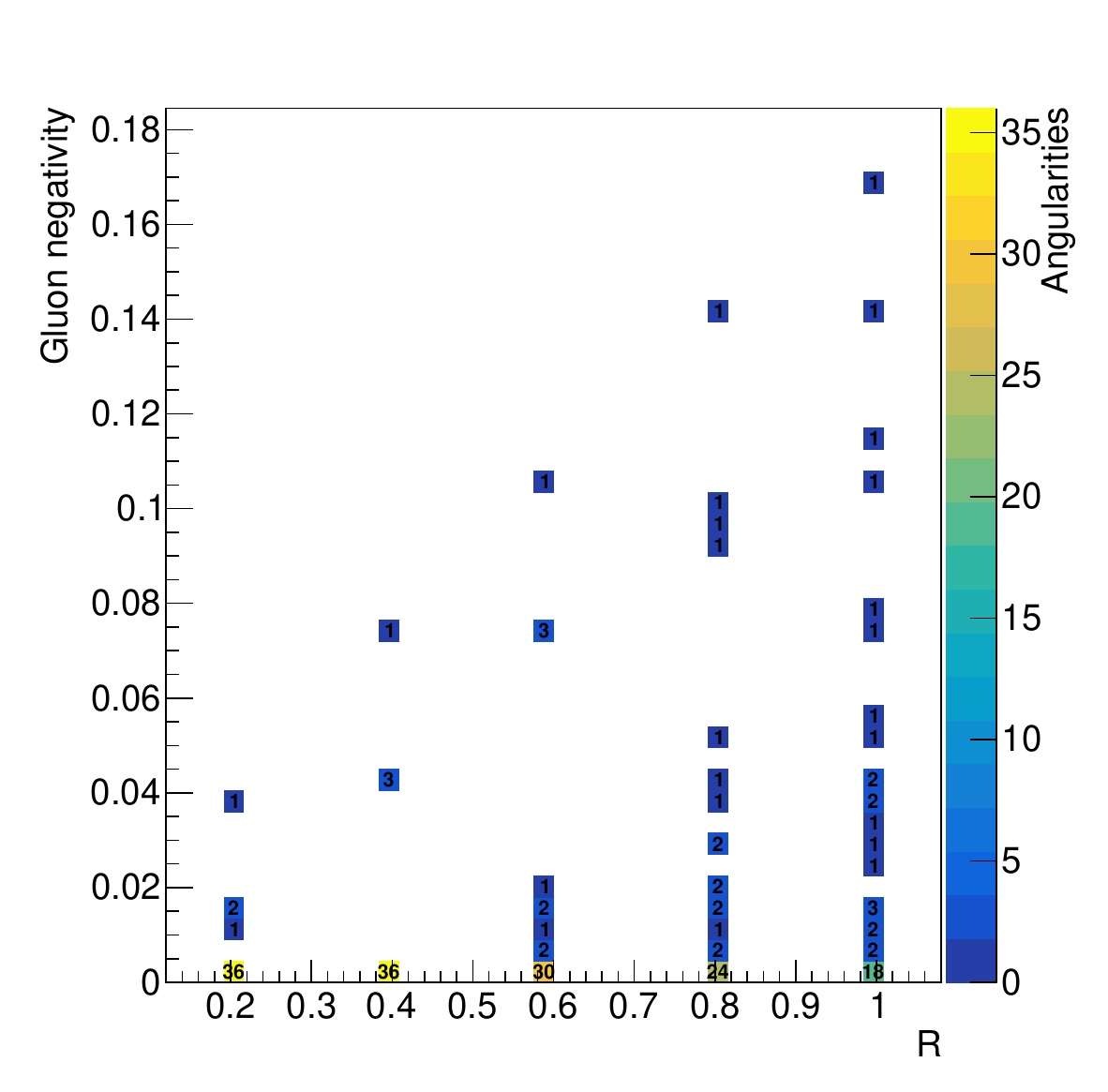}
    \caption{Fourth/fifth column quark (top) and gluon negativity (bottom) as a function of jet radius.}
   \label{fig:negativity_R_neg}
\end{figure}

\begin{figure}[ht!]
    \includegraphics[width=8cm]{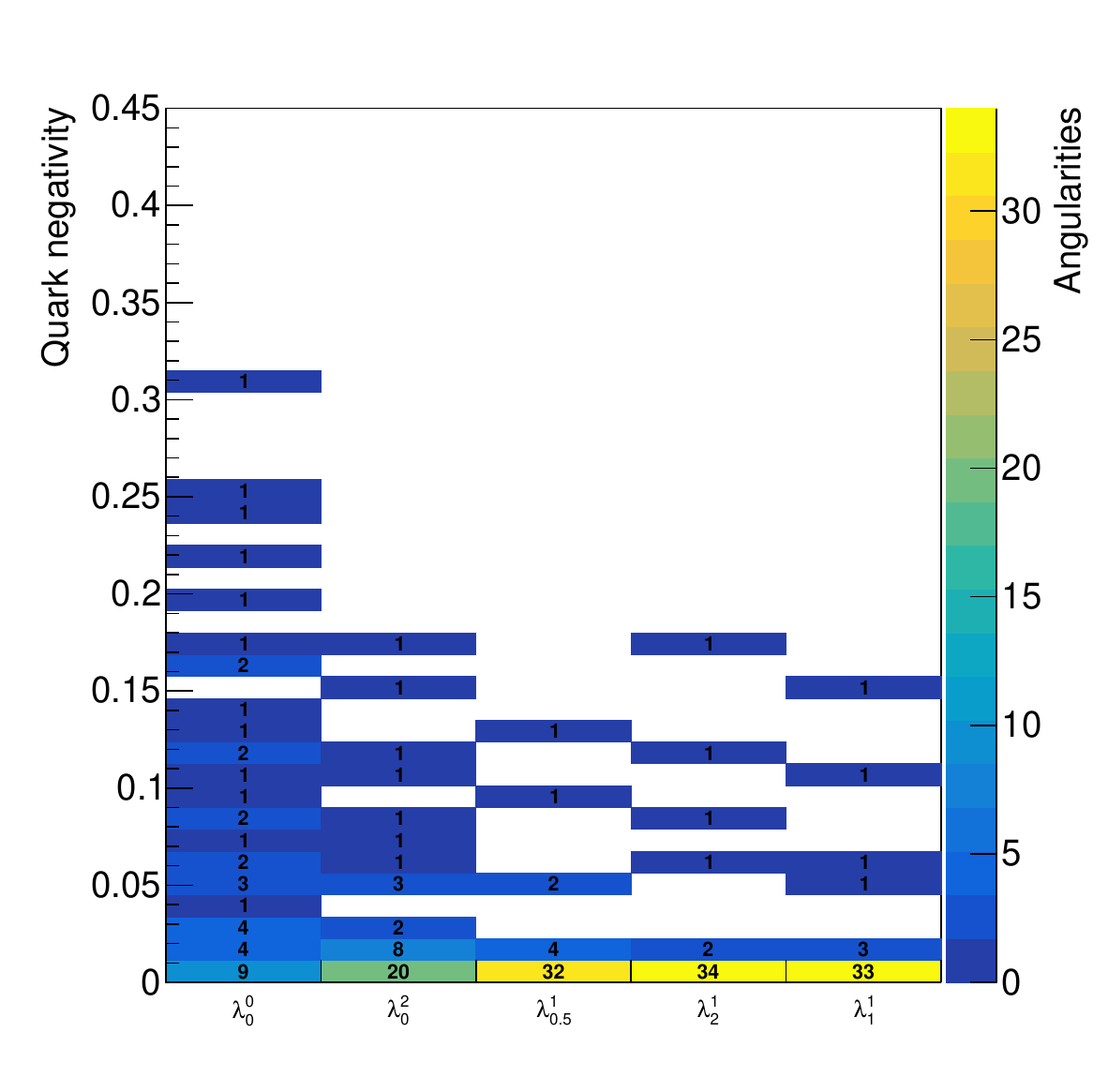}
    \includegraphics[width=8cm]{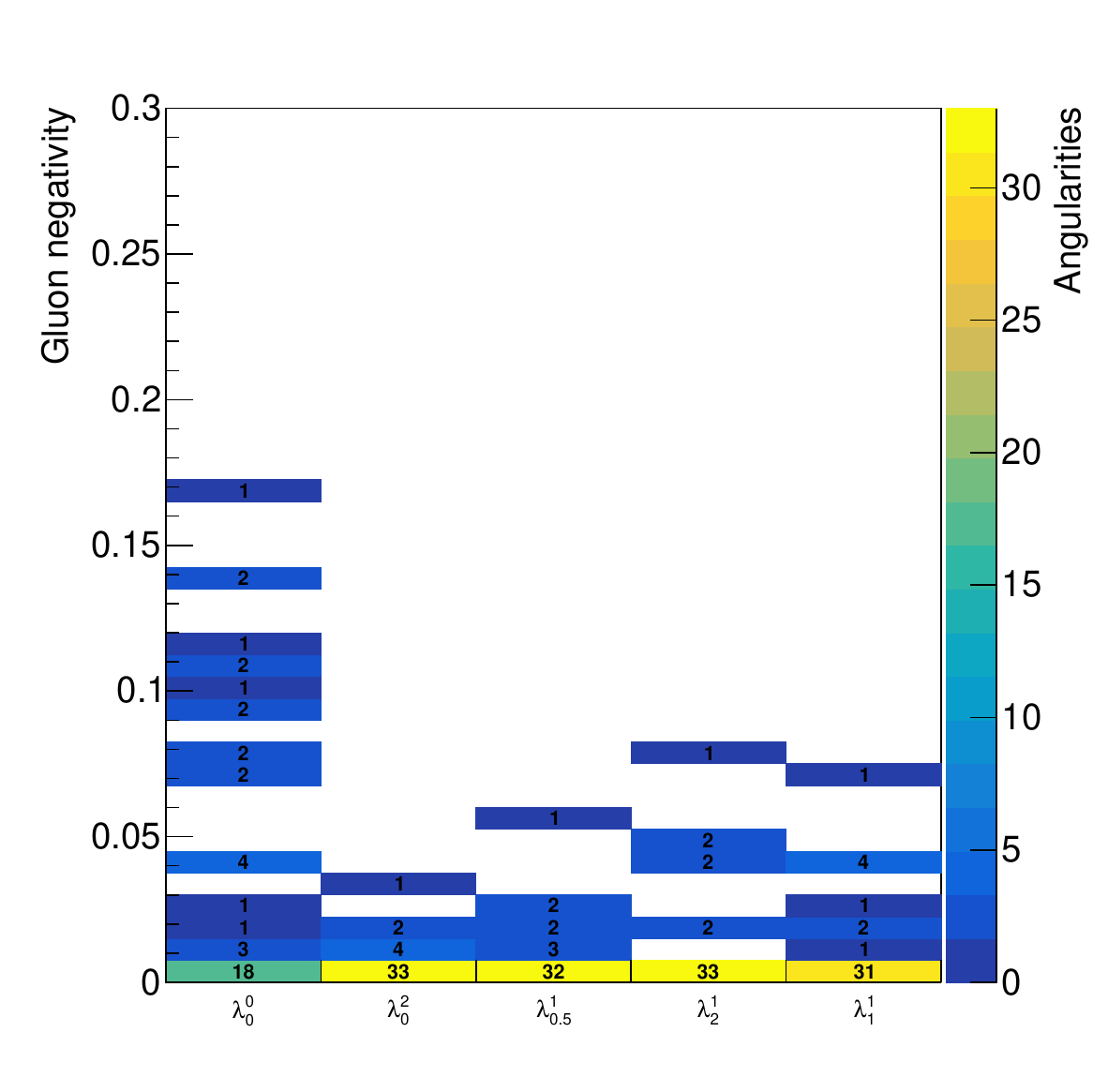}
    \caption{Fourth/fifth column quark (top) and gluon negativity (bottom) as a function of angularities.}
    \label{fig:negativity_ang_neg}
\end{figure}

\begin{figure}[ht!]
    \includegraphics[width=8cm]{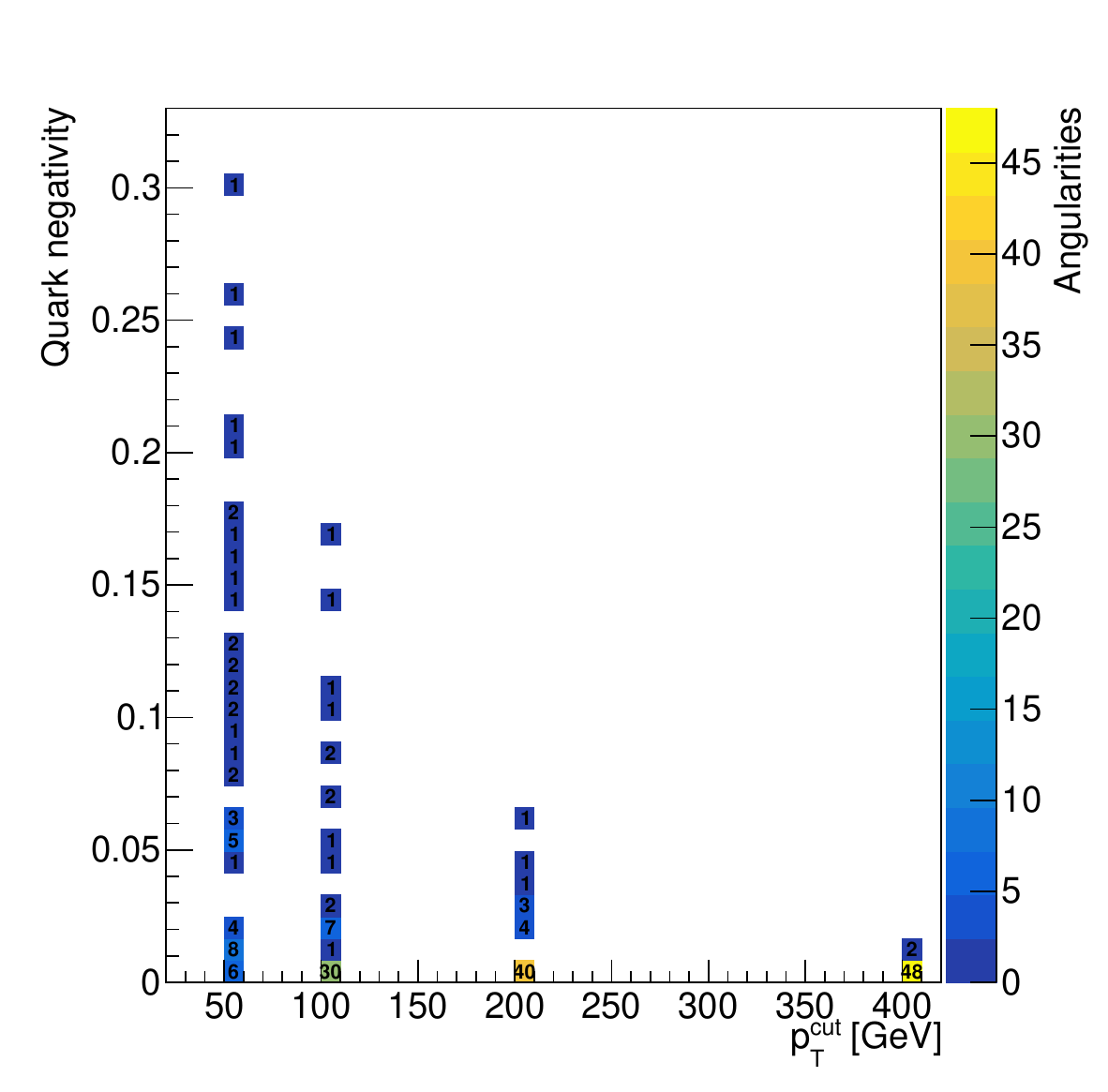}
    \includegraphics[width=8cm]{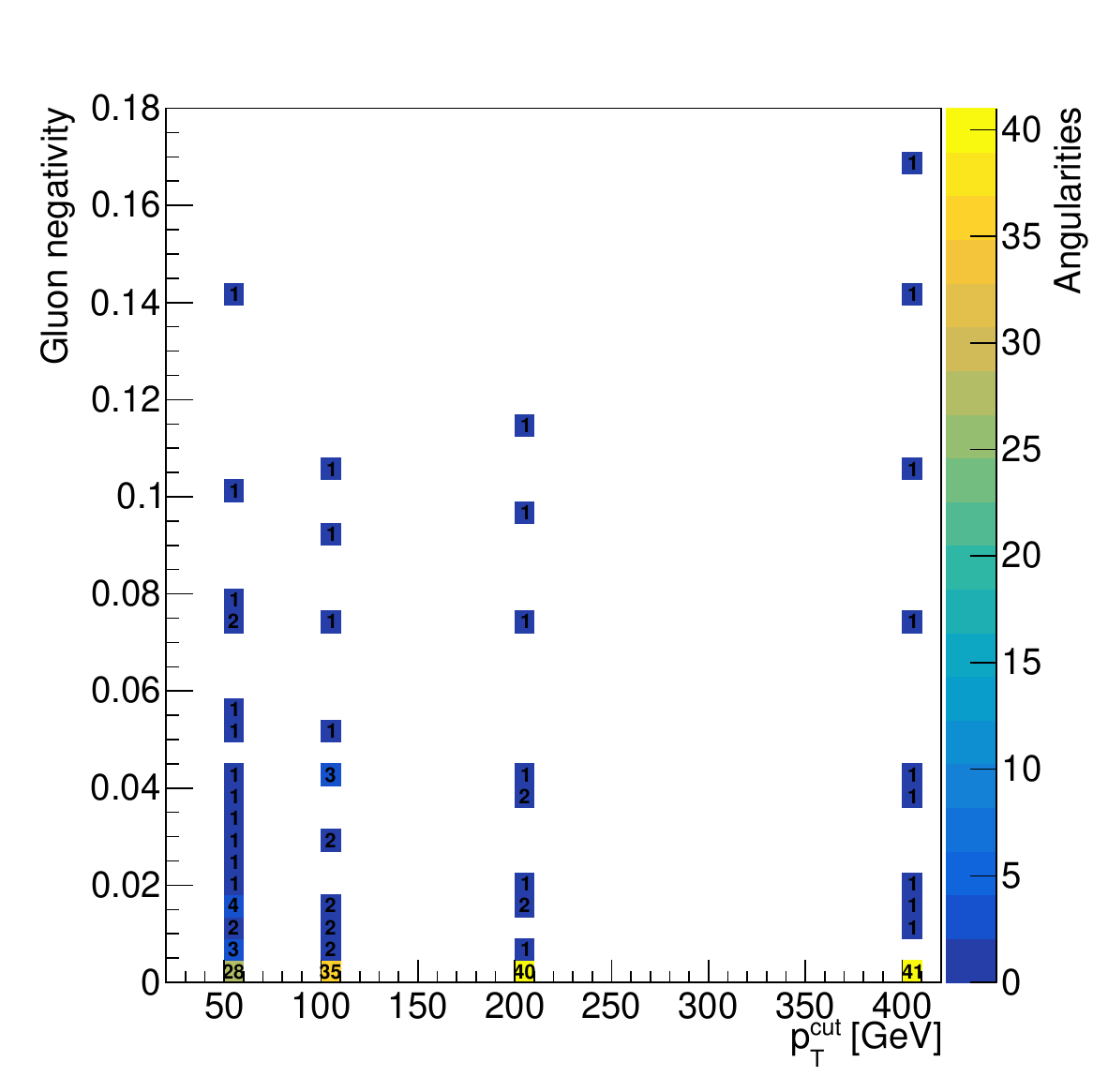}
    \caption{Fourth/fifth column quark (top) and gluon negativity (bottom) as a function of $p_{T}^{\mathrm{cut}}$.}
    \label{fig:negativity_Q_neg}
\end{figure}

\begin{figure}[ht!]
    \includegraphics[width=8cm]{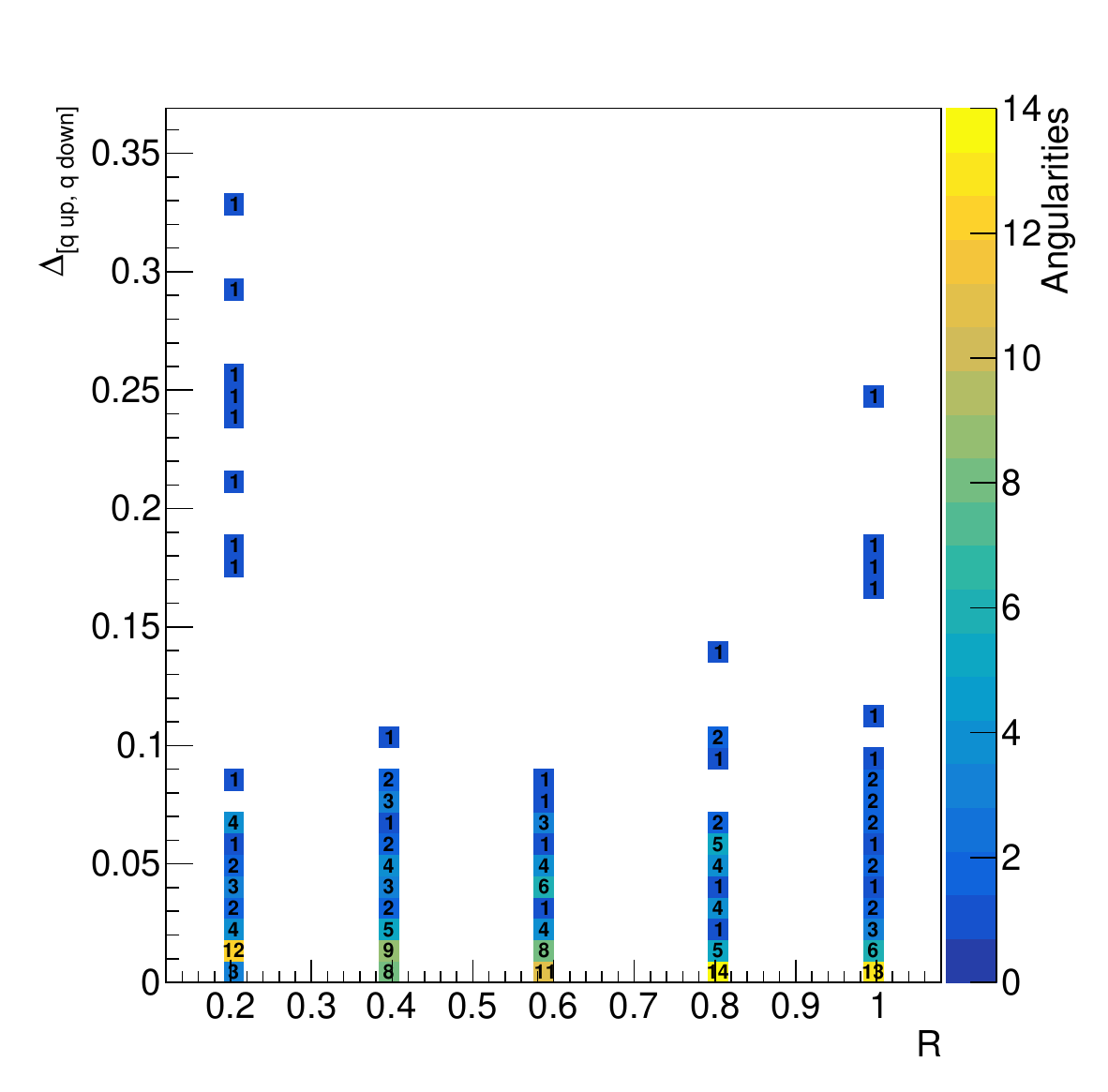}
    \includegraphics[width=8cm]{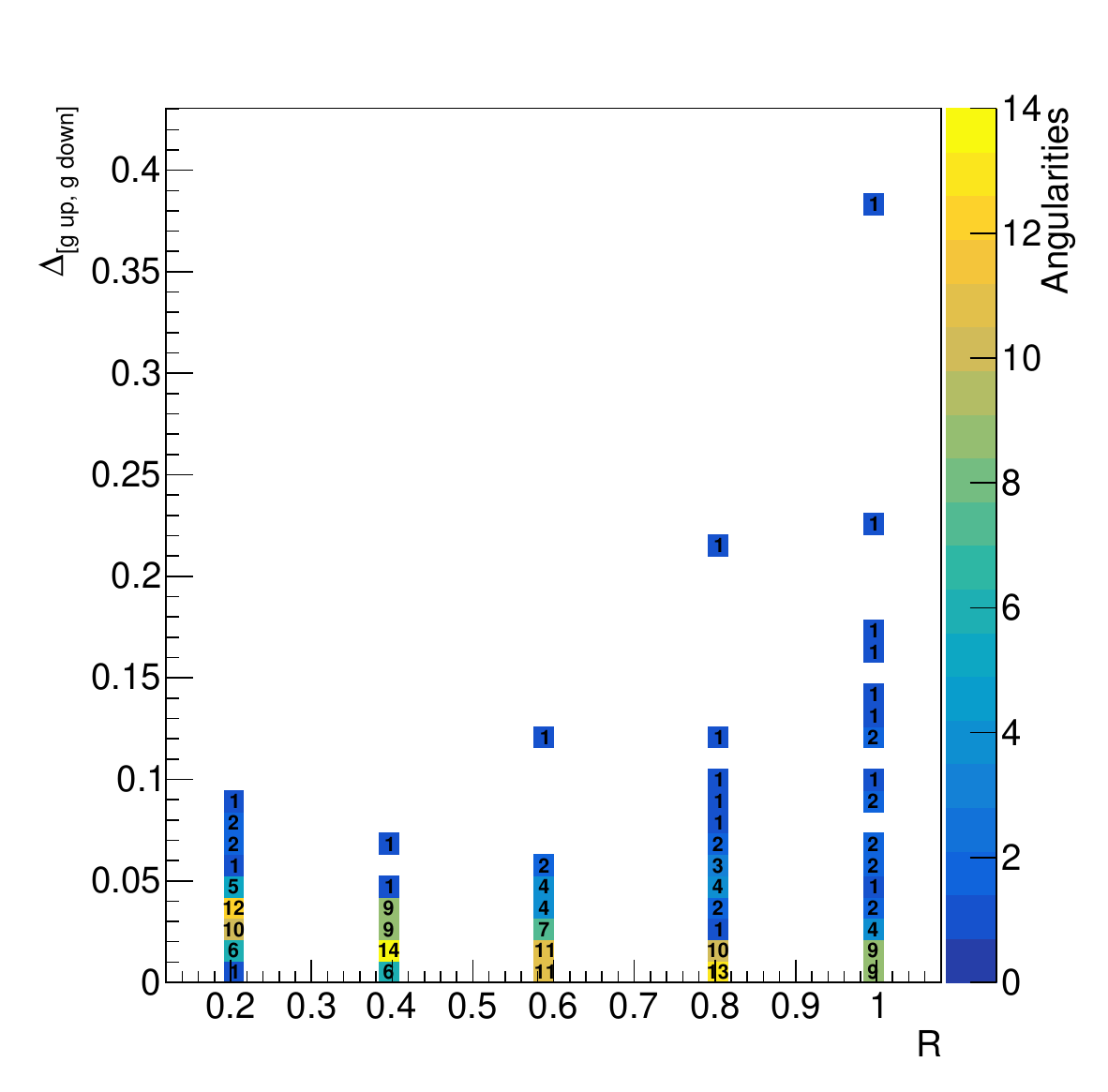}
    \caption{Sixth/Seventh column quark $\Delta_{[q~\mathrm{down},q~\mathrm{up}]}$ (top) and gluon $\Delta_{[g~\mathrm{down},g~\mathrm{up}]}$ (bottom) as a function of jet radius.}
   \label{fig:negativity_R_E}
\end{figure}

\begin{figure}[ht!]
    \includegraphics[width=8cm]{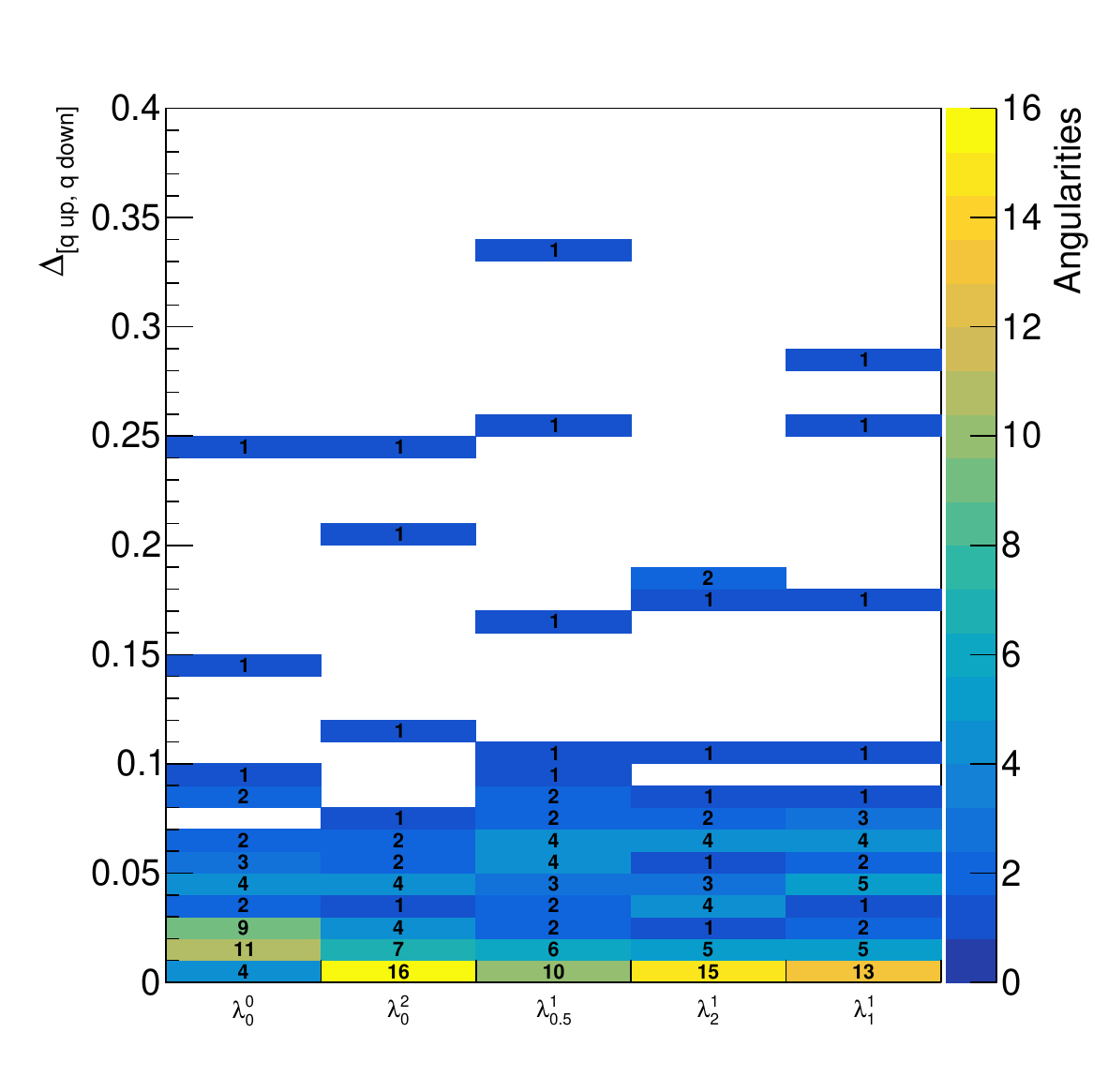}
    \includegraphics[width=8cm]{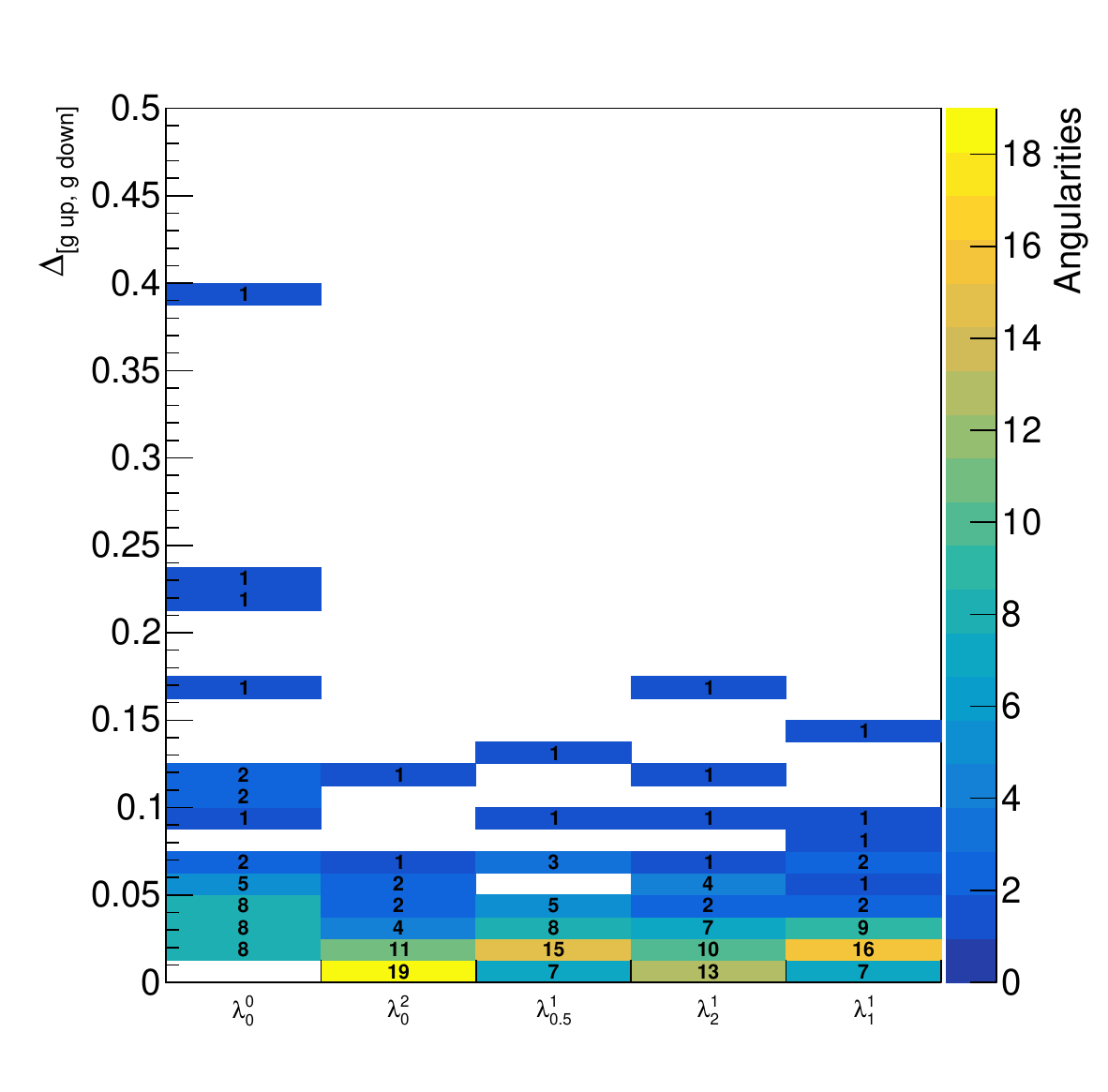}
    \caption{Sixth/Seventh column quark $\Delta_{[q~\mathrm{down},q~\mathrm{up}]}$ (top) and gluon $\Delta_{[g~\mathrm{down},g~\mathrm{up}]}$ (bottom) as a function of angularities.}
    \label{fig:negativity_ang_E}
\end{figure}

\begin{figure}[ht!]
    \includegraphics[width=8cm]{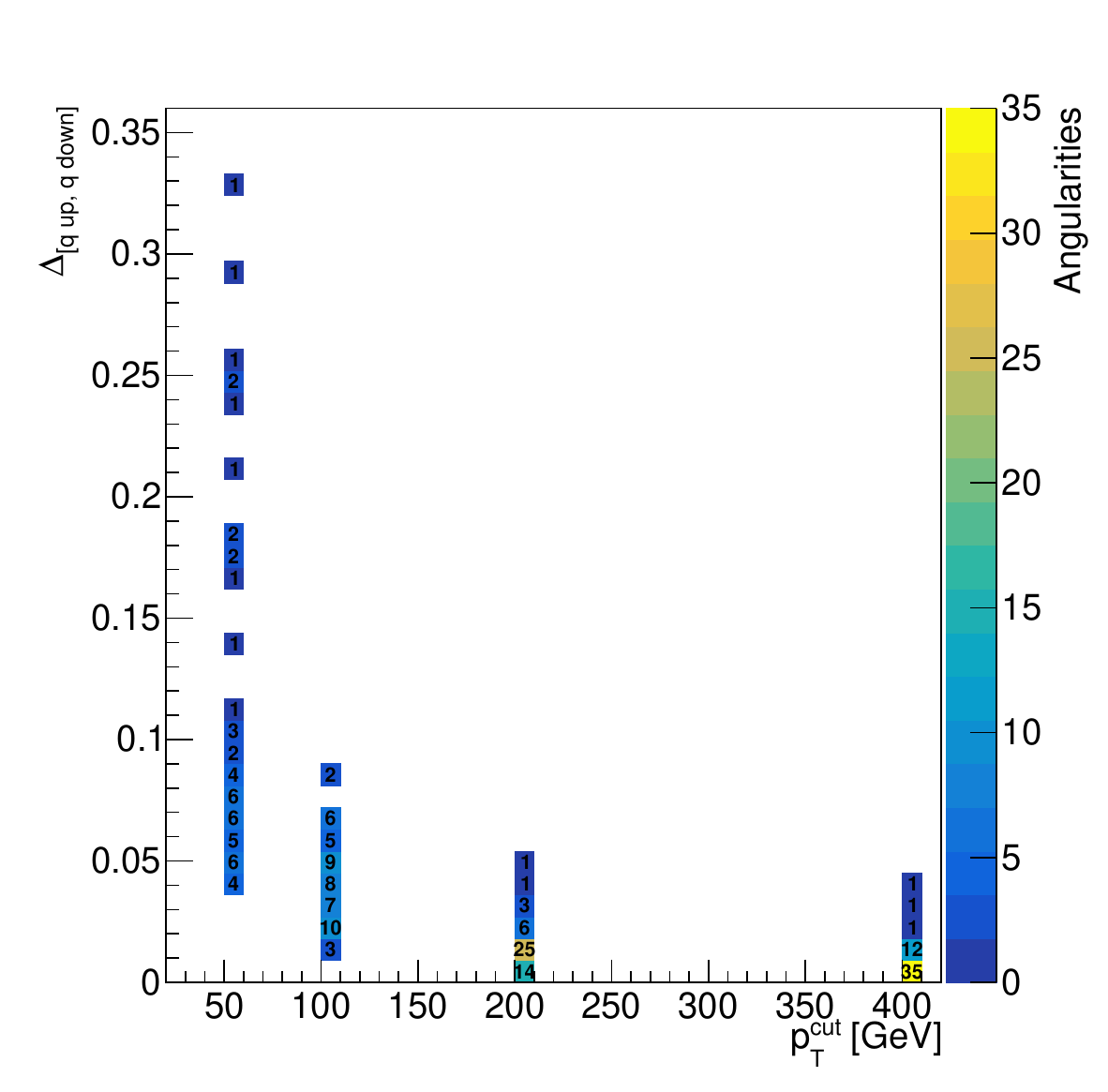}
    \includegraphics[width=8cm]{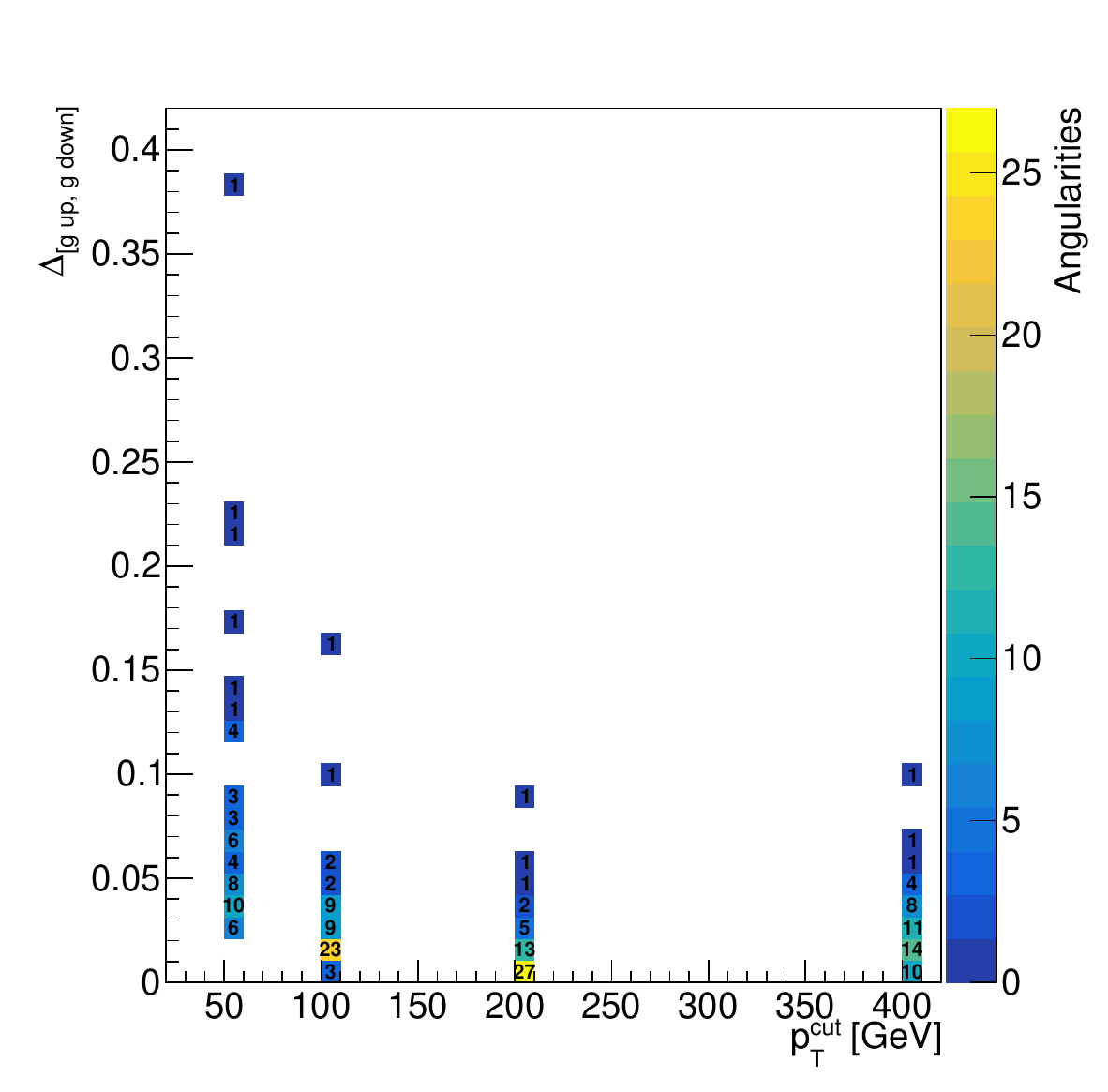}
    \caption{Sixth/Seventh column quark $\Delta_{[q~\mathrm{down},q~\mathrm{up}]}$ (top) and gluon $\Delta_{[g~\mathrm{down},g~\mathrm{up}]}$ (bottom) as a function of $p_{T}^{\mathrm{cut}}$.}
    \label{fig:negativity_Q_E}
\end{figure}

Since we will be testing many variants of observables, we need a way to
quantify the quark/gluon separation power in a robust way that can
be easily summarised by a single number. For example, in~\cite{Larkoski:2014pca},
the authors quantify discrimination performance in the context of quark/gluon jets 
using classifier separation $\Delta_{[q,g]}$  (a default output of the TMVA package \cite{2007physics...3039H}):
\begin{equation}
\label{eq:delta}
    \Delta_{[q,g]} =\frac12 \sum_{i=1}^{N} \frac{(\lambda_{q_{i}}-\lambda_{g_{i}})^2}{\lambda_{q_{i}}+\lambda_{g_{i}}}.    
\end{equation}
Here $N$ denotes the number of bins, $\lambda_{q_{i}}$  ($\lambda_{g_{i}}$) is the 
$i$-th bin content\footnote{As we saw earlier the resulting distributions can have negative bin values. This can lead to large $\Delta_{[q,g]}$ values since the denominator can be equal to or close to zero. Therefore, in the case of negative bins in eq.~\ref{eq:delta}, we set their value to zero. When the bin values for both quark and gluon angularities are equal to zero, these bins are neglected in the sum.} of the probability distribution for the quark jet (gluon jet) 
sample as a function of the classifier $\lambda$. $\Delta_{[q,g]} = 0$ corresponds 
to no discrimination power (the distributions are exactly the same) while 
$\Delta_{[q,g]} = 1$ corresponds to perfect discrimination power.
\begin{figure}[ht!]
    \centering
    \includegraphics[width=8cm]{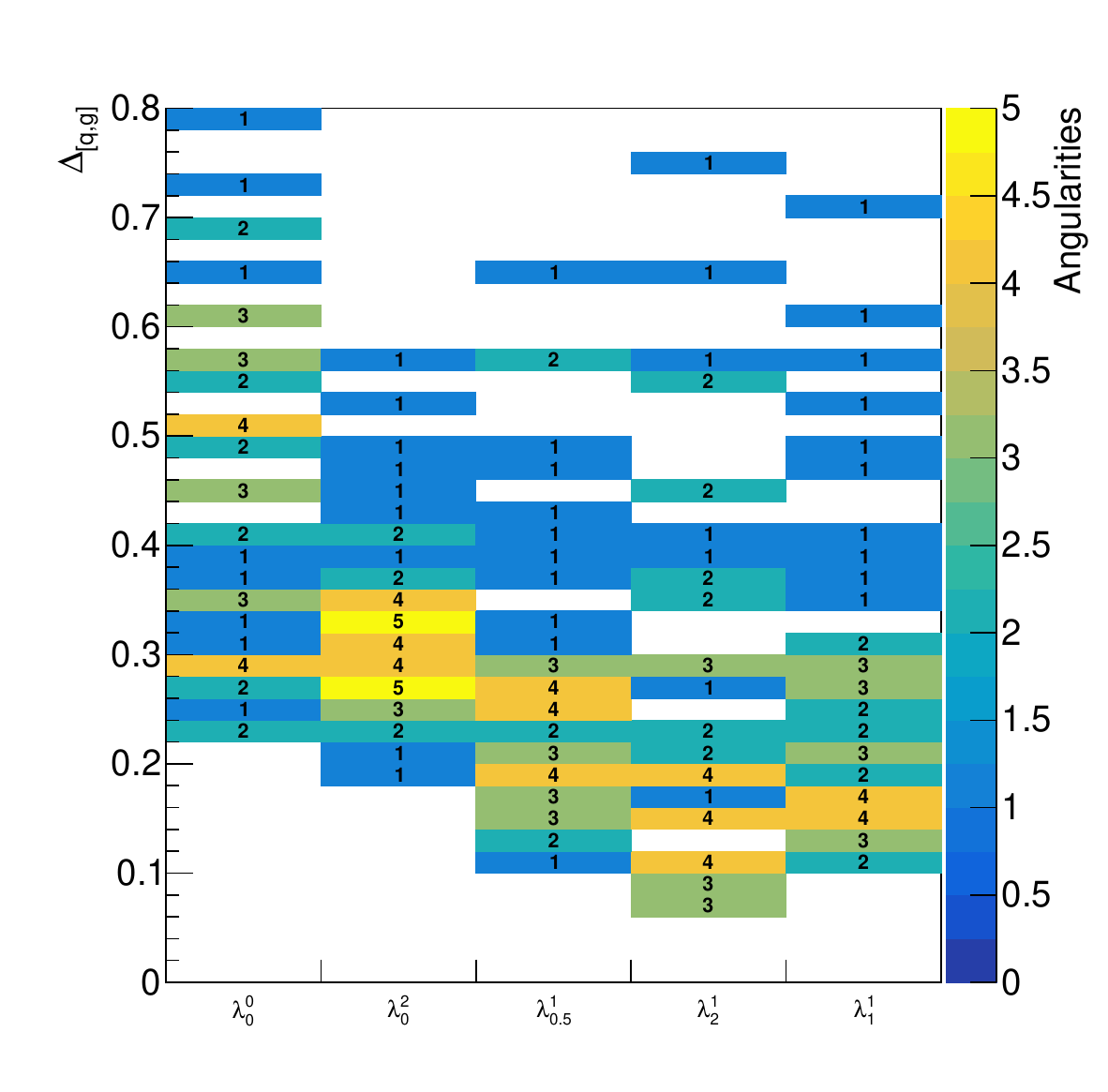} 
    \caption{Classifier separation $\Delta_{[q,g]}$ as a function of angularities.}
    \label{fig:separation_ang}
\end{figure}
In Fig.~\ref{fig:separation} (left panel), we show the separation of the classifier $\Delta_{[q,g]}$ 
as a function of the jet radius for all the energy-averaged angularities studied. We see that the separation power increases with increasing jet radius. This can be intuitively understood, since in larger jets more information about the radiation pattern is contained, which should be different for quark and gluon jets. Similarly, in Fig.~\ref{fig:separation}  (right panel) we see that the separation power decreases with increasing \ptcutwb. Finally, it is clear from Fig.~\ref{fig:separation_ang} that
although several individual cases of other angularities have high separation, it is
multiplicity that scores the highest
in most cases.
\begin{figure*}[ht!]
    \centering
    \includegraphics[width=\textwidth]{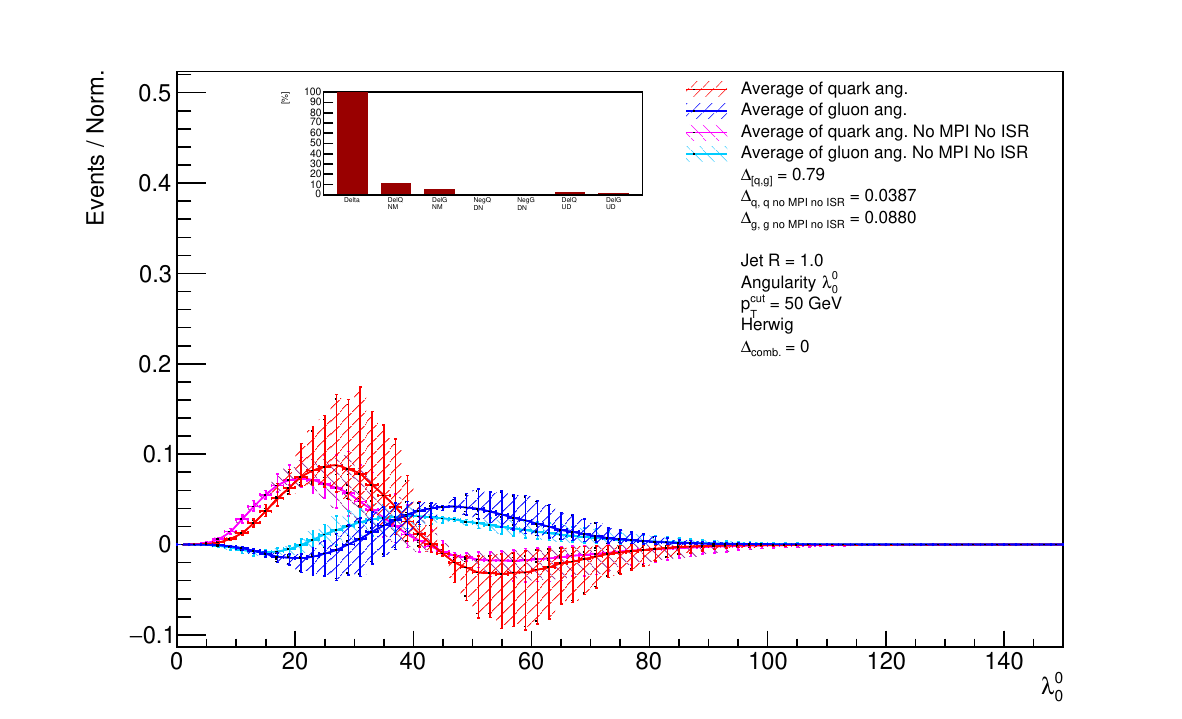} 
    \caption{Quark and gluon averaged angularities $\lambda_{0}^{0}$, $R = 1.0$ with score $\Delta_{\mathrm{comb}}=0$. Using \herwig{} event generator, with \ptcut{50}, using the average of 6 energy combinations 
900--2360, 900--7000, 900--13000, 2360--7000, 2360--13000, 7000--13000~GeV.
In the subpad, the columns from left to right show Delta (1st column),  DelQ NM and DelG NM (2nd and 3rd columns),  $\mathrm{NegQ~DN}$ and $\mathrm{NegG~DN}$  (4th and 5th columns), DelQ UD and DelG UD
(6th and 7th columns). 
}
    \label{fig:multi_bad_example}
\end{figure*}
According to this measure, the best observable is the multiplicity $\lambda_0^0$ 
with $R=1.0$ for \ptcut{50} that is shown in Fig.~\ref{fig:multi_bad_example}. 
\begin{figure*}[ht!]
    \centering
    \includegraphics[width=\textwidth]{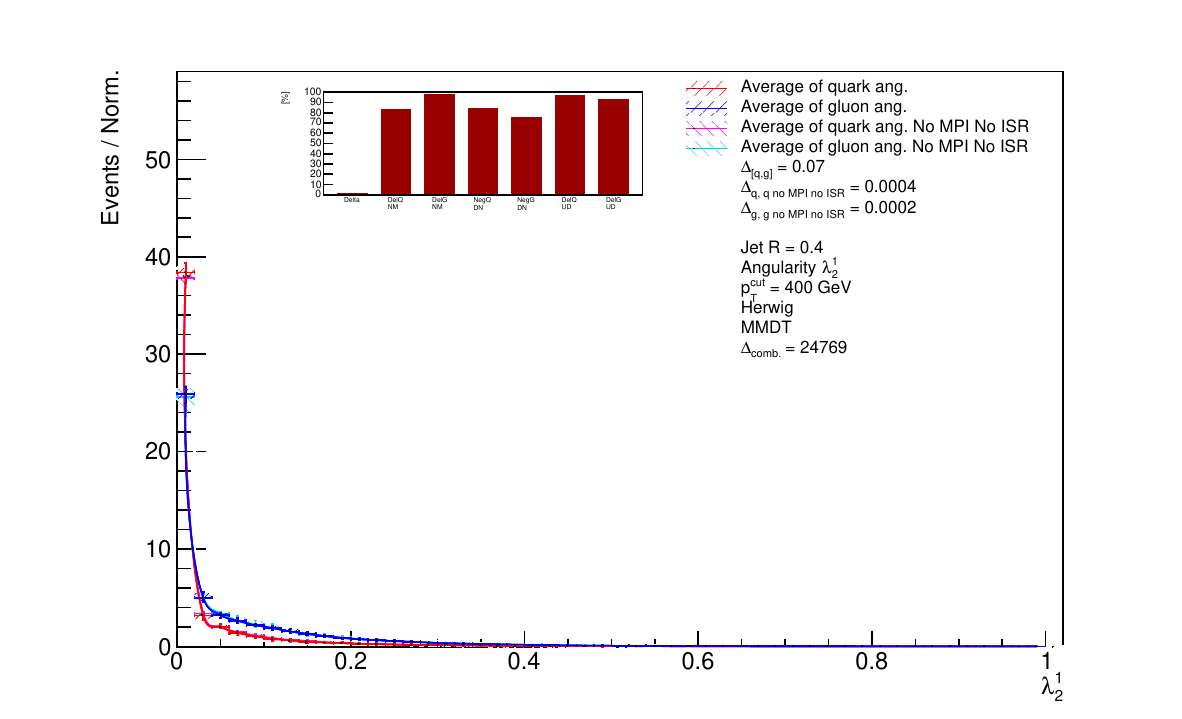} 
    \caption{Quark and gluon averaged angularities $\lambda_{2}^{1}$, $R = 0.4$ with score $\Delta_{\mathrm{comb}}=24769$. Using \herwig{} event generator, with \ptcut{400}, using the average of 6 energy combinations 
900--2360, 900--7000, 900--13000, 2360--7000, 2360--13000, 7000--13000~GeV. In the subpad, the columns from left to right show Delta (1st column),  DelQ NM and DelG NM (2nd and 3rd columns),  $\mathrm{NegQ~DN}$ and $\mathrm{NegG~DN}$  (4th and 5th columns), DelQ UD and DelG UD
(6th and 7th columns). }
    \label{fig:mass_bad_separation}
\end{figure*}
As we can see from the figure,  despite a large separation power, this observable suffers from various problems. First, it is not robust to MPI and ISR effects, and is also very energy dependent, which leads to unphysical negative bins of the probability distributions.  For this reason, it is clear that this single measure is not suitable for our problem; therefore, we introduce additional measures to evaluate the robustness to the important systematic effects. 
To check the robustness of an observable to MPI and ISR effects, we calculated the separation power between the quark (gluon) angularity obtained with and without MPI and ISR i.e.: $\Delta_{[q, \;q\mathrm{~no~MPI~no~ISR}]}$ ($\Delta_{[g,\;g\mathrm{~no~MPI~no~ISR}]}$). If these values are close to zero, then the observable is not sensitive to MPI and ISR effects. Similarly, to determine the extent to which the observable is energy independent, we calculated its separation power between the upper boundary and the lower boundary of the energy envelope of an angularity (see, for example, the energy envelope in Fig.~\ref{fig:multi_bad_example}). For quark (gluon) angularity, 
we denote this measure by $\Delta_{[q(s)_{\mathrm{UP}}, \;q(s)_{\mathrm{DOWN}}]}$
($\Delta_{[g(s)_{\mathrm{UP}}, \;g(s)_{\mathrm{DOWN}}]}$) and
its value close to zero means that the observable is energy independent. Finally, to measure whether the 
reconstructed observable suffers from the fact that part of it is negative, 
we calculate the percentage of negative area 
of down variation of the uncertainty band of quark (gluon) angularity and 
call it quark (gluon) negativity.
We show the distributions of all of these measures in the Appendix (including repetition of Figs.~\ref{fig:separation} and~\ref{fig:separation_ang} for completeness). We see that the MPI and ISR affect larger radius jets more than smaller (Fig.~\ref{fig:negativity_R_MPI}), as would be expected, and multiplicity much more than the other angularities (Fig.~\ref{fig:negativity_ang_MPI}). Its effect gets somewhat less important at higher \ptcutwb\ (Fig.~\ref{fig:negativity_Q_MPI}). The energy-dependence shows a rather similar behaviour (Figs.~\ref{fig:negativity_R_E}--\ref{fig:negativity_Q_E}), except that the smallest jet radius, 0.2, shows a strong dependence (Fig.~\ref{fig:negativity_R_E}). The energy-dependence of the multiplicity distributions is again strong (Fig.~\ref{fig:negativity_ang_E}). A strong energy dependence or MPI and ISR dependence also leads to a significant amount of negativity in the distributions, and the negativity tends to follow these same patterns (Figs.~\ref{fig:negativity_R_neg}--\ref{fig:negativity_Q_neg}).

Scores for a given observable of each metric, including \qgsp, are shown as red columns in the subpad, see, for example, Figure~\ref{fig:multi_bad_example}.
Each column represents the percentiles of a measure of angularities in all \ptcutwb\ regions 50, 100, 200, 
and 400~GeV. 
The higher the column, the better the performance of the characteristic; for example, if the 
first column denoted by Delta
in Figure~\ref{fig:multi_bad_example} is the highest (100\%) it means that no other angularity has a higher separation power $\Delta_{[q,g]}$.
Similarly, if the other columns are high, it means that the corresponding measures are good, i.e.\
have low values.
Successively, the columns from left to right show the percentiles of $\Delta_{[q,g]}$ (1st column -- Delta), the percentiles of $\Delta_{[q,~q~\mathrm{no~MPI~no~ISR}]}$ 
and $\Delta_{[g,~g~\mathrm{no~MPI~no~ISR}]}$ (2nd -- DelQ NM and 3rd -- DelG NM columns), the percentiles of quark and gluon negativity (4th -- NegQ~DN and 5th -- NegG~DN columns), 
the percentiles of $\Delta_{[q(s)_{\mathrm{UP}}, \;q(s)_{\mathrm{DOWN}}]}$ and $\Delta_{[g(s)_{\mathrm{UP}}, \;g(s)_{\mathrm{DOWN}}]}$
(6th - DelQ UD and 7th - DelG UD columns). 
From the subpanel of Fig.~\ref{fig:multi_bad_example}, we can read that 
the multiplicity $\lambda_0^0$ 
with $R=1.0$ for \ptcut{50} has the best separation power (1st column) but
the fact that all other columns (2--7) are very low show that this is amongst the worst observables on these metrics.
On the other hand, we can also have examples of observables that are very robust to all systematic effects, but have minimal separation power; see, for example, Fig.\ref{fig:mass_bad_separation}.
Therefore, in the next section, we will provide selections of plots which have strong robustness to systematic effects and have a high separation power.
\section{Results}
\label{sec:results}
In studying these different observables, we have generated a huge number of distributions, since we consider all combinations of:
\begin{itemize}
    \item 5 -- angularities $\lambda_0^0$, $\lambda_{0.5}^1$, $\lambda_1^1$, $\lambda_0^2$, $\lambda_2^1$ 
    \item 2 -- using groomed (MMDT) / not groomed jets 
    \item 5 -- jet radii $R = 0.2, 0.4, 0.6, 0.8, 1.0$
    \item 4 -- regions - dijet average \ptcut{50}, 100, 200, and 400~GeV
    \item 2 -- quark/gluon
    \item 2 -- MPI and ISR switched on/off
    \item 6 -- energy combinations: 900--2360, 900--7000, 900--13000, 2360--7000, 2360--13000, 7000--13000~GeV
    \item 2 -- event generators \herwig{} and \pythia{}
\end{itemize}
This results in 200 plots for each of the two generators, each containing four distributions, each with an envelope of six different energy combinations, or 9600 distributions in total. Our aim is to sort through these to find the best performing combinations. Clearly to do so, we need a quantitative quality measure.

With the aim of picking the best candidates, maximising the separation power while also considering the other quality measures, we define the combined
measure $\Delta_{\mathrm{comb}}$ as:
\begin{eqnarray}
    \Delta_{\mathrm{comb}} = 1000 \cdot \ln \Big [ 1 +(\mathrm{Delta})^{3}\cdot (\mathrm{DelQ~NM}) \cdot \nonumber\\ 
     (\mathrm{DelG~NM}) \cdot (\mathrm{NegQ~DN}) \cdot (\mathrm{NegG~DN}) \cdot \nonumber\\  
     (\mathrm{DelQ~UD}) \cdot (\mathrm{DelG~UD}) \Big ]
\end{eqnarray}
where the power of three enhances the separation power $\Delta_{[q,g]}$. Each of the inputs into this formula is the percentile of the corresponding variable, i.e.\ the red bars in the inset. The addition of 1 is mainly only relevant to ensure that if a given observable is the worst on any single measure, it will be given an overall score of zero, and is otherwise unimportant. If there was one observable that was the best on every criterion, it would score the maximum possible value\footnote{Note that we do not ascribe any importance to the absolute value of $\Delta_{\mathrm{comb}}$, it is purely a means to rank the observables.} of $\Delta_{\mathrm{comb}}=41447$.
We focus initially on the results from \herwig{} and return to the comparison with \pythia{} in the next section.

\begin{figure}[ht!]
\centering
\includegraphics[width=0.5\textwidth]{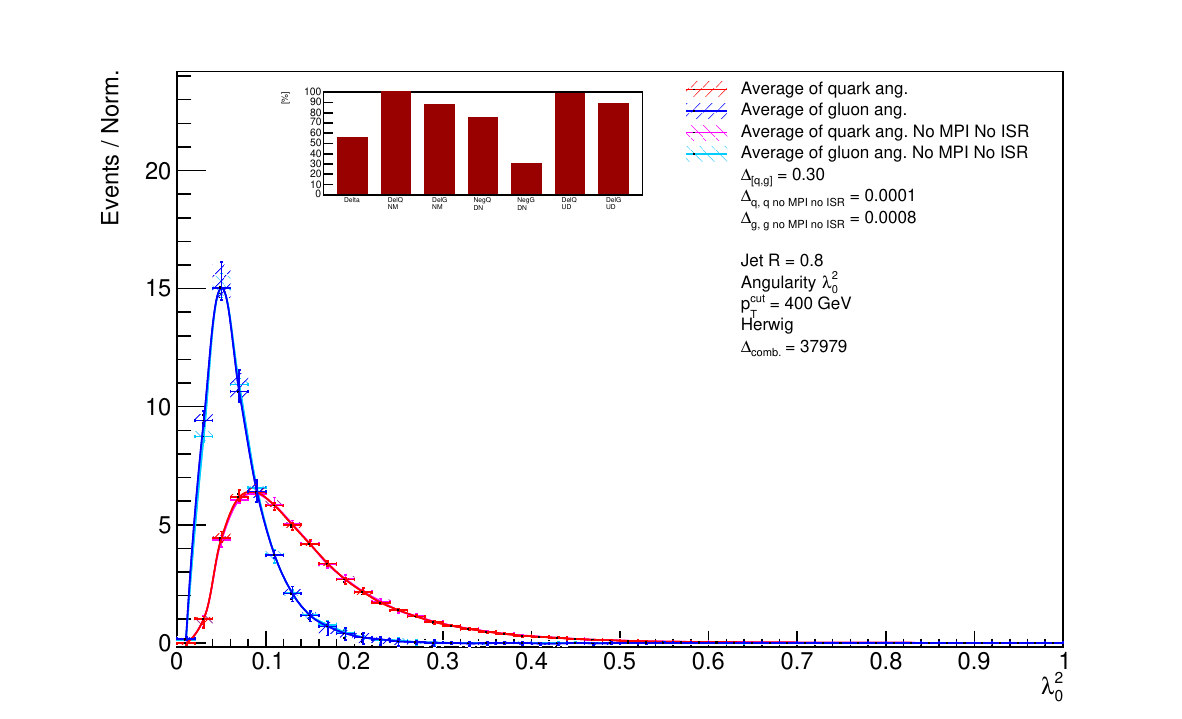} 
\caption{Quark and gluon averaged angularities $\lambda_{0}^{2}$, $R = 0.8$ with highest score $\Delta_{\mathrm{comb}}=37979$. Using \herwig{} event generator, with \ptcut{400}, using the average of 6 energy combinations 
900--2360, 900--7000, 900--13000, 2360--7000, 2360--13000, 7000--13000~GeV.}
\label{fig:best1}
\end{figure}

\begin{figure}[ht!]
\centering
\includegraphics[width=0.5\textwidth]{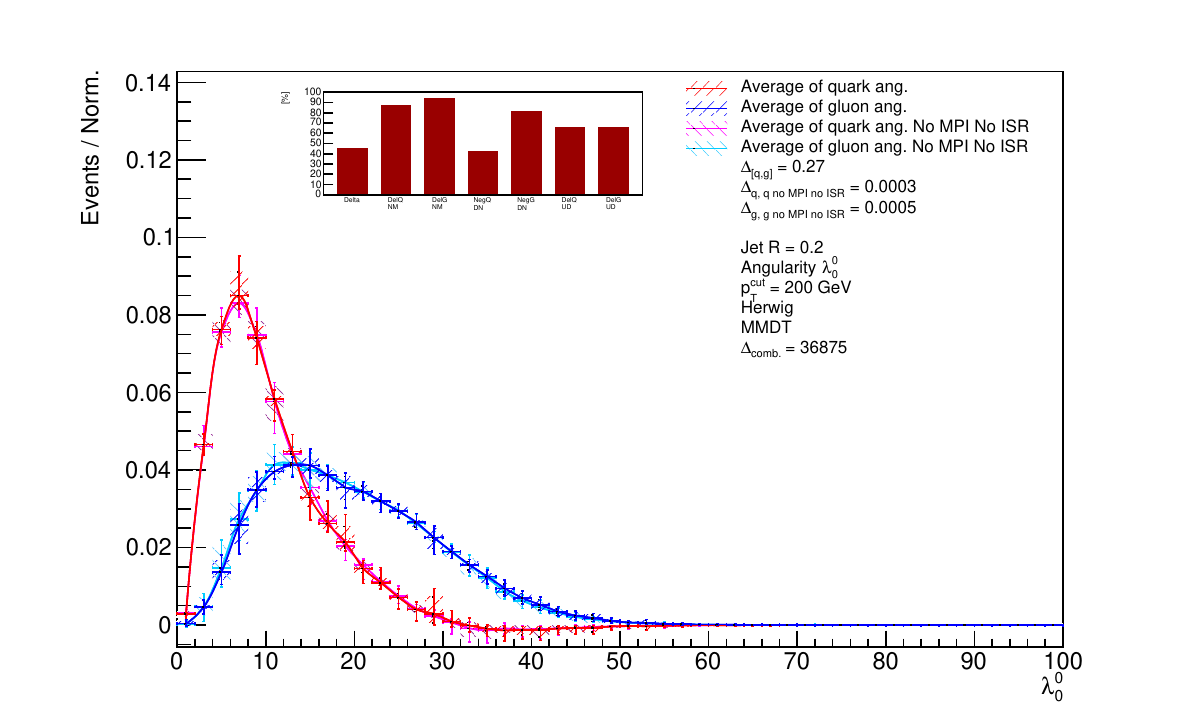} 
\caption{Quark and gluon averaged angularities $\lambda_{0}^{0}$, $R = 0.2$ with score $\Delta_{\mathrm{comb}}=36875$. Using \herwig{} event generator, with \ptcut{200}, using the average of 6 energy combinations 
900--2360, 900--7000, 900--13000, 2360--7000, 2360--13000, 7000--13000~GeV.}
\label{fig:best2}
\end{figure}

\begin{figure}[ht!]
\centering
\includegraphics[width=0.5\textwidth]{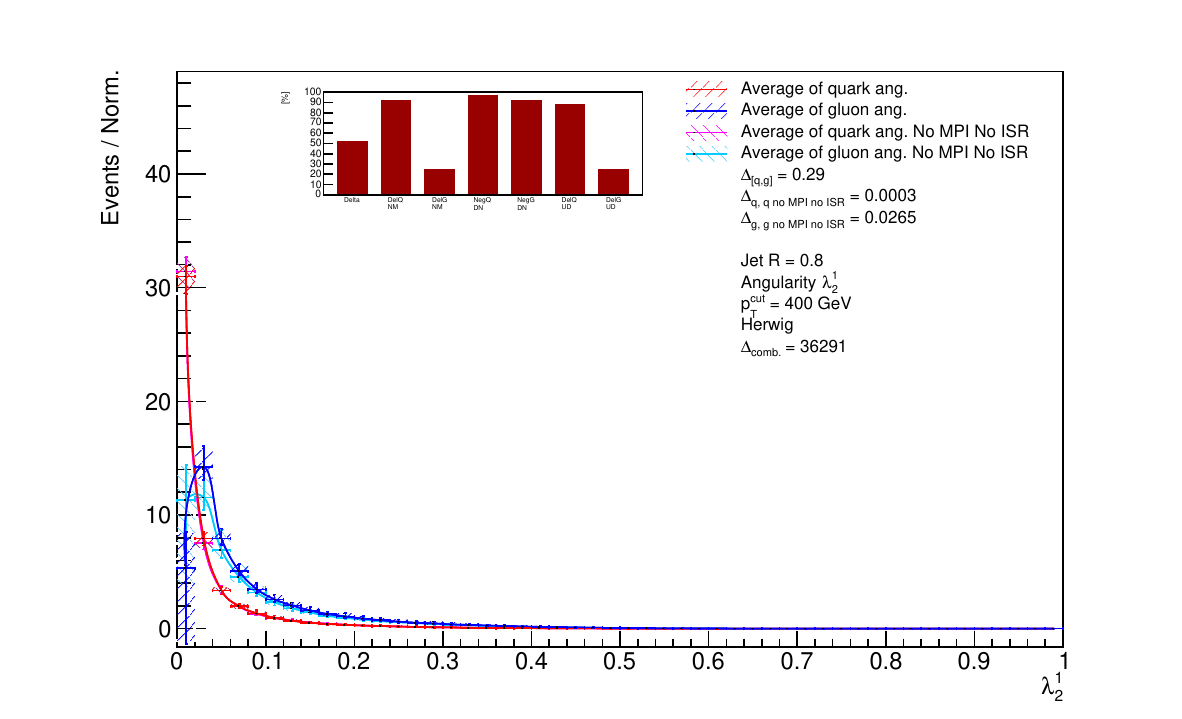} 
\caption{Quark and gluon averaged angularities $\lambda_{2}^{1}$, $R = 0.8$ with score $\Delta_{\mathrm{comb}} = 36291$. Using \herwig{} event generator, with \ptcut{400}, using the average of 6 energy combinations 
900--2360, 900--7000, 900--13000, 2360--7000, 2360--13000, 7000--13000~GeV.}
\label{fig:best3}
\end{figure}

\begin{figure}[ht!]
\centering
\includegraphics[width=0.5\textwidth]{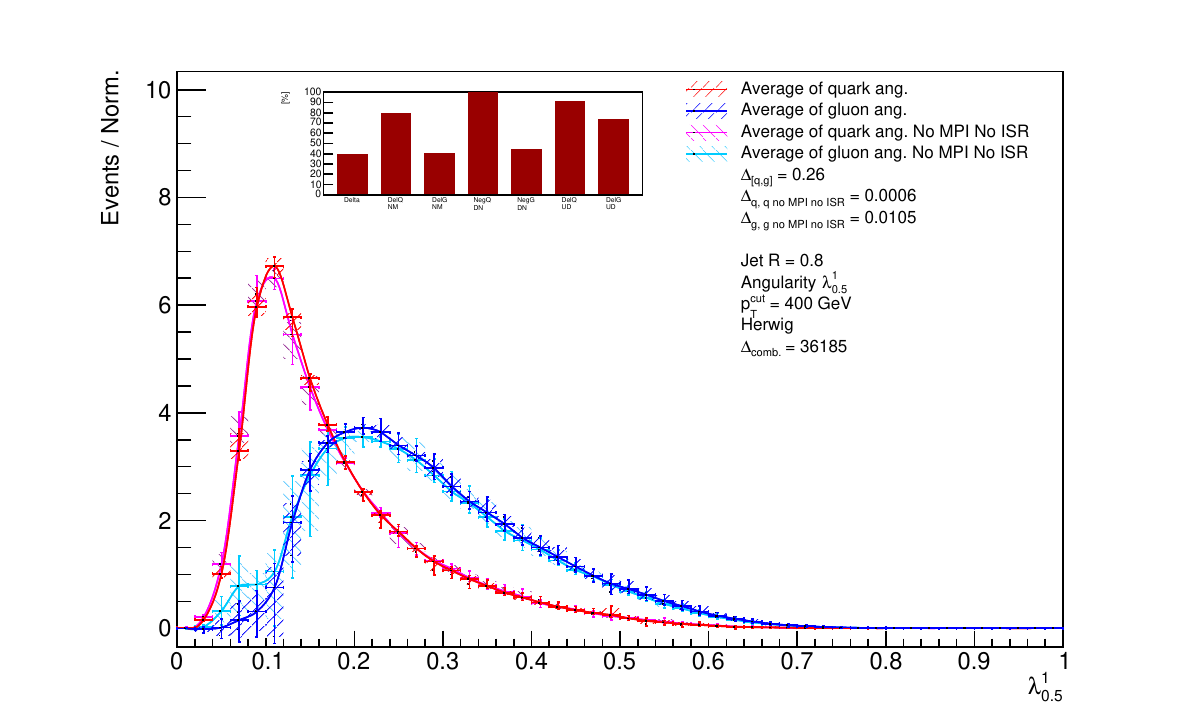} 
\caption{Quark and gluon averaged angularities MMDT $\lambda_{0.5}^{1}$, $R = 0.8$ with score $\Delta_{\mathrm{comb}} = 36185$. Using \herwig{} event generator, with \ptcut{400}, using the average of 6 energy combinations 
900--2360, 900--7000, 900--13000, 2360--7000, 2360--13000, 7000--13000~GeV.}
\label{fig:best4}
\end{figure}

\begin{figure}[ht!]
\centering
\includegraphics[width=0.5\textwidth]{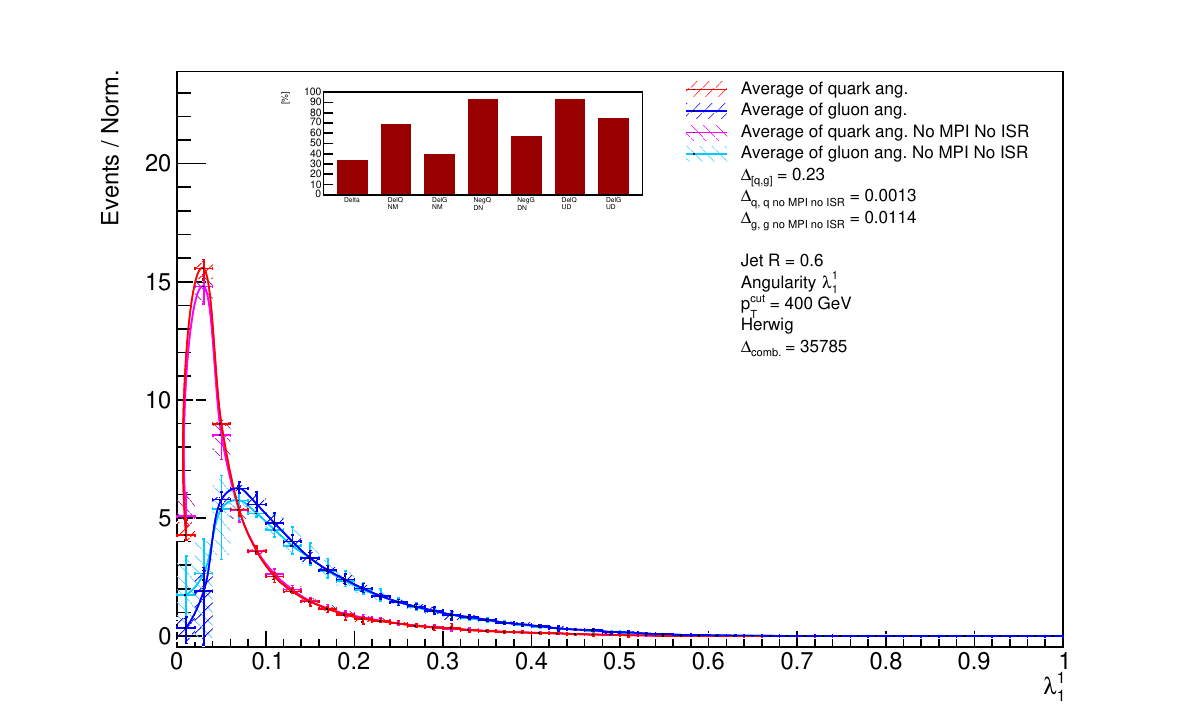} 
\caption{Quark and gluon averaged angularities $\lambda_{1}^{1}$, $R = 0.6$ with score $\Delta_{\mathrm{comb}} = 35785$. Using \herwig{} event generator, with \ptcut{400}, using the average of 6 energy combinations 
900--2360, 900--7000, 900--13000, 2360--7000, 2360--13000, 7000--13000~GeV.}
\label{fig:best5}
\end{figure}

Figures~\ref{fig:best1},~\ref{fig:best2},~\ref{fig:best3},~\ref{fig:best4} and \ref{fig:best5} represent 
the best selection based on $\Delta_{\mathrm{comb}}$ score for each type of angularity.

\begin{figure}[ht!]
\centering
\includegraphics[width=0.5\textwidth]{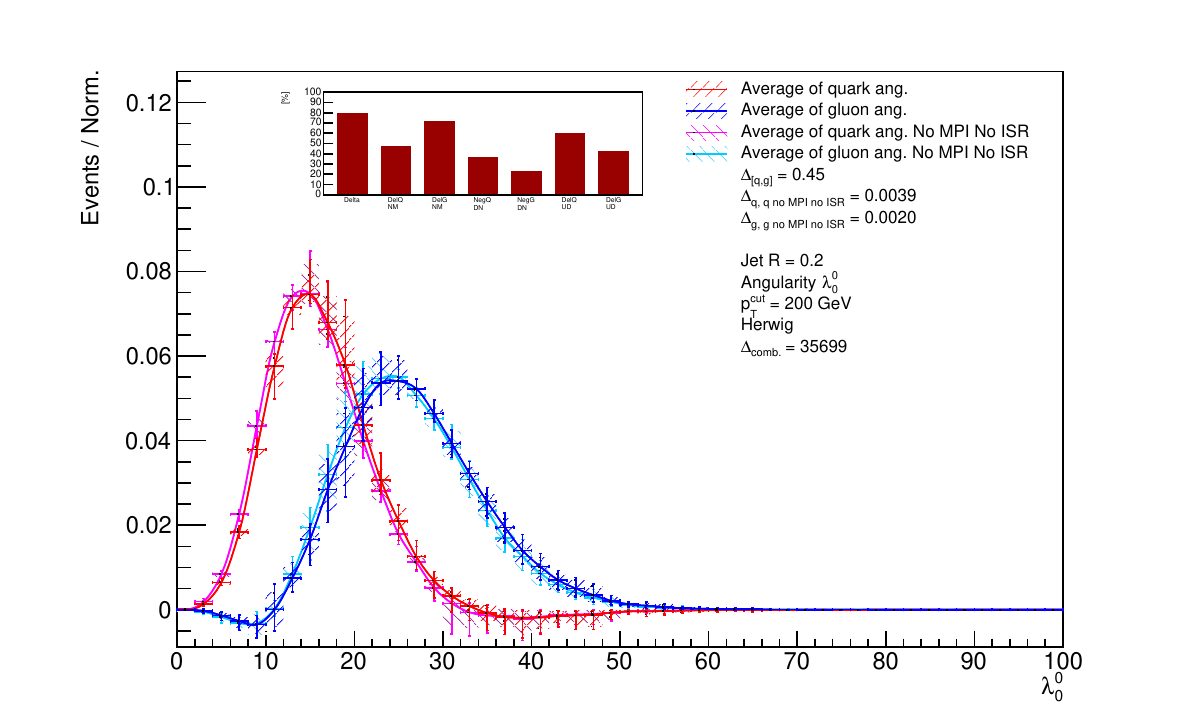} 
\caption{Quark and gluon averaged angularities $\lambda_{0}^{0}$, $R = 0.2$ with score $\Delta_{\mathrm{comb}}=35699$. Using \herwig{} event generator, with \ptcut{200}, using the average of 6 energy combinations 
900--2360, 900--7000, 900--13000, 2360--7000, 2360--13000, 7000--13000~GeV.}
\label{fig:wildcard2}
\end{figure}

\begin{figure}[ht!]
    \centering
    \includegraphics[width=0.5\textwidth]{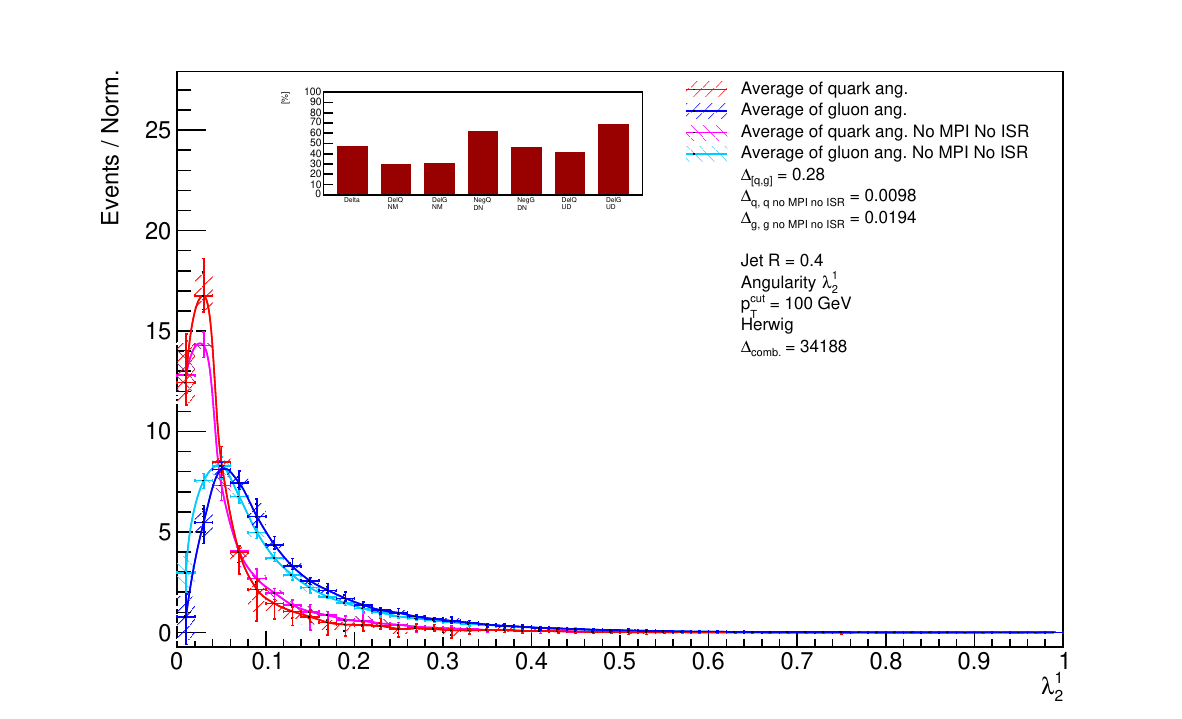} 
    \caption{Quark and gluon averaged angularities $\lambda_{2}^{1}$, $R = 0.4$ with score $\Delta_{\mathrm{comb}} = 34188$. Using \herwig{} event generator, with \ptcut{100}, using the average of 6 energy combinations 
    900--2360, 900--7000, 900--13000, 2360--7000, 2360--13000, 7000--13000~GeV.}
    \label{fig:wildcard3}
    \end{figure}

\begin{figure}[ht!]
    \centering
    \includegraphics[width=0.5\textwidth]{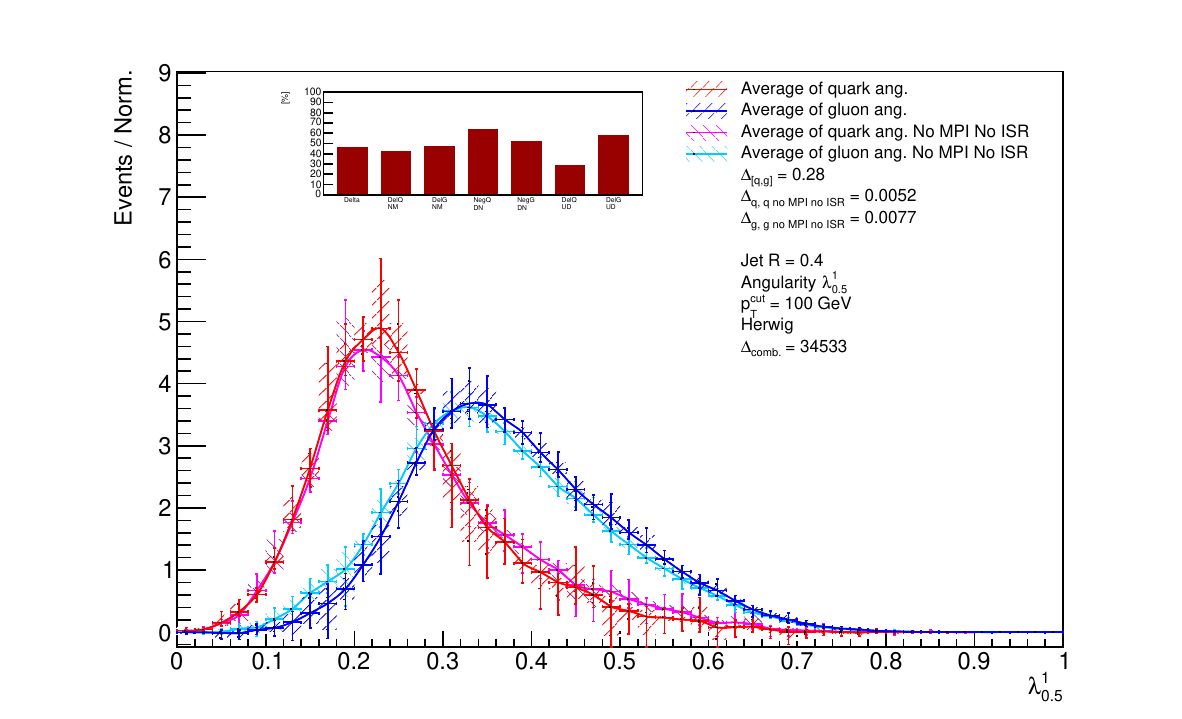} 
    \caption{Quark and gluon averaged angularities MMDT $\lambda_{0.5}^{1}$, $R = 0.4$ with score $\Delta_{\mathrm{comb}} = 34533$. Using \herwig{} event generator, with \ptcut{100}, using the average of 6 energy combinations 
    900--2360, 900--7000, 900--13000, 2360--7000, 2360--13000, 7000--13000~GeV.}
    \label{fig:wildcard4}
    \end{figure}

\begin{figure}[ht!]
    \centering
    \includegraphics[width=0.5\textwidth]{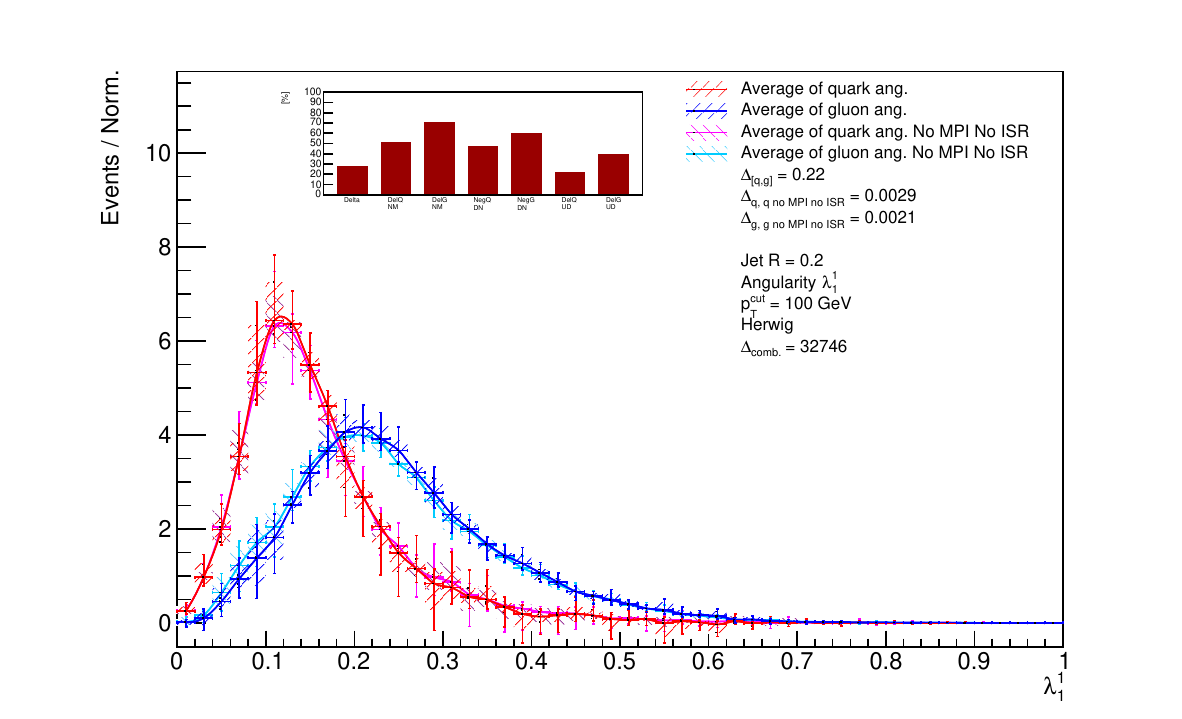} 
    \caption{Quark and gluon averaged angularities $\lambda_{1}^{1}$, $R = 0.2$ with score $\Delta_{\mathrm{comb}} = 32746$. Using \herwig{} event generator, with \ptcut{100}, using the average of 6 energy combinations 
    900--2360, 900--7000, 900--13000, 2360--7000, 2360--13000, 7000--13000~GeV.}
    \label{fig:wildcard5}
    \end{figure}

In addition to these best of each type, we have also selected four others, shown in
Figures~\ref{fig:wildcard2},~\ref{fig:wildcard3},~\ref{fig:wildcard4} and~\ref{fig:wildcard5} based more on giving a representation of a typical range of good results, even though the score $\Delta_{\mathrm{comb.}}$ is not the highest. They generally show a good quark/gluon jet separation, even if they
suffer from lower robustness to variations without MPI and ISR or negativity, etc.

The high-scored angularities presented in the plots provide compelling evidence supporting the assumption that quark and gluon angularities remain independent of collision energy. This conclusion is drawn from the relatively narrow envelope of the filled area, which represents angularities derived at different energy combinations. The consistency observed across these plots strongly suggests that quark and gluon angularities are not significantly affected by changes in collision energy. However, it is essential to acknowledge that this assumption may not be entirely valid for all angularities, as shown in Figure~\ref{fig:multi_bad_example} where the filled area is broader. Therefore, it is important to consider this uncertainty when interpreting the results.

\subsection{Comparison with \pythia{}}
\begin{figure*}[ht!]
    \centering
    \includegraphics[width=0.49\textwidth]{./figures/pubplots/FastJets08PtLam400__herwig_average.pdf}
    \includegraphics[width=0.49\textwidth]{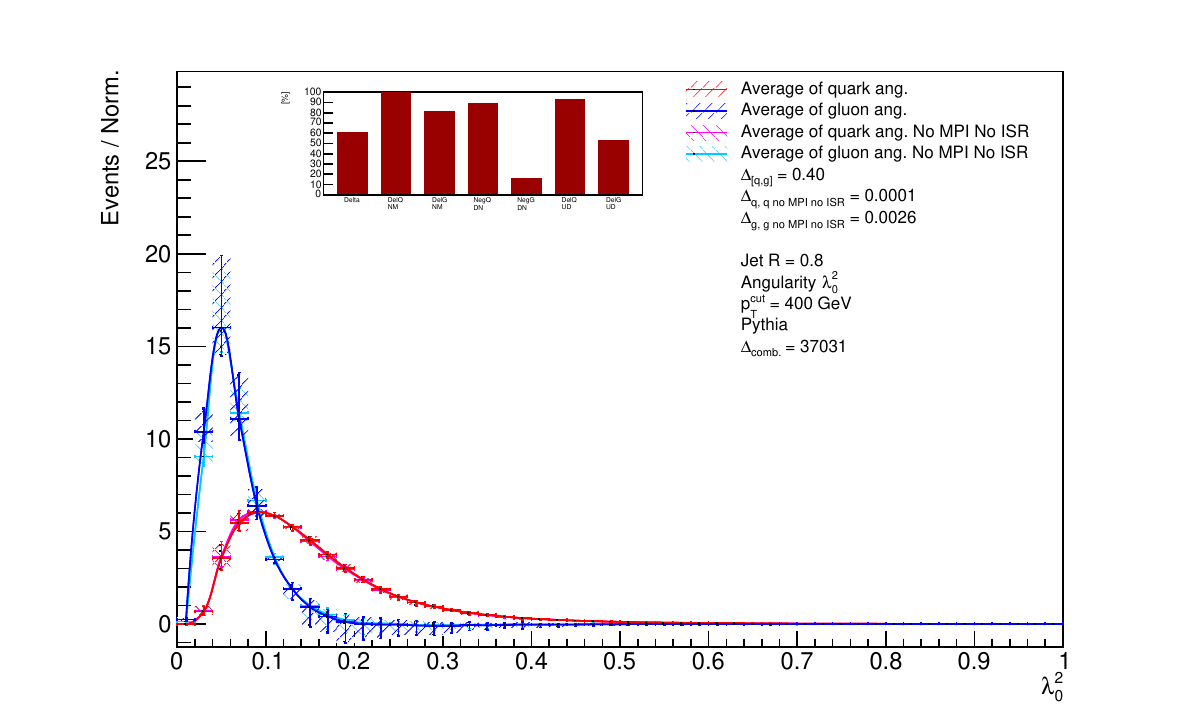} 
    \caption{Quark and gluon averaged angularities $\lambda_{0}^{2}$, $R = 0.8$ with highest score $\Delta_{\mathrm{comb}}=37979$ using \herwig{} event generator (left) and $\Delta_{\mathrm{comb}}=37031$ using \pythia{} event generator (right), with \ptcut{400}, using the average of 6 energy combinations 
    900--2360, 900--7000, 900--13000, 2360--7000, 2360--13000, 7000--13000~GeV.}
    \label{fig:best1_pythia}
\end{figure*}

We have rerun the preceding analysis using the \pythia{} event generator in place of \herwig{}. The results are very similar in almost all cases. As an example, we show the angularity that has the highest $\Delta_{\mathrm{comb}}$ score in Figure~\ref{fig:best1_pythia}.

\begin{figure*}[ht!]
\centering
\includegraphics[width=0.49\textwidth]{./figures/pubplots/FastJets08LhaLam400__herwig_average.pdf} 
\includegraphics[width=0.49\textwidth]{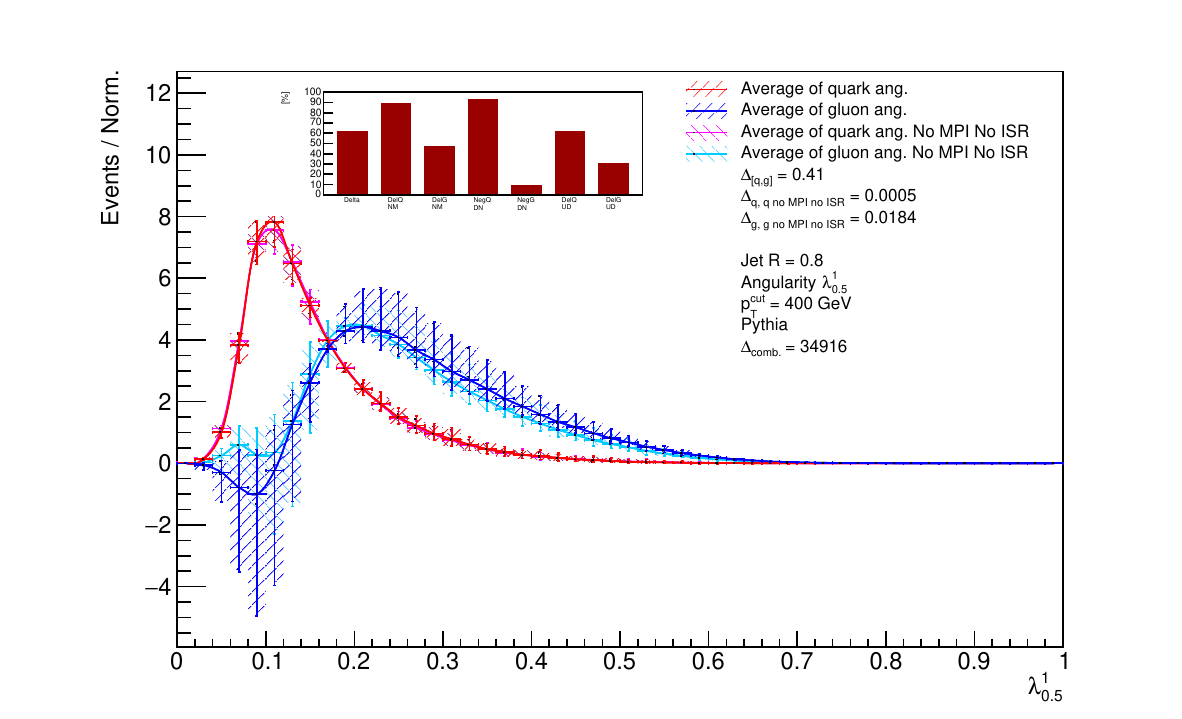} 
\caption{Quark and gluon averaged angularities MMDT $\lambda_{0.5}^{1}$, $R = 0.8$ with score $\Delta_{\mathrm{comb}} = 36185$ using \herwig{} event generator (left) and $\Delta_{\mathrm{comb}} = 34916$ using \pythia{} event generator (right), with \ptcut{400}, using the average of 6 energy combinations 
900--2360, 900--7000, 900--13000, 2360--7000, 2360--13000, 7000--13000~GeV.}
\label{fig:best4_pythia}
\end{figure*}

We also show, in Figure~\ref{fig:best4_pythia}, the example from all those we have studied that shows the biggest difference between \herwig{} and \pythia{}.
In general, and most noticeably in Fig.~\ref{fig:best4_pythia}, \pythia{} shows slightly more energy- and MPI/ISR-dependence than \herwig{}, and as a result slightly more negativity. Although for most of the robust observables this is a small effect, it is possible that it could also be used to constrain the underlying models, by studying observables where it is significant.

\section{Conclusion}
\label{sec:summary}
We have shown that several jet angularity observables can be robust in measuring the properties of jets and successful in yielding significantly different distributions for quark and gluon jets. And, moreover, that the energy-dependence of the distributions can be used at the LHC to separate the two on a statistical basis.
The method relies crucially on the assumption that the angularities for quark and gluon jets are separately independent of $\sqrt{s}$, and we have shown this to be the case, particularly for higher $p_T$ jets. Only for multiplicity in large-radius jets are the uncertainties too high to be useful.

Of course, the LHC will not make special runs at different energies only to conduct the proposed measurement. Therefore, we should use data already recorded, or data that will soon be measured at the LHC. During its early startup phase, CMS recorded some minimum-bias events of proton-proton collisions at energies $\sqrt{s} = 900$~GeV and $\sqrt{s} = 2360$~GeV. In the publication~\cite{CMS:2010bva}, they presented properties of inclusive jets and dijet events measured in these samples.
However, the number of events with jet $p_T > 8$~GeV or $10$~GeV was
less then $1000$ or $200$, at $\sqrt{s} = 900$~GeV and $\sqrt{s} = 2360$~GeV respectively.  
Due to the low statistics and the low $p_T$ jet cut, these data samples are not optimal for using our method.
For this reason, carrying out the proposed measurement at the LHC at energies of $7$ and $13$ TeV, where low statistics will not be an issue, would be a much better strategy. It is also worth mentioning that the ALICE experiment has also measured jets at energies
$\sqrt{s} = 2360$~GeV  and $5.02$~TeV. Jet measurements at these energies and in particular at $\sqrt{s} = 5.02$ TeV where the number of jets measured is high could serve to check the energy independence of the quark/gluon jet measurement. 
Moreover, recently ALICE published results~\cite{ALICE:2021njq,Lesser:2022lpc} 
of jet substructure including jet angularities carried out by the experiment using data recorded at the LHC from pp collisions at $\sqrt{s} = 5.02$ or $13$ TeV. It is possible that these data could
be reanalysed to obtain the first results using the method proposed here.
Another possibility could be to use CERN Open Data~\cite{CERNopenData} similarly to what was done for jet topics~\cite{Komiske:2022vxg}. 

One interesting extension of this research would be to use the recently developed 
IRC-safe flavoured jet algorithms~\cite{Caletti:2022glq,Banfi:2006hf,Caletti:2022hnc,Czakon:2022wam,Gauld:2022lem,Caola:2023wpj}.
Knowing the flavour of jets could be used to construct the fraction of jets after the evolution of the parton shower,
which could help to trace the origin of the jets through hadronisation and parton showering,
and finally be used to develop a Machine Learning method to optimise the q/g classification strategy. 
Another interesting extension would be to study different jet production processes, such as vector boson plus jet, at different collision energies. Not only is the flavour mix different in these processes, but also the colour structure.
Yet another intriguing possibility would be to apply the variable collision energy samples to the above-mentioned jet topics. 

\section*{Acknowledgements}
MS gratefully acknowledges funding from the UK Science and Technology Facilities Council (grant number ST/T001038/1). The work of AS and PB is funded by grant no. 2019/34/E/ST2/00457 of the National Science Centre, Poland. PB is also supported by the Priority Research Area Digiworld under the program Excellence Initiative – Research University at the Jagiellonian University in Cracow and by the Polish National Agency for Academic Exchange NAWA  under the Programme STER Internationalisation of doctoral schools, Project no. PPI/STE/2020/1/00020.

\section{Appendix}

\subsection{Comparison against ``truth level'' distributions}
In this section, we compare the quark and gluon distributions we extract against ``truth level'' distributions, which are shown as the black lines. We obtain these by assigning each reconstructed jet the flavour of the hard parton to which it is closest in direction (smallest $\Delta R$).
We see that for our most robust angularities, there is very good agreement, while for the distributions that have significant energy-dependence or negativity, the agreement is poorer.

\begin{figure}[ht!]
    \centering
    \includegraphics[width=0.5\textwidth]{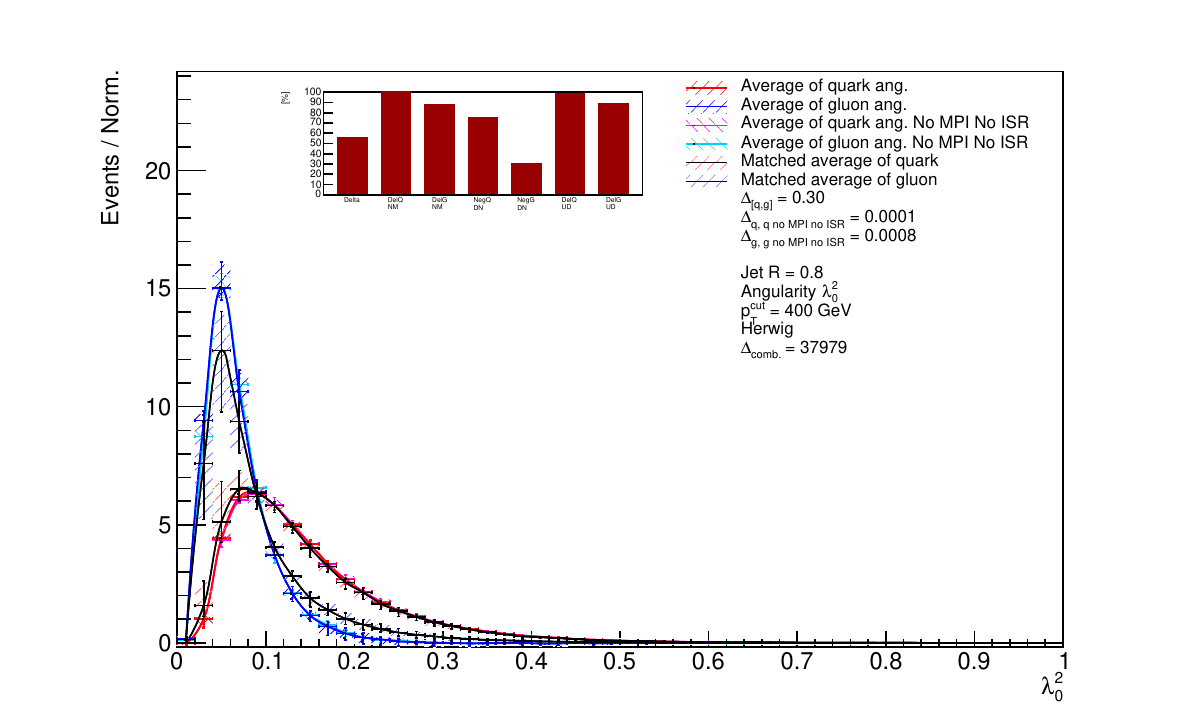} 
    \caption{Quark and gluon averaged angularities $\lambda_{0}^{2}$, $R = 0.8$ with highest score $\Delta_{\mathrm{comb}}=37979$. Using \herwig{} event generator, with \ptcut{400}, using the average of 6 energy combinations 
    900--2360, 900--7000, 900--13000, 2360--7000, 2360--13000, 7000--13000~GeV.}
    \label{fig:best1b}
    \end{figure}
    
    \begin{figure}[ht!]
    \centering
    \includegraphics[width=0.5\textwidth]{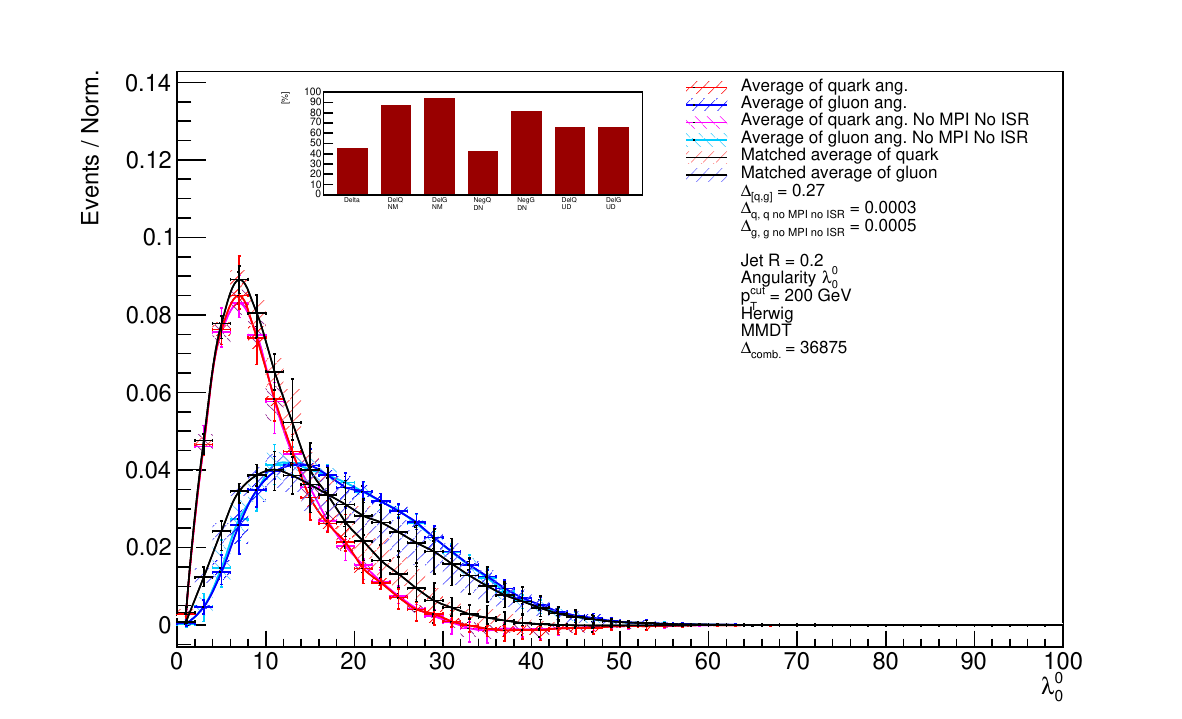} 
    \caption{Quark and gluon averaged angularities $\lambda_{0}^{0}$, $R = 0.2$ with score $\Delta_{\mathrm{comb}}=36875$. Using \herwig{} event generator, with \ptcut{200}, using the average of 6 energy combinations 
    900--2360, 900--7000, 900--13000, 2360--7000, 2360--13000, 7000--13000~GeV.}
    \label{fig:best2b}
    \end{figure}
    
    \begin{figure}[ht!]
    \centering
    \includegraphics[width=0.5\textwidth]{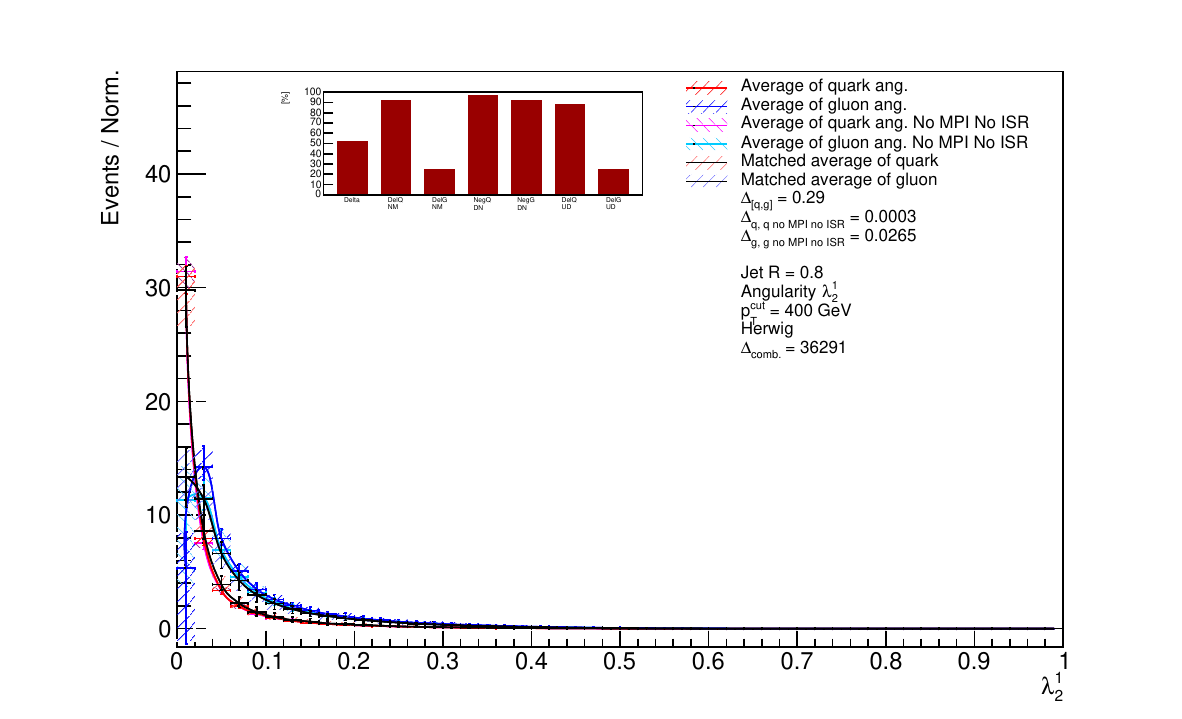} 
    \caption{Quark and gluon averaged angularities $\lambda_{2}^{1}$, $R = 0.8$ with score $\Delta_{\mathrm{comb}} = 36291$. Using \herwig{} event generator, with \ptcut{400}, using the average of 6 energy combinations 
    900--2360, 900--7000, 900--13000, 2360--7000, 2360--13000, 7000--13000~GeV.}
    \label{fig:best3b}
    \end{figure}
    
    \begin{figure}[ht!]
    \centering
    \includegraphics[width=0.5\textwidth]{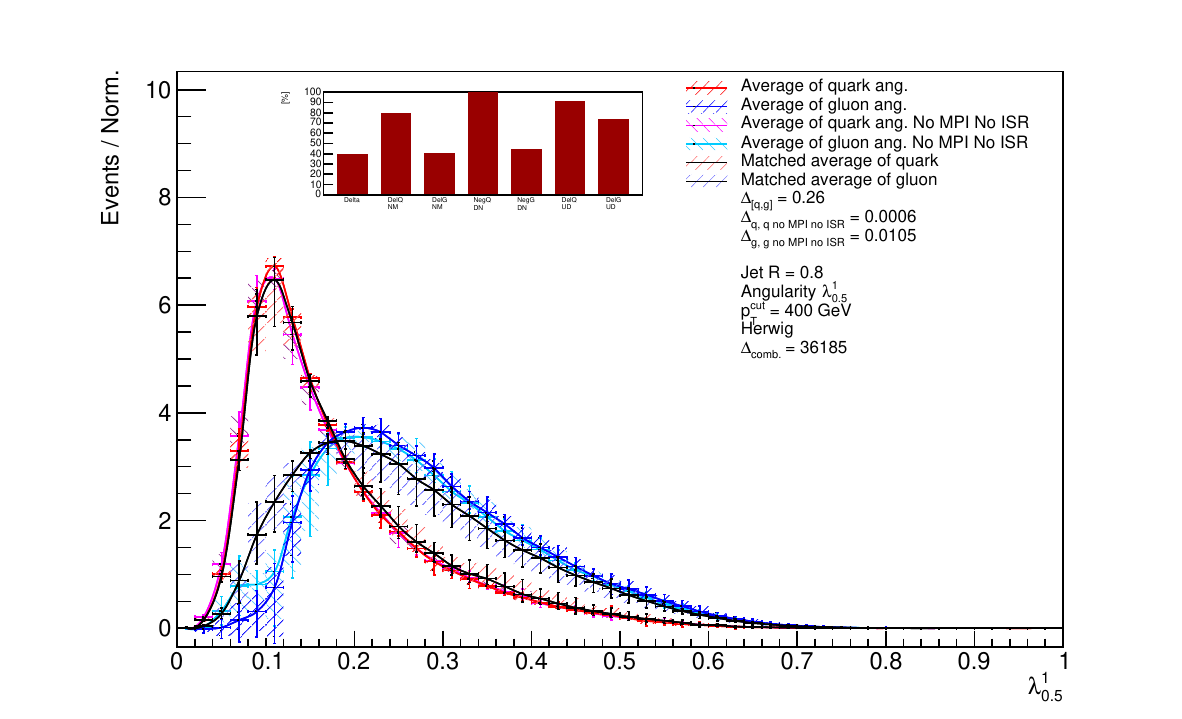} 
    \caption{Quark and gluon averaged angularities MMDT $\lambda_{0.5}^{1}$, $R = 0.8$ with score $\Delta_{\mathrm{comb}} = 36185$. Using \herwig{} event generator, with \ptcut{400}, using the average of 6 energy combinations 
    900--2360, 900--7000, 900--13000, 2360--7000, 2360--13000, 7000--13000~GeV.}
    \label{fig:best4b}
    \end{figure}
    
    \begin{figure}[ht!]
    \centering
    \includegraphics[width=0.5\textwidth]{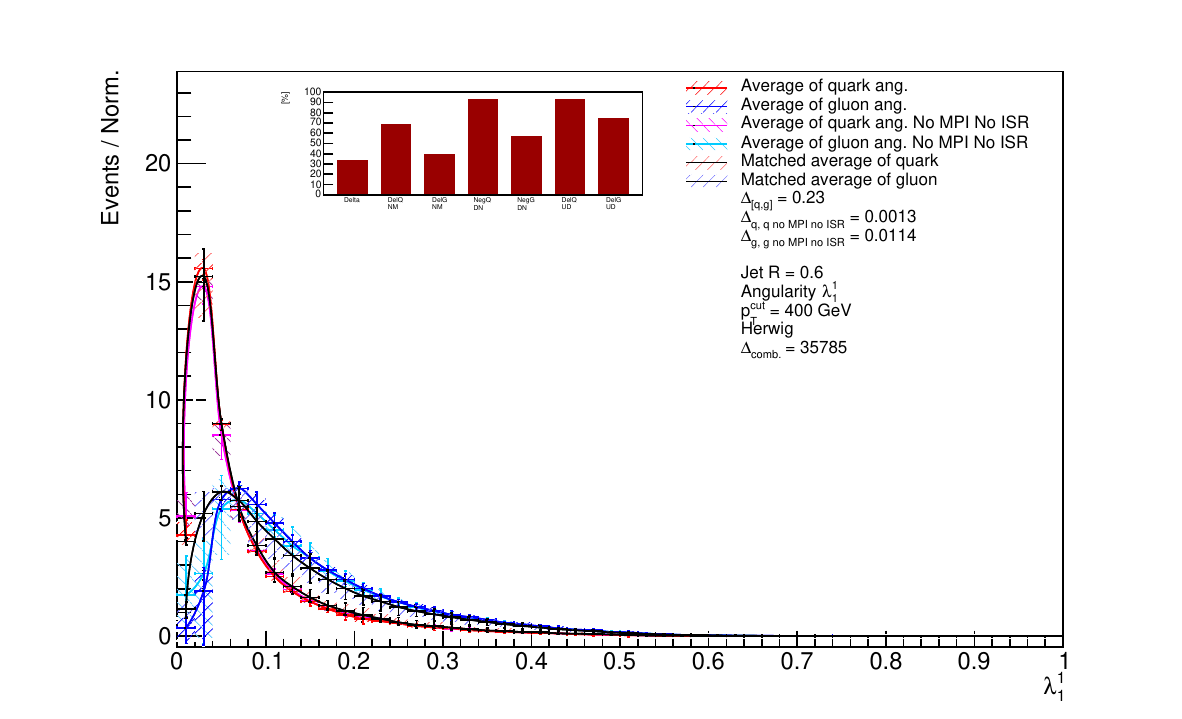} 
    \caption{Quark and gluon averaged angularities $\lambda_{1}^{1}$, $R = 0.6$ with score $\Delta_{\mathrm{comb}} = 35785$. Using \herwig{} event generator, with \ptcut{400}, using the average of 6 energy combinations 
    900--2360, 900--7000, 900--13000, 2360--7000, 2360--13000, 7000--13000~GeV.}
    \label{fig:best5b}
    \end{figure}
        
    \begin{figure}[ht!]
    \centering
    \includegraphics[width=0.5\textwidth]{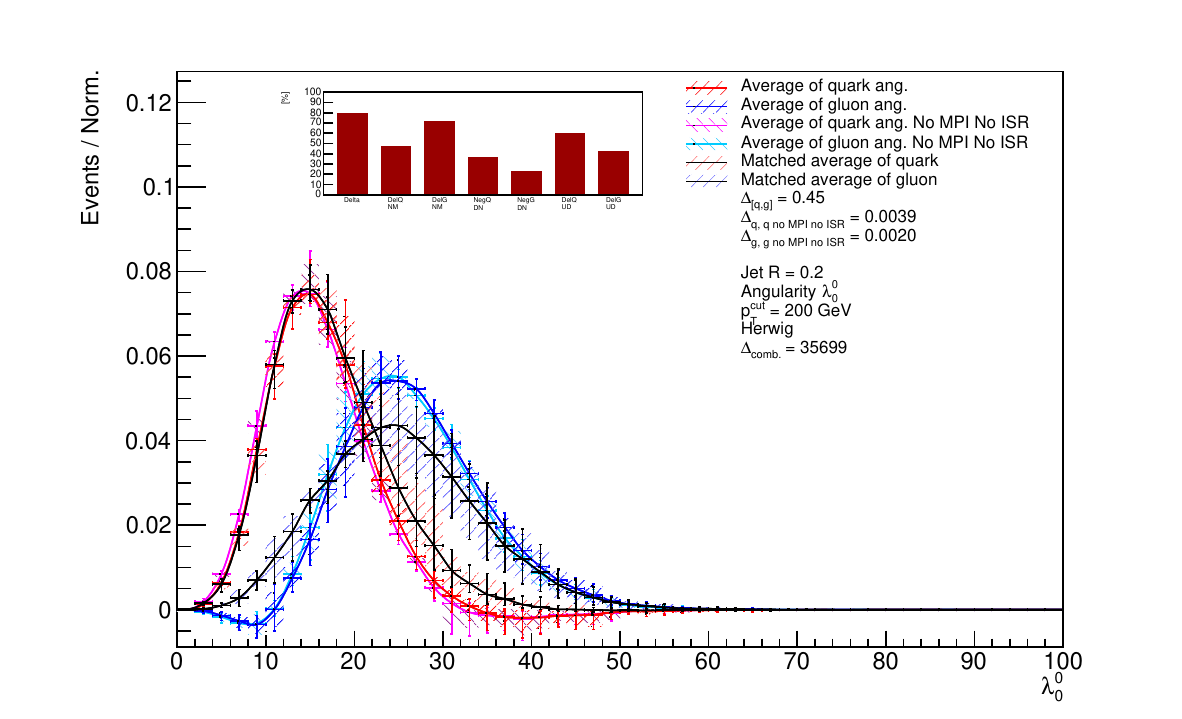} 
    \caption{Quark and gluon averaged angularities $\lambda_{0}^{0}$, $R = 0.2$ with score $\Delta_{\mathrm{comb}}=35699$. Using \herwig{} event generator, with \ptcut{200}, using the average of 6 energy combinations 
    900--2360, 900--7000, 900--13000, 2360--7000, 2360--13000, 7000--13000~GeV.}
    \label{fig:wildcard2b}
    \end{figure}

        \begin{figure}[ht!]
        \centering
        \includegraphics[width=0.5\textwidth]{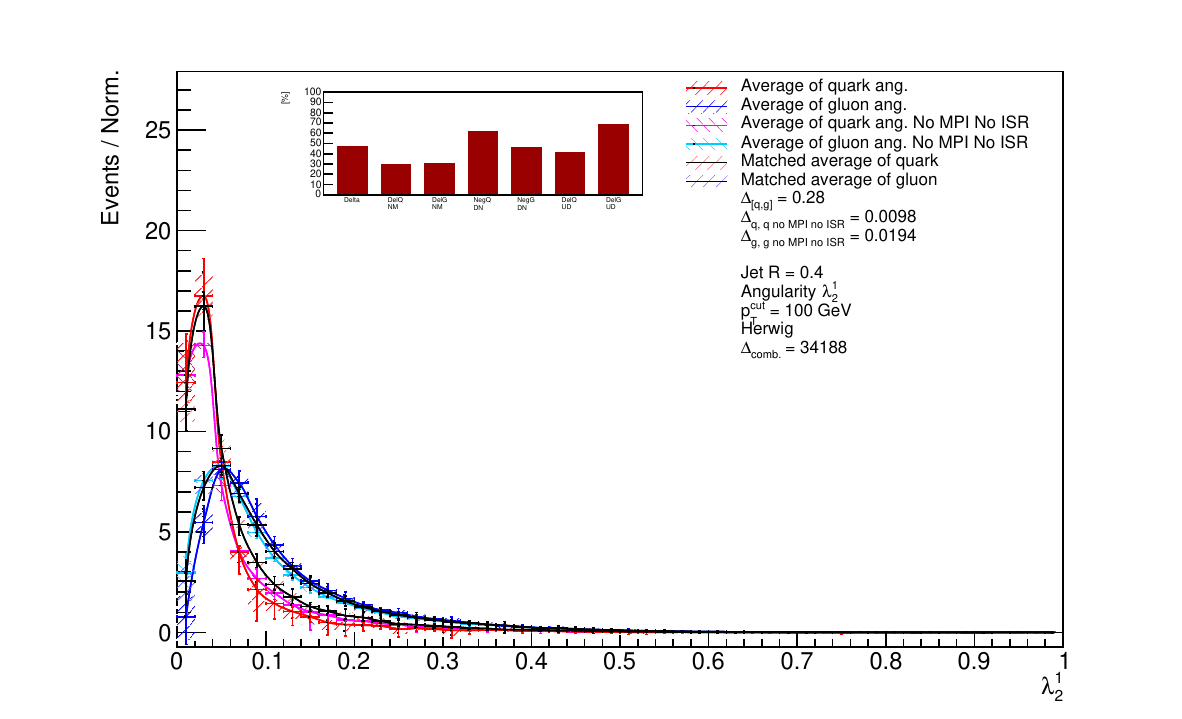} 
        \caption{Quark and gluon averaged angularities $\lambda_{2}^{1}$, $R = 0.4$ with score $\Delta_{\mathrm{comb}} = 34188$. Using \herwig{} event generator, with \ptcut{100}, using the average of 6 energy combinations 
        900--2360, 900--7000, 900--13000, 2360--7000, 2360--13000, 7000--13000~GeV.}
        \label{fig:wildcard3b}
        \end{figure}
    
\vspace*{1cm}
\subsection{Fractions using different PDF sets}
We have studied the PDF dependence of our results by comparing the results extracted using the gluon fractions from the NNPDF23 LO QED with $\alpha_s=0.130$~\cite{Ball:2013hta}, NNPDF31 LO with $\alpha_s=0.118$~\cite{NNPDF:2017mvq} and CT14lo~\cite{Dulat:2015mca} sets and find generally very small differences. In this section, we illustrate this for the CT14lo PDF set, which gives the results plotted by black lines over our default results in red and blue.
        
        \begin{figure}[ht!]
        \centering
        \includegraphics[width=0.5\textwidth]{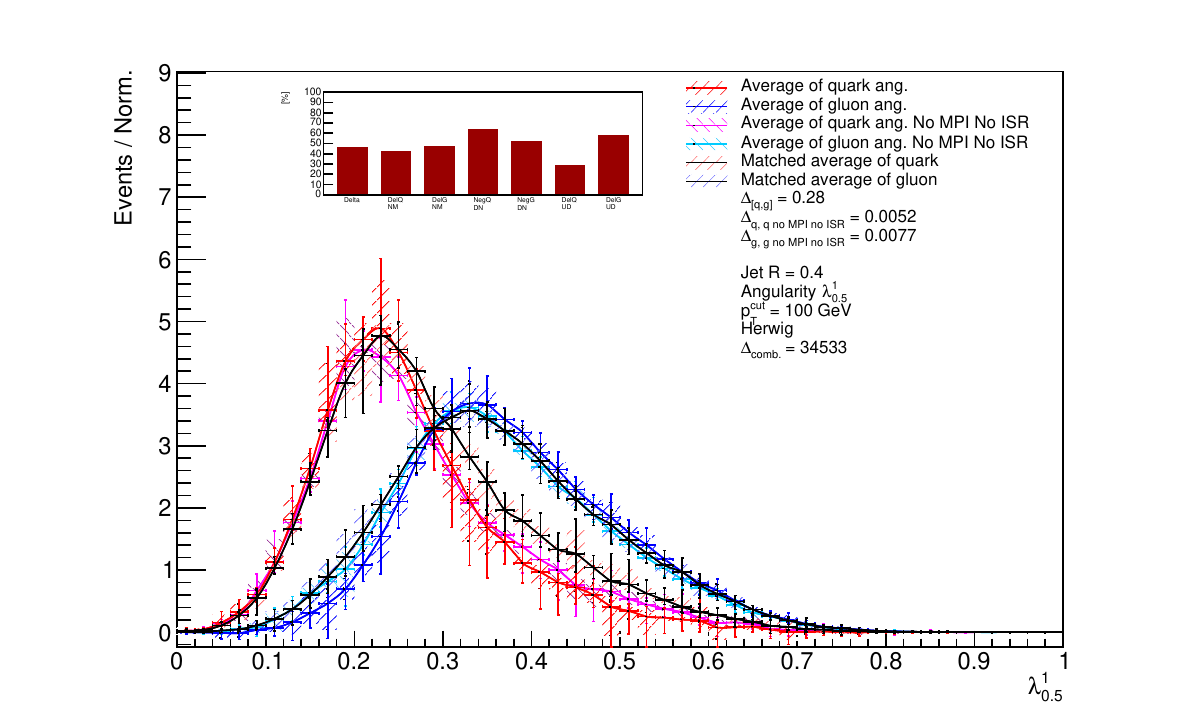} 
        \caption{Quark and gluon averaged angularities MMDT $\lambda_{0.5}^{1}$, $R = 0.4$ with score $\Delta_{\mathrm{comb}} = 34533$. Using \herwig{} event generator, with \ptcut{100}, using the average of 6 energy combinations 
        900--2360, 900--7000, 900--13000, 2360--7000, 2360--13000, 7000--13000~GeV.}
        \label{fig:wildcard4b}
        \end{figure}
        \begin{figure}[ht!]
        \centering
        \includegraphics[width=0.5\textwidth]{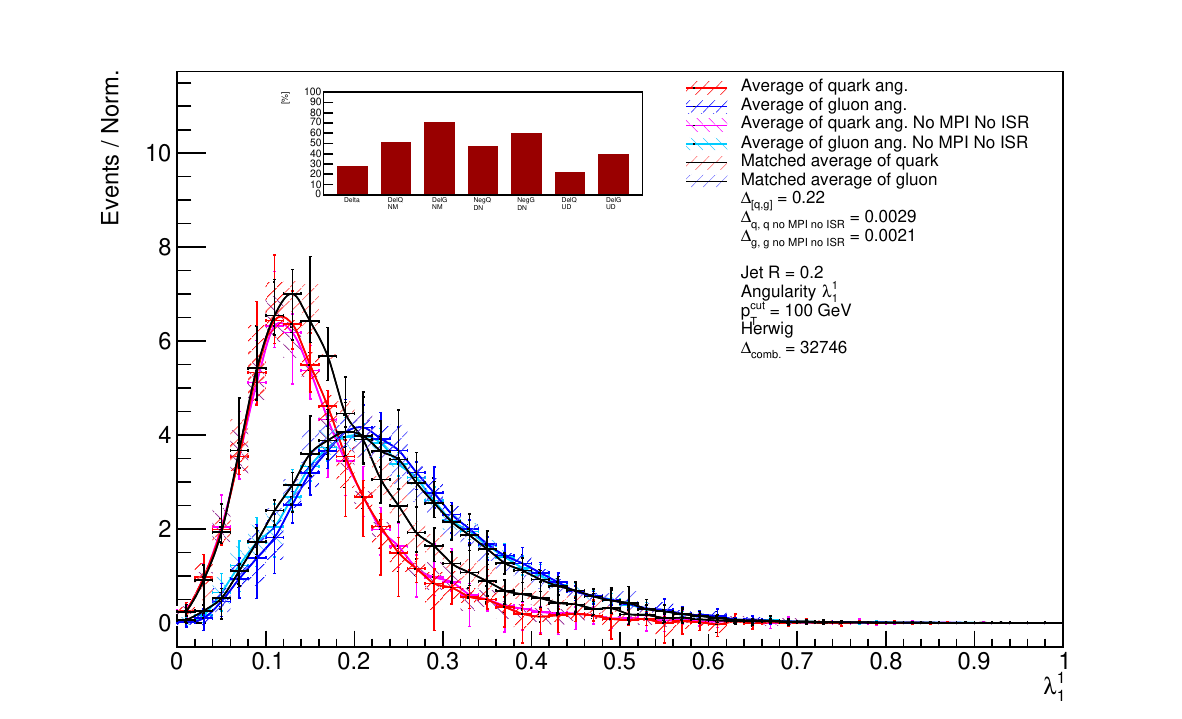} 
        \caption{Quark and gluon averaged angularities $\lambda_{1}^{1}$, $R = 0.2$ with score $\Delta_{\mathrm{comb}} = 32746$. Using \herwig{} event generator, with \ptcut{100}, using the average of 6 energy combinations 
        900--2360, 900--7000, 900--13000, 2360--7000, 2360--13000, 7000--13000~GeV.}
        \label{fig:wildcard5b}
        \end{figure}
\begin{figure*}[ht!]
        \vspace*{-3mm}
        \centering
        \includegraphics[width=0.76\textwidth]{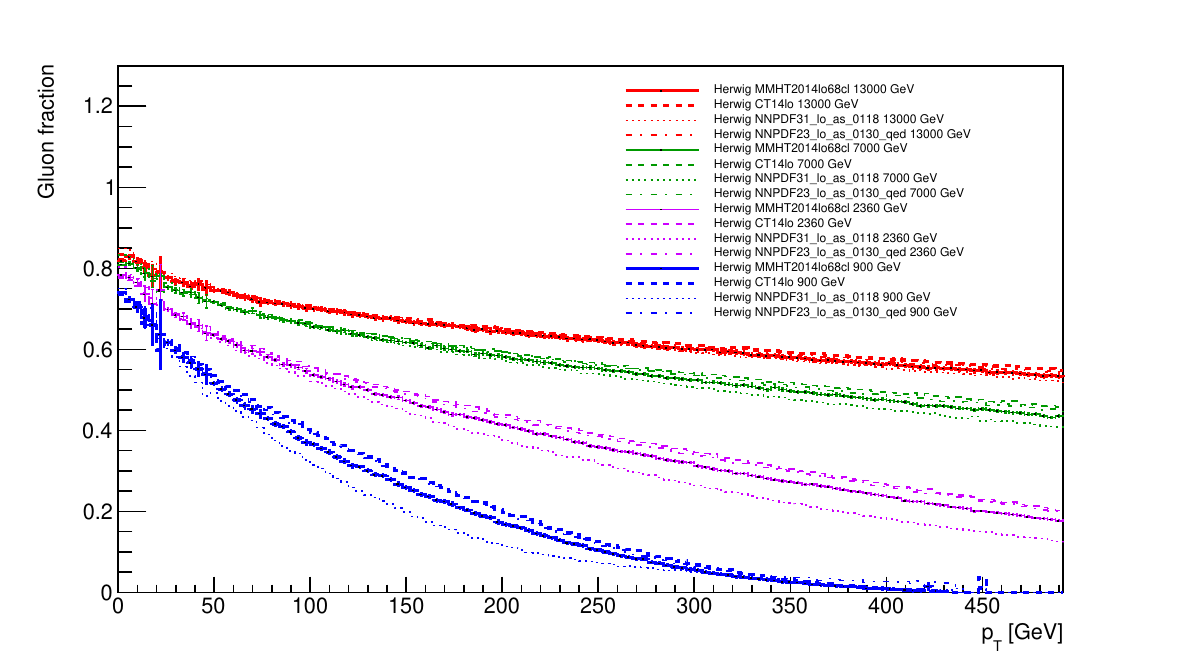}
        \caption{Gluon fractions obtain from \herwig{'s} simulation of 
        proton-proton dijet process without hadronization and parton showering at 
        $\sqrt{s}= 900, 2360, 7000,~$and~$13000$~GeV using different PDF sets.}
        \label{fig:pdf3}
    \end{figure*}
\begin{figure}[ht!]
    \centering
    \includegraphics[width=0.5\textwidth]{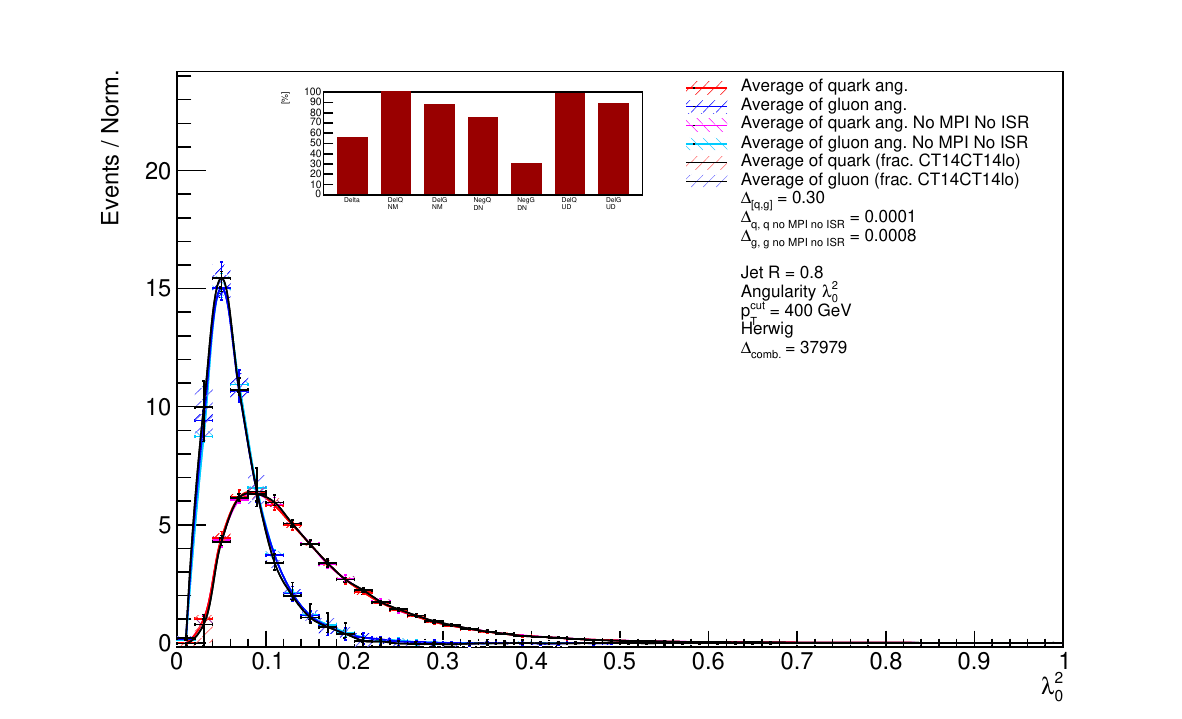} 
    \caption{Quark and gluon averaged angularities $\lambda_{0}^{2}$, $R = 0.8$ with highest score $\Delta_{\mathrm{comb}}=37979$. Using \herwig{} event generator, with \ptcut{400}, using the average of 6 energy combinations 
    900--2360, 900--7000, 900--13000, 2360--7000, 2360--13000, 7000--13000~GeV.}
    \label{fig:best1b}
    \end{figure}
    
    \begin{figure}[ht!]
    \centering
    \includegraphics[width=0.5\textwidth]{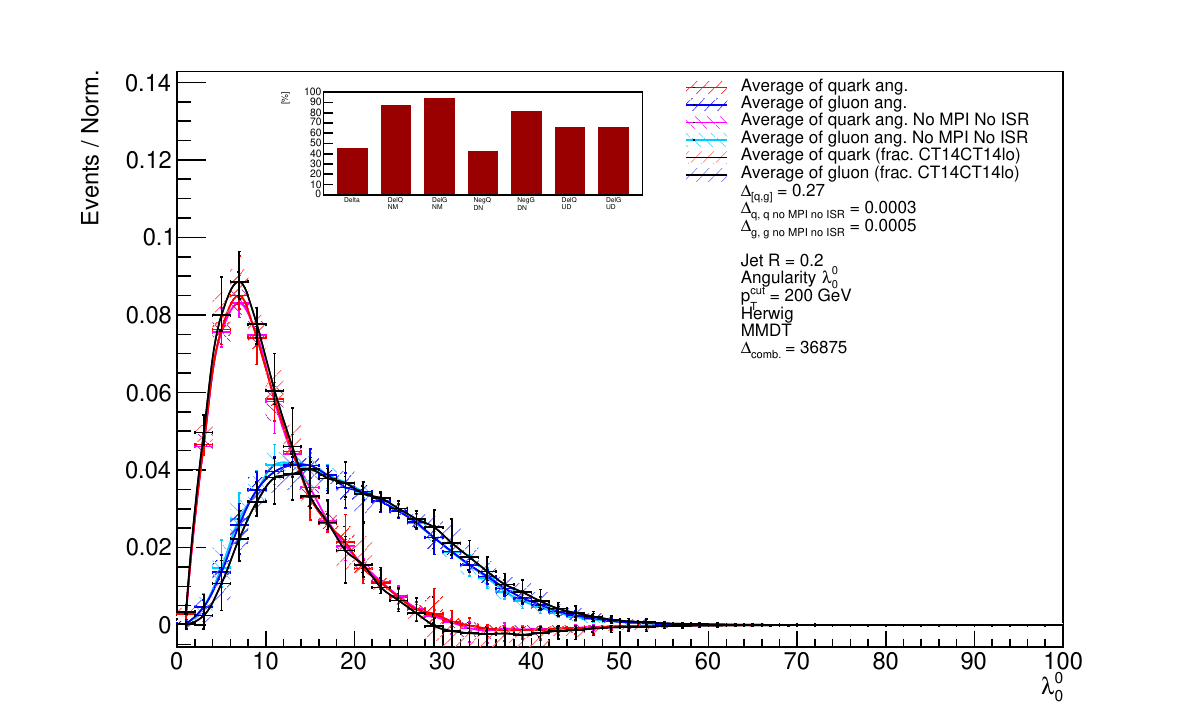} 
    \caption{Quark and gluon averaged angularities $\lambda_{0}^{0}$, $R = 0.2$ with score $\Delta_{\mathrm{comb}}=36875$. Using \herwig{} event generator, with \ptcut{200}, using the average of 6 energy combinations 
    900--2360, 900--7000, 900--13000, 2360--7000, 2360--13000, 7000--13000~GeV.}
    \label{fig:best2b}
    \end{figure}
    
    \begin{figure}[ht!]
    \centering
    \includegraphics[width=0.5\textwidth]{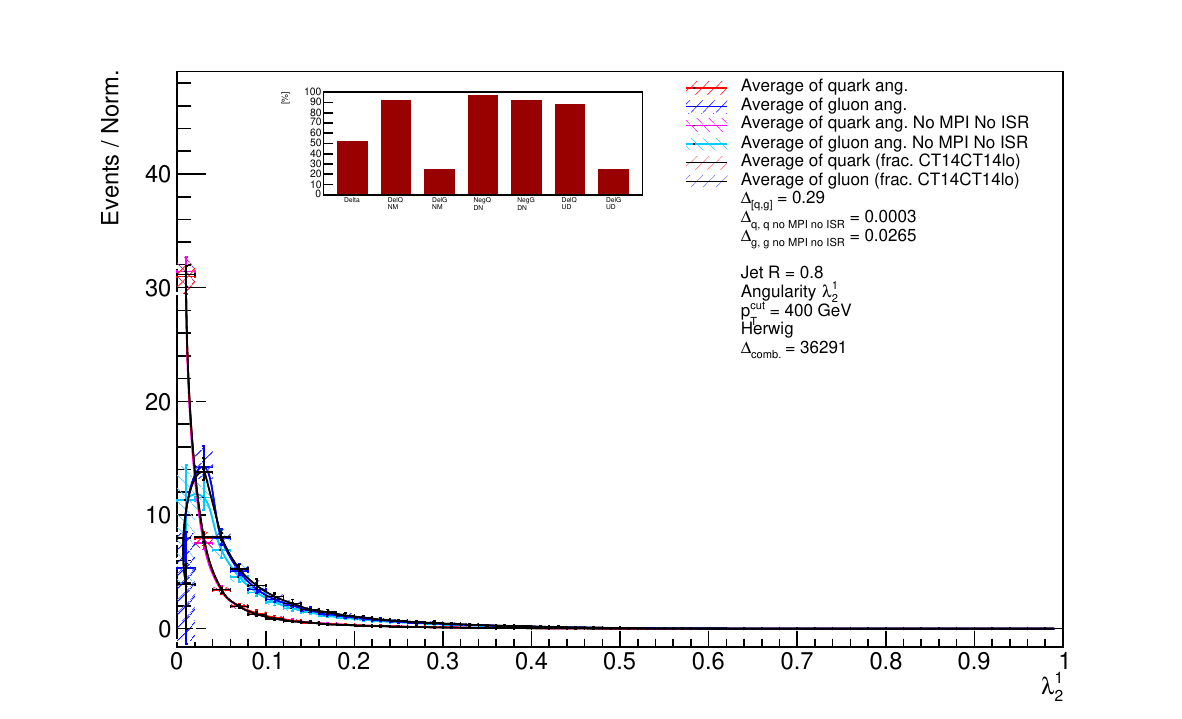} 
    \caption{Quark and gluon averaged angularities $\lambda_{2}^{1}$, $R = 0.8$ with score $\Delta_{\mathrm{comb}} = 36291$. Using \herwig{} event generator, with \ptcut{400}, using the average of 6 energy combinations 
    900--2360, 900--7000, 900--13000, 2360--7000, 2360--13000, 7000--13000~GeV.}
    \label{fig:best3b}
    \end{figure}
    
    \begin{figure}[ht!]
    \centering
    \includegraphics[width=0.5\textwidth]{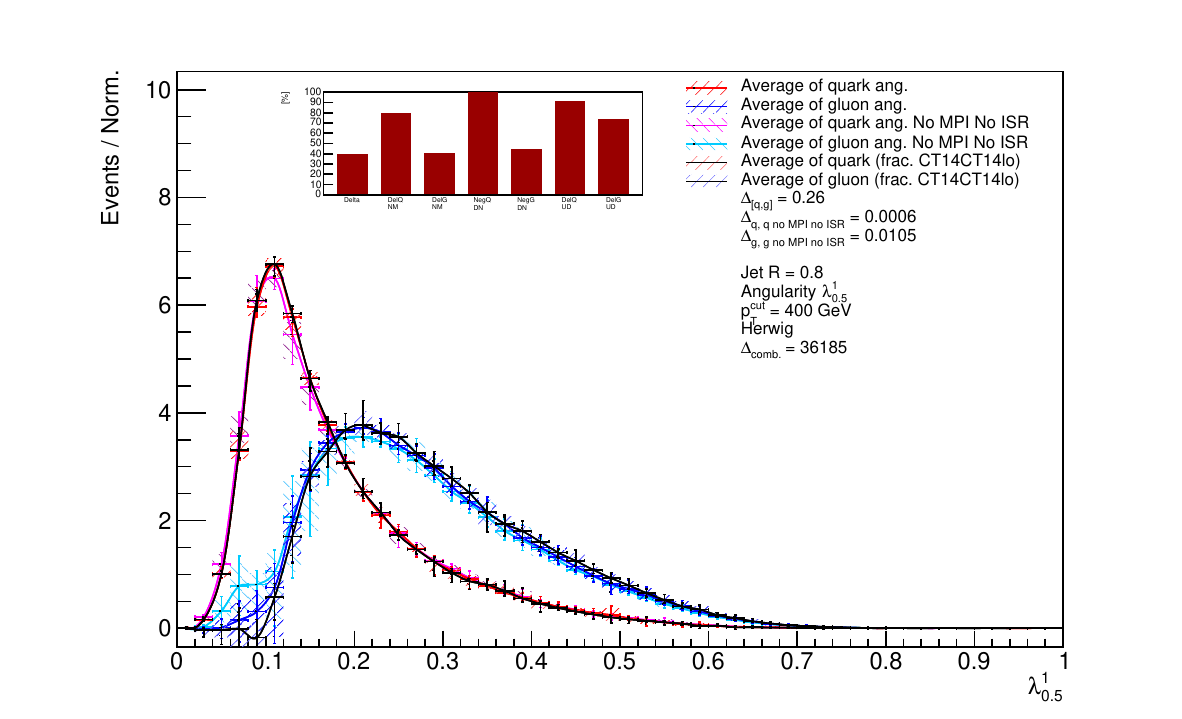} 
    \caption{Quark and gluon averaged angularities MMDT $\lambda_{0.5}^{1}$, $R = 0.8$ with score $\Delta_{\mathrm{comb}} = 36185$. Using \herwig{} event generator, with \ptcut{400}, using the average of 6 energy combinations 
    900--2360, 900--7000, 900--13000, 2360--7000, 2360--13000, 7000--13000~GeV.}
    \label{fig:best4b}
    \end{figure}
    
    \begin{figure}[ht!]
    \centering
    \includegraphics[width=0.5\textwidth]{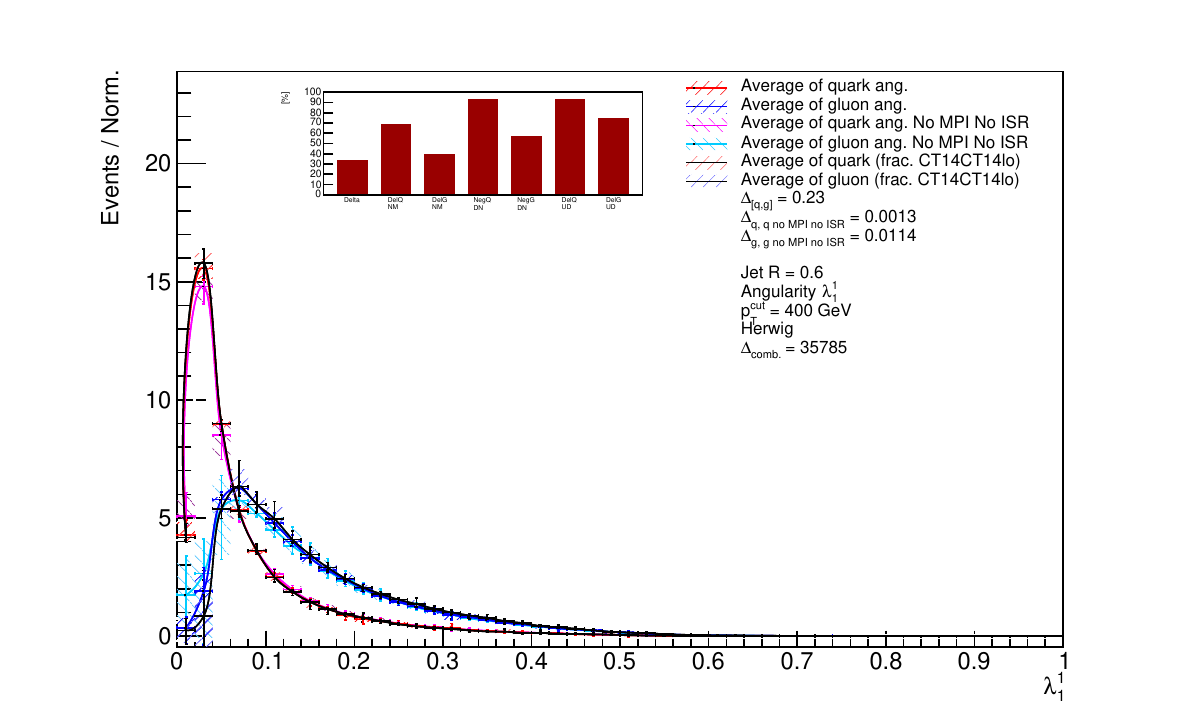} 
    \caption{Quark and gluon averaged angularities $\lambda_{1}^{1}$, $R = 0.6$ with score $\Delta_{\mathrm{comb}} = 35785$. Using \herwig{} event generator, with \ptcut{400}, using the average of 6 energy combinations 
    900--2360, 900--7000, 900--13000, 2360--7000, 2360--13000, 7000--13000~GeV.}
    \label{fig:best5b}
    \end{figure}
        
    \begin{figure}[ht!]
    \centering
    \includegraphics[width=0.5\textwidth]{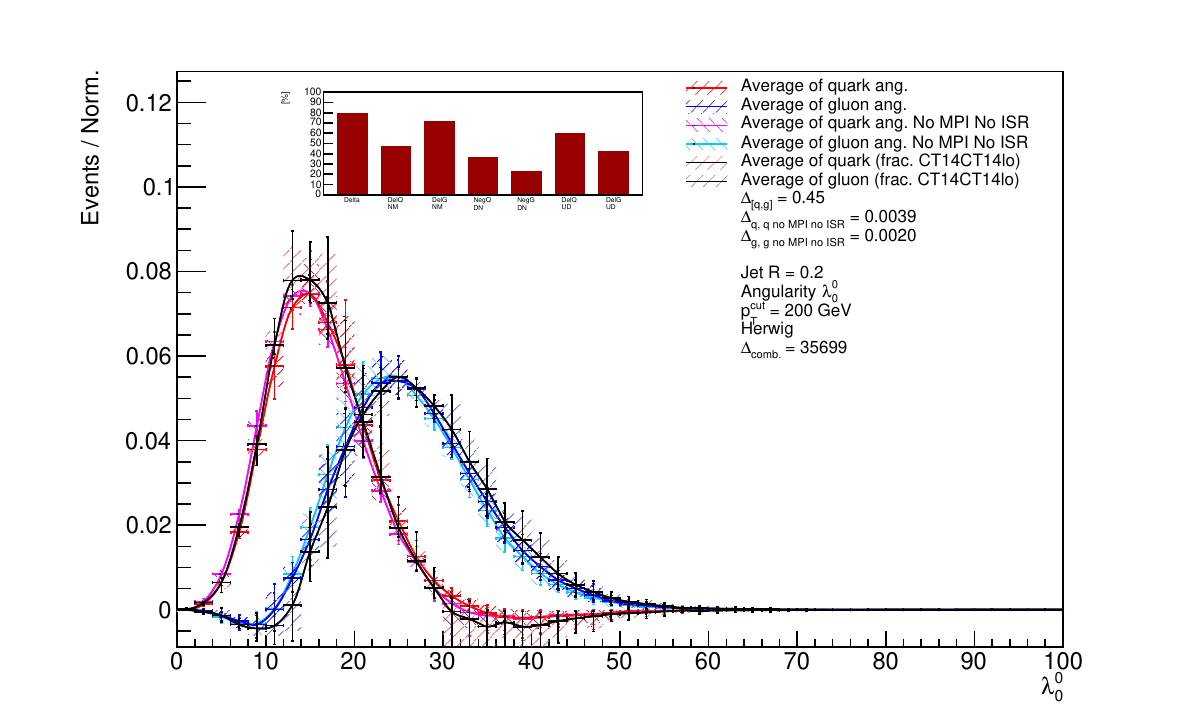} 
    \caption{Quark and gluon averaged angularities $\lambda_{0}^{0}$, $R = 0.2$ with score $\Delta_{\mathrm{comb}}=35699$. Using \herwig{} event generator, with \ptcut{200}, using the average of 6 energy combinations 
    900--2360, 900--7000, 900--13000, 2360--7000, 2360--13000, 7000--13000~GeV.}
    \label{fig:wildcard2b}
    \end{figure}

        \begin{figure}[ht!]
        \centering
        \includegraphics[width=0.5\textwidth]{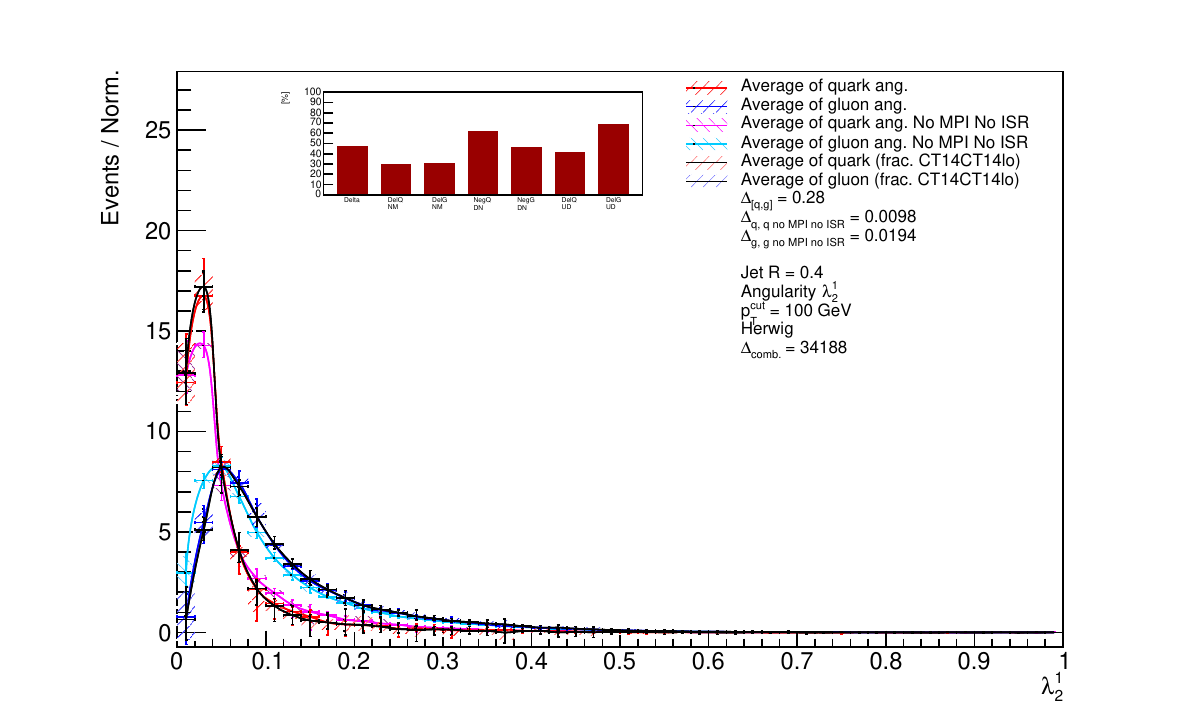} 
        \caption{Quark and gluon averaged angularities $\lambda_{2}^{1}$, $R = 0.4$ with score $\Delta_{\mathrm{comb}} = 34188$. Using \herwig{} event generator, with \ptcut{100}, using the average of 6 energy combinations 
        900--2360, 900--7000, 900--13000, 2360--7000, 2360--13000, 7000--13000~GeV.}
        \label{fig:wildcard3b}
        \end{figure}

        \begin{figure}[ht!]
        \centering
        \includegraphics[width=0.5\textwidth]{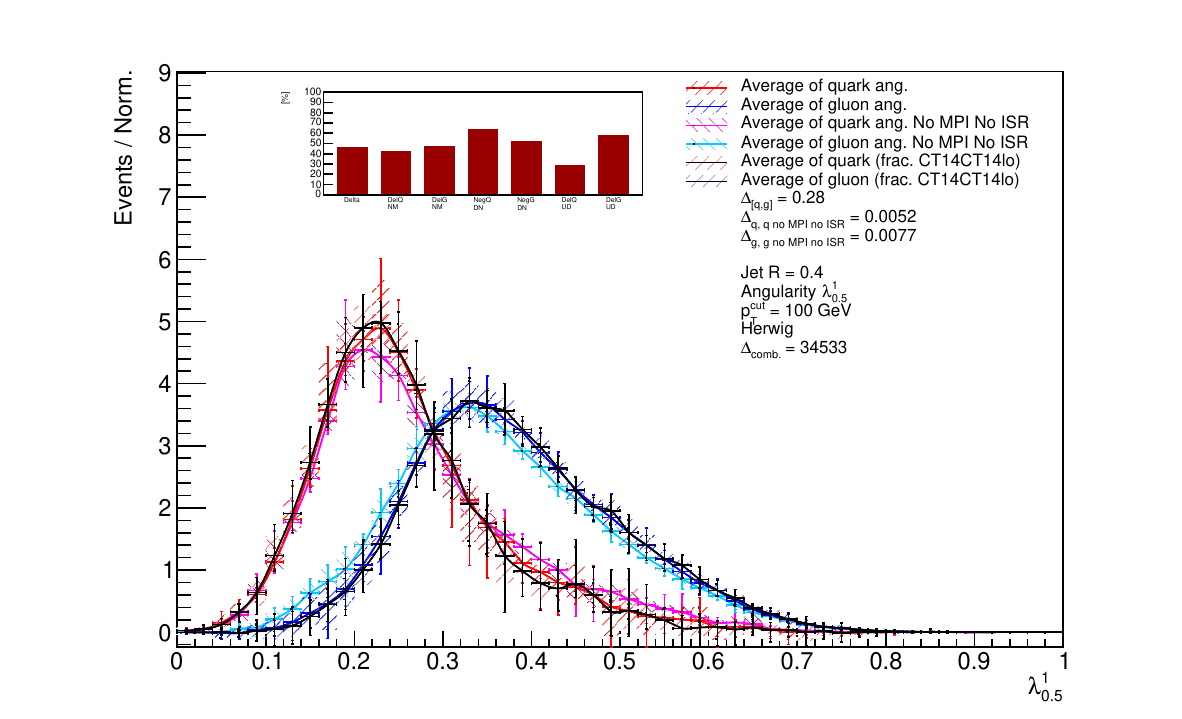} 
        \caption{Quark and gluon averaged angularities MMDT $\lambda_{0.5}^{1}$, $R = 0.4$ with score $\Delta_{\mathrm{comb}} = 34533$. Using \herwig{} event generator, with \ptcut{100}, using the average of 6 energy combinations 
        900--2360, 900--7000, 900--13000, 2360--7000, 2360--13000, 7000--13000~GeV.}
        \label{fig:wildcard4b}
        \end{figure}

        \begin{figure}[ht!]
        \centering
        \includegraphics[width=0.5\textwidth]{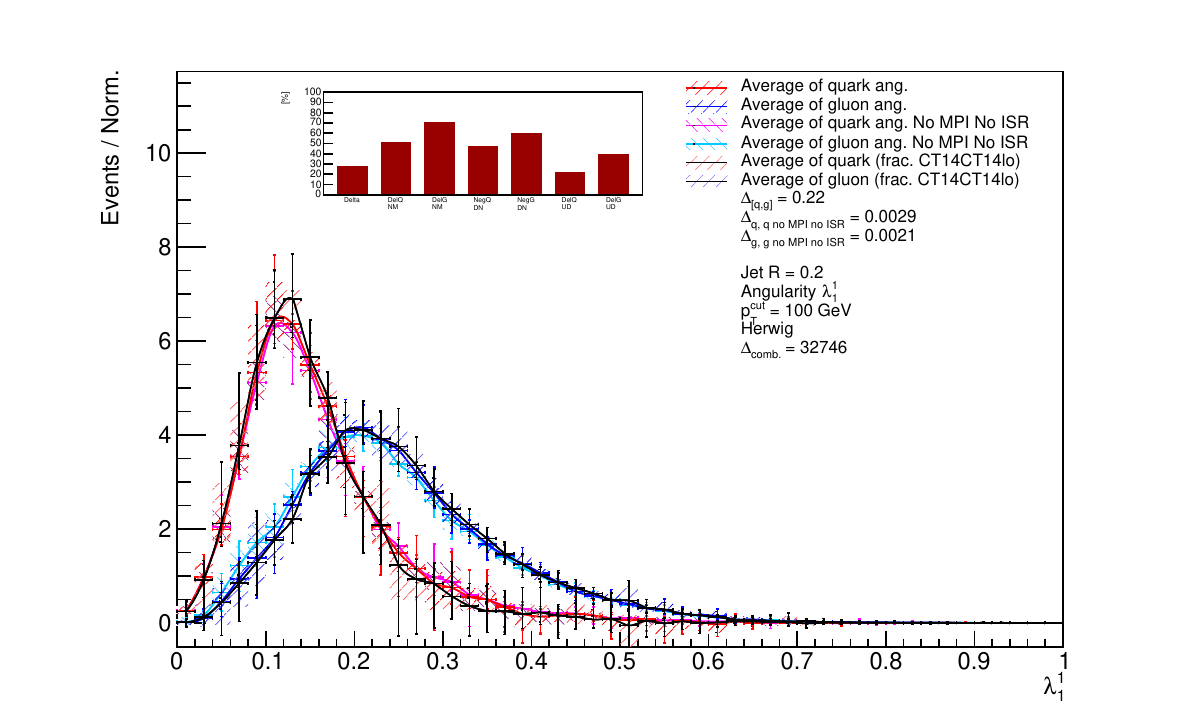} 
        \caption{Quark and gluon averaged angularities $\lambda_{1}^{1}$, $R = 0.2$ with score $\Delta_{\mathrm{comb}} = 32746$. Using \herwig{} event generator, with \ptcut{100}, using the average of 6 energy combinations 
        900--2360, 900--7000, 900--13000, 2360--7000, 2360--13000, 7000--13000~GeV.}
        \label{fig:wildcard5b}
        \end{figure}
\clearpage
\subsection{Scatter Plots}
In this appendix, we show scatter plots of the absolute values of the measures $\Delta_{[q,g]}$, $\Delta_{[q,~q~\mathrm{no~MPI~no~ISR}]}$ 
and $\Delta_{[g,~g~\mathrm{no~MPI~no~ISR}]}$, quark and gluon negativity, 
$\Delta_{[q(s)_{\mathrm{UP}}, \;q(s)_{\mathrm{DOWN}}]}$ and $\Delta_{[g(s)_{\mathrm{UP}}, \;g(s)_{\mathrm{DOWN}}]}$ as functions of radius $R$, angularities, and jet \ptcutwb. The columns in the subpad of the results plots of quark and gluon angularities are calculated as percentiles of these absolute values. 

\begin{figure}[ht!]
    \includegraphics[width=8cm]{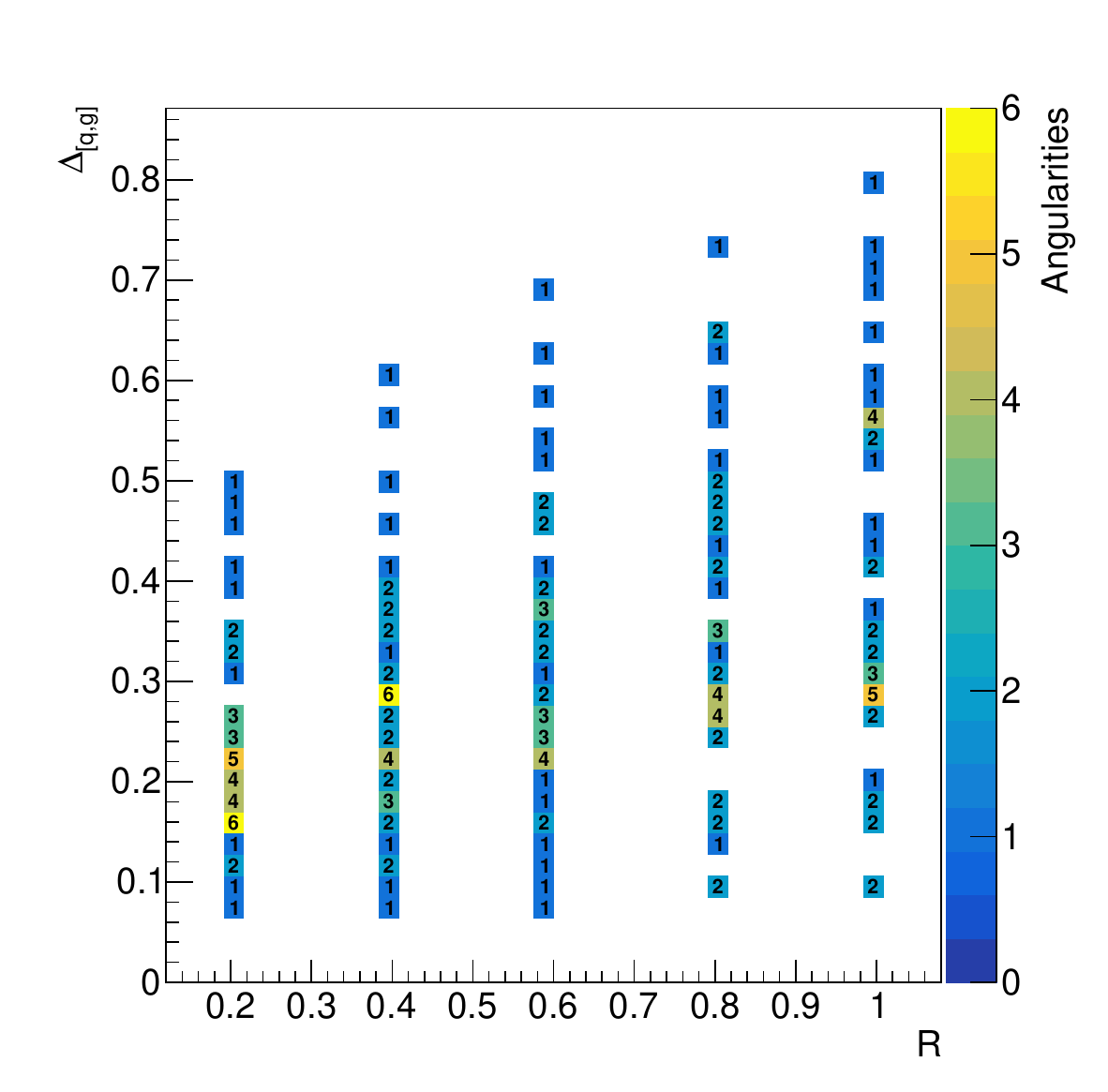}
    \caption{First column scatter plot of $\Delta_{[q,g]}$ as a function of jet radius.}
   \label{fig:negativity_R}
\end{figure}

\begin{figure}[ht!]
    \includegraphics[width=8cm]{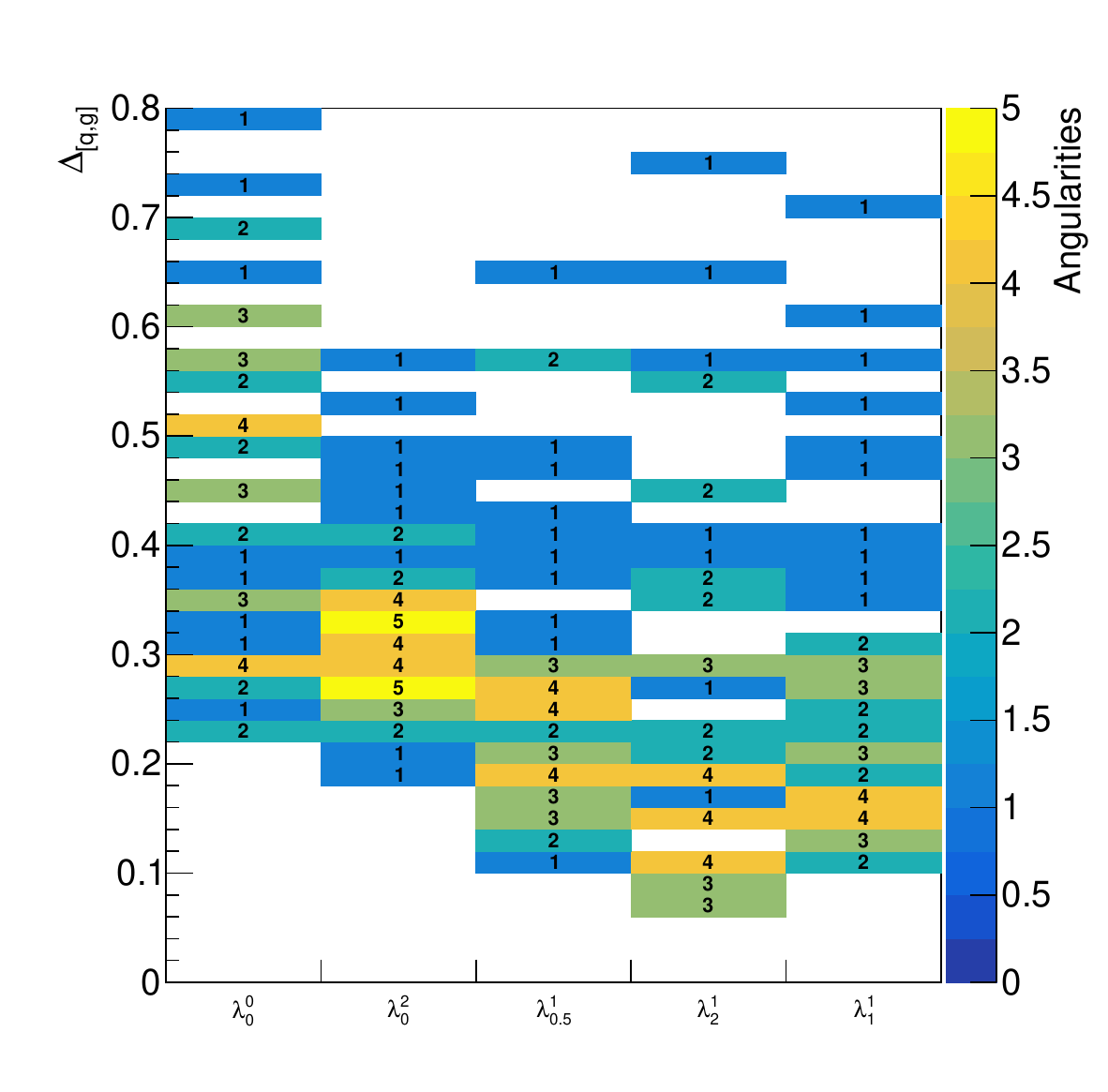}
    \caption{First column scatter plot of $\Delta_{[q,g]}$ as a function of jet angularity.}
    \label{fig:negativity_ang}
\end{figure}

\begin{figure}[ht!]
    \includegraphics[width=8cm]{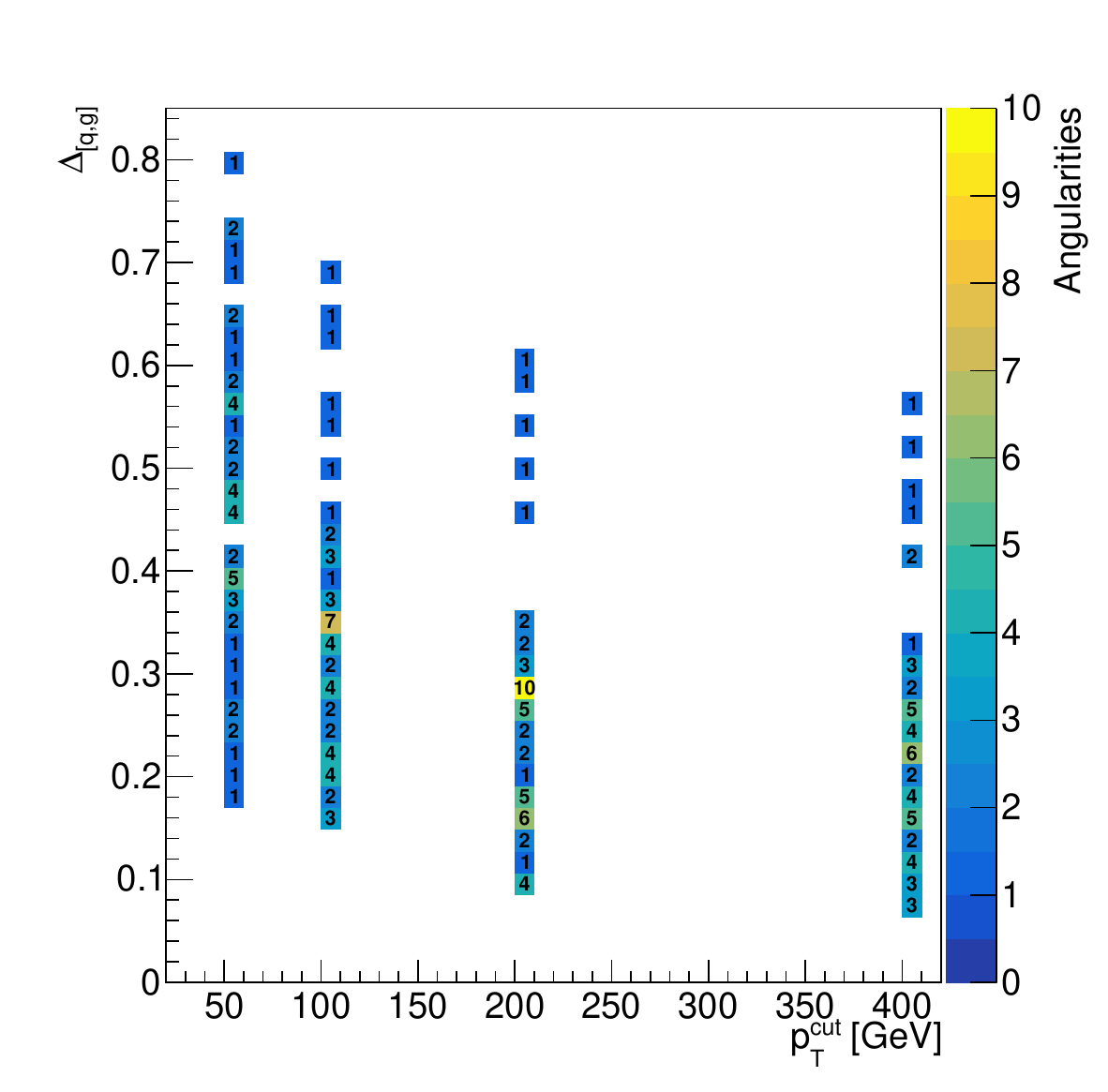}
    \caption{First column scatter plot of $\Delta_{[q,g]}$ as a function of $p_{T}^{\mathrm{cut}}$.}
    \label{fig:negativity_Q}
\end{figure}

\clearpage

\providecommand{\href}[2]{#2}\begingroup\raggedright\endgroup
    
\end{document}